\documentclass[%
 aip,
 amsmath,amssymb,
reprint,%
floatfix,]{revtex4-1}

\usepackage{graphicx}% Include figure files
\usepackage{dcolumn}% Align table columns on decimal point
\usepackage{bm}% bold math
\usepackage[utf8]{inputenc}
\usepackage[T1]{fontenc}
\usepackage{mathptmx}
\usepackage{etoolbox}

\makeatletter
\def\@email#1#2{%
 \endgroup
 \patchcmd{\titleblock@produce}
  {\frontmatter@RRAPformat}
  {\frontmatter@RRAPformat{\produce@RRAP{*#1\href{mailto:#2}{#2}}}\frontmatter@RRAPformat}
  {}{}
}%
\makeatother
\begin{document}

\preprint{AIP/123-QED}

\title{Large/small eddy simulations: A high-fidelity method for studying high-Reynolds number turbulent flows}% Force line breaks with \\
%Large/small eddy simulations: A multi-fidelity method for obtaining DNS-level accuracy from unresolved simulations
%L/SES: A DNS equivalent for high-fidelity simulations of high Reynolds number flows

\author{Arnab Moitro}
\email{arnab.moitro@uconn.edu}
\author{Sai Sandeep Dammati}
%\email{saisandeepdammati@uconn.edu}
\author{Alexei Y. Poludnenko}%
\altaffiliation[Also at ]{Department of Aerospace Engineering, Texas A\&M University, College Station, TX, USA}
\email{alexei.poludnenko@uconn.edu }
\affiliation{%
 School of Mechanical, Aerospace and Manufacturing Engineering, University of Connecticut,
Storrs, CT 06269, USA\\
}%

\date{\today}% It is always \today, today,

\begin{abstract}
% NEW ABSTRACT 250 words
Direct numerical simulations (DNS) are one of the main ab initio tools to study turbulent flows. However, due to their considerable computational cost, DNS are primarily restricted to canonical flows at moderate Reynolds numbers, in which turbulence is isolated from the realistic, large-scale flow dynamics. In contrast, lower fidelity techniques, such as large eddy simulations (LES), are employed for modelling real-life systems. Such approaches rely on closure models that make multiple assumptions, including turbulent equilibrium, small-scale universality, etc., which require prior knowledge of the flow and can be violated. We propose a method, which couples a lower-fidelity, unresolved, time-dependent calculation of an entire system (LES) with an embedded Small-Eddy Simulation (SES) that provides a high-fidelity, fully resolved solution in a sub-region of interest of the LES. Such coupling is achieved by continuous replacement of the large SES scales with a low-pass filtered LES velocity field. The method is formulated in physical space, makes no assumptions of equilibrium, small-scale structure, and boundary conditions. A priori tests of both steady and unsteady homogeneous, isotropic turbulence are used to demonstrate the method accuracy in recovering turbulence properties, including spectra, probability density functions of the intermittent quantities, and sub-grid dissipation. Finally, SES is compared with two alternative approaches: one embedding a high-resolution region through static mesh refinement and a generalization of the traditional volumetric spectral forcing. Unlike these methods, SES is shown to achieve DNS-level accuracy at a fraction of the cost of the full DNS, thus opening the possibility to study high-Re flows.

\end{abstract}

\maketitle

%\tableofcontents

% \section{\label{sec:level1}First-level heading:\protect\\ The line
% break was forced \lowercase{via} \textbackslash\textbackslash}
\section{Introduction}\label{sec:Introduction}

Turbulence is one of the most ubiquitous physical processes in the Universe: it controls the flow of nutrients in the living organisms \citep{ghosal2000turbulent}, enables efficient combustion in engines \citep{sadykova2021influence}, drives mixing and circulation in the planetary oceanic and atmospheric processes \citep{d2014turbulence}, and plays one of the central roles virtually in all astrophysical contexts, from the birth of stars in turbulent interstellar clouds \citep{mac2004control} to their deaths in violent supernovae explosions \citep{poludnenko2019unified}. Highly nonlinear dynamics of turbulence is often further complicated by the presence of other, tightly coupled physical processes. As a result, detailed understanding of such systems requires first-principles modeling by solving the governing equations for the full range of scales, which are dynamically important in a given system. In most cases, such range of scales is large, with the overall cost of a fully-resolved computation increasing with the Reynolds number~\citep{jimenez2000large} as $O(\mathrm{Re}^3)$, thus placing first-principles modeling of realistic systems outside the reach of modern computing.

%Despite the challenges in driving turbulence, DNS remains  a high-fidelity method for studying the fundamental physical properties of turbulence. However, its computational cost rapidly increases  rendering the method prohibitively expensive for problems of engineering relevance such as the design and optimization of aircraft engine combustors. 

Traditionally, there have been two paths to resolve this difficulty. The first path is to forego the resolution of small scales, focusing the computational resources on the system-specific large-scale dynamics, while capturing the effects of small scales with the appropriate subgrid-scale (SGS) models. This gave rise to a multitude of reduced-order methods, including Reynolds Averaged Navier Stokes (RANS), Large Eddy Simulations (LES), etc. \citep{pope2001turbulent,sagaut2005large,tennekes1972first,spalart2009detached,zhiyin2015large,corson2009industrial} The second path concentrates the limited computational resources on fully resolving the small-scale flow structure in the context of Direct Numerical Simulations (DNS) \citep{moin1998direct}. Traditionally, DNS and the low-order methods have remained largely disconnected, even though their conceptual synergy is critical as the first-principles understanding provided by the DNS enables the development of the accurate SGS models, which are central to the reduced-order methods.

In order for the DNS to be able to both capture the dissipation range and also develop an inertial range within a computationally limited range of scales, the energy-generating large scales must be excluded from a simulation. This, however, immediately leads to two key limitations. First, when the DNS are disconnected from a realistic large-scale flow, they are typically restricted to more idealized, canonical configurations. Second, aside from a more specialized case of decaying turbulence, such idealized DNS must include a mechanism for generating and sustaining turbulence, thus mimicking the effect of the missing large scales that feed the energy to smaller scales. At the same time, a particular forcing method may not necessarily be physically equivalent to the action of large scales, and furthermore it can critically affect the structure of the resulting turbulence \citep{quadrio2016does,dhariwal2022forcing,das2022effect,federrath2010comparing,john2019solenoidal}. As a result, any such external forcing must be treated carefully.

%Turbulent flows are ubiquitous both in Nature and in practical systems, from energy generation, propulsion, and chemical processing applications \citep{REFS} to the dynamics of oceans and the atmosphere \citep{REFS} to astrophysical systems, such as star forming regions \citep{REFS} and stellar interiors \citep{REFS}. Direct numerical simulations (DNS) are high-fidelity methods for the computational modelling of turbulent flows. These simulations directly solve the Navier-Stokes equations from first principles on a grid that resolves all the dynamically relevant scales of turbulent motion. However this leads to a very high computational cost, and DNS studies often sacrifice the realism of large scales to focus on studying the fundamental properties of the small scales of turbulence. These simulations are generally restricted to low Reynolds number flows in simplified geometries, such as shear flows or homogeneous isotropic turbulence (HIT) in periodic domains \citep{moureau2011large}. 

%As turbulence continuously dissipates its kinetic energy, pure unforced DNS simulations are unable to reach a statistically steady state, and the range of Reynolds numbers accessible in such cases is further reduced. A majority of DNS studies overcome this problem by adding an external forcing term to the momentum equation. 

%\subsection{Turbulence forcing methods}

Over the years, a wide range of approaches for turbulence forcing have been introduced. Broadly, these can be divided into two classes: (i) generalized methods for canonical flows; and (ii) system-specific methods, which aim to incorporate the large-scale flow information. The first class treats forcing as a computational technique, which is intended to inject kinetic energy into the domain at a prescribed rate to sustain turbulence. Such methods are the simplest and most widely used, however the resulting forcing is idealized in nature, whose sole objective is to compensate the energy loss due to turbulent dissipation. As a result, they implement forcing without a connection to any particular large-scale flow. Thus, they crucially rely on the assumption of the universality of small scales, which are independent of the details of a large-scale flow.
%The resulting small-scale turbulent flow may only be a subset of the myriad of possibilities arising from the variety of conditions possible at the larger scales. 
%However these methods are applied to specific large scale flows, and are not very versatile to capture an arbitrary large scale flow.  

Such generalized forcing techniques involve two major sub-classes: volumetric and boundary forcing. One of the earliest methods for direct volumetric forcing is based on turbulence stirring in spectral space. A method proposed by \citet{eswaran1988examination} (also see \citet{Lemaster2009}) is based on the stochastic forcing of the smallest wavenumbers of the flow. Similar approach has also been used, for example, by \citet{seror2001radiated}, which is based on replenishing the dissipated energy back at the large scales, by \citet{siggia1978intermittency}, which freezes the Fourier coefficients of velocity at low wavenumbers, and by \citet{carati1995representation}, which apply forcing proportional to the velocity in the wavenumber space. Broadband spectral-space forcing methods have also been proposed \citep{witkowska1997numerical}, which multiply the Fourier coefficients of the velocity in a chosen wavenumber range by a scalar to compensate for the energy dissipation.

The main advantage of such spectral forcing techniques is their ability to target specific scales for energy injection with a precisely defined energy injection spectrum. While such approaches are natural for solvers operating in the Fourier space, for physical-space solvers they introduce significant cost of the Fourier transforms, and their implementation can be non-trivial in the domains with non-periodic boundaries or complex geometry.
%Anisotropic driving has also been proposed by \citet{alvelius1999random} which uses a prescribed force spectrum in a way that gives control over the anisotropy.
These limitations are partially addressed by the methods, which apply volumetric forcing directly in physical space. For instance, linear forcing, proposed by \citet{lundgren2003linearly} for incompressible flows, and later extended by \citet{rosales2005linear}, applies a forcing term proportional to the velocity at every point in the domain. This approach is simple to implement but it energizes all wavenumbers, which at least in principle implies that the inertial range dynamics is no longer independent of driving. An extension of linear forcing has also been proposed \cite{lundgren2003linearly}, which applies forcing only to the small wavenumbers, however, it requires low-pass filtering of the velocity field. Linear forcing method was further extended to compressible flows by \citet{petersen2010forcing}, who introduced separate coefficients to force the solenoidal and dilatational components of the velocity field. This, however, also requires the Fourier decomposition of the velocity field, and thus it sacrifices the simplicity of linear forcing, restricting the method to periodic domains. Moreover, for all types of linear forcing, the resultant driving is isotropic, and extensions to inhomogeneous flows and non-periodic domains are not straightforward.

The second sub-class of the generalized forcing methods involves approaches, which rely on the boundary, rather than volumetric, forcing. These techniques generate shear-driven turbulence by superposing small fluctuations over an initial flow and allowing turbulence to evolve \citep{stanley2002study,hawkes2012petascale} in the presence of a globally imposed velocity gradient. Similar methods, known as the shearing box approximation, are also used to model turbulence in various differentially rotating astrophysical systems, e.g., accretion disks \citep{hawley1995local} and stellar interiors \citep{prat2016shear}. All such methods, however, provide less control over the resulting turbulence properties compared to the volumetric forcing, and they also introduce by design significant mean shear, which may not be always desirable. Furthermore, such shear-driven turbulence may require significant amount of time to develop, which can restrict the accessible range of Reynolds numbers and exacerbate the computational cost.% Also, these methods consider a simple shear flow where the imposed fluctuations are not connected to any spectrum or large-scale structures.

The generalized forcing techniques described above form the cornerstone of the existing studies of the canonical flows, in particular homogeneous, isotropic (HIT) and shear-driven/wall-bounded turbulence. At the same time, they suffer from two key limitations. First, they are restricted to just a few idealized flow configurations, which may not represent the full richness of possible turbulent flow conditions present in realistic systems. Second, such DNS can achieve only moderate Reynolds numbers with the computational cost growing rapidly as $O(\mathrm{Re}^3)$. For instance, largest DNS to-date of the HIT reach the Taylor-scale Reynolds number\citep{buaria2020b} of $Re_{\lambda}=1300$, which requires an extremely large grid $N^3=12,288^3$. Similarly, the largest DNS of a turbulent channel flow has a friction Reynolds number, $Re_\tau=5200$ and the grid size of $10,240\times7,680\times1,536$ \citep{lee2015direct}. Such turbulent conditions, which are at the limit of what is possible with modern computing resources, are still far from the realistic turbulent flows even in engineering systems, let alone atmospheric or astrophysical environments. As a result, modern DNS are not able to address many open questions regarding the nature of high-Re turbulence.

% thereby incorporating the physics of the cascade of energy from the large to the small scales.
Contrary to the methods described above, which are decoupled from any external flow, the second major class of turbulence forcing techniques aims to either (i) drive the fully resolved flow solution using a realistic model of the large scales of a particular flow of interest %based on the theoretical, experimental, or numerical results, 
or (ii) by directly embedding the high-fidelity solution in a large-scale flow.
%Such approaches connect forcing with the properties of a specific external flow.
An example of the former is a method proposed by \citet{rah2018derivation}, which focuses on the flow in turbulent round jets, with the forcing term based on the mean centerline jet velocity. Although this method achieves forcing, which is anisotropic and is based on an experimentally determined external flow field, it is only applicable in triply periodic domains for a particular region of the jet, and the forcing parameters must be determined individually for other regions of interest or for different flow configurations. An example of a direct embedding approach was described by \citet{elnahhas2020flow}, who applied a DNS patch in the near-wall region of an LES of an incompressible channel flow to serve as the wall model. Statistical quantities from the LES were used to provide the large-scale flow information for the DNS, while the DNS data were used to obtain the wall shear stress for the LES. %Such approach, however, is applicable only near the walls in a wall-bounded flow.
Overall, both of these methods are developed with the aim of capturing a specific flow, whether it is a jet or a wall-bounded flow. %and not for any arbitrary boundary conditions.  %Additionally, all these methods require some prior understanding of the flow.

Hybrid LES-DNS methods, analogous to the approach of \citet{elnahhas2020flow}, have also been proposed for planar jets and decaying turbulence, where a certain sub-region in an LES is refined to the DNS resolution using static (SMR) or adaptive mesh refinement (AMR) techniques \citep{macart2021embedded,sirignano2023dynamic}. The embedded high resolution sub-region is not forced in any special way with the kinetic energy from large scales in the LES entering it via the boundaries, thus effectively forcing this region. These studies, however, primarily aim at training machine learning algorithms to obtain LES closures, and thus they focus on the large-scale or mean-flow behavior, such as the resolved kinetic energy spectra, mean dissipation, or mean streamwise jet velocity, rather than on the accuracy of the small scales in the refined region. Other studies \citep{towery2020scaling,christopher2022high} have also described such direct AMR-based embedding of a high-fidelity solution, in particular focusing on exploring optimal characteristics of the mesh refinement approaches, such as the interpolation and integration schemes or the coarse-fine transition-zone depth.

This second class of turbulence forcing methods partially addresses one of the major limitations of the generalized forcing techniques described above by directly injecting the large-scale flow information in the small-scale calculation, and thus at least potentially allowing one to treat arbitrary large-scale flows. %This, however, is done either using an analytical model for the large scales introduced volumetrically, as in the method of \citet{rah2018derivation}, which lacks the flexibility since such model has to be established for each particular flow of interest. Or, alternatively, such forcing can be achieved by injecting the large-scale flow through the boundaries of the directly embedded high-fidelity computation, as in the approach of \citet{elnahhas2020flow} or the SMR/AMR-based techniques \citep{macart2021embedded,sirignano2023dynamic,towery2020scaling,christopher2022high}.
At the same time, the ability of such methods to provide DNS-level solution accuracy in the general case, especially in terms of the higher-order metrics, and most importantly, their ability to reach high Reynolds numbers beyond what can be achieved in the traditional DNS, has not been established.

%It may also be challenging to extend this procedure to other types of flows such as driven turbulence or reacting flows. %However, none of these methods can ensure a high-level of fidelity for the small scales or compute exact closures for an LES.  

%A further shortcoming of all the above approaches is that the largest scales of the flow being studied are confined to the largest dimensions of the simulation domain, i.e. the domain needs to capture the largest scales of the flow. This restricts the range of Reynolds numbers that can be accessed in the simulation.
It is also worth noting that atmospheric modeling community has employed a range of multi-fidelity approaches, which also aim to couple a low-fidelity, large-scale flow information with a higher fidelity small-scale calculation, albeit not a DNS. For instance, spectral nudging techniques \citep{waldron1996sensitivity,von2000spectral,radu2008spectral} use the large-scale data obtained from a global reanalysis model for forcing specific wavenumbers in the interior of a smaller regional model. Regional modelling is treated as a downscaling problem, as it incorporates large-scale information while allowing small-scale dynamics to develop. Similar method has also been applied outside of weather modelling, for example to infer physical parameters in turbulent flows \citep{di2018inferring}. While providing a great deal of flexibility in incorporating the large-scale flow information, spectral nudging also has several important limitations. First, it operates in spectral space, and extensions to physical space on the non-periodic domains are not straightforward due to the requirement of spectral decomposition of various modes. Furthermore, this approach generally does not aim to achieve DNS-level fidelity of the small-scale solution, and its ability to provide such level of accuracy has largely not been explored. Finally, in recent years, attempts have been made to extend nudging techniques, as a form of data assimilation, from the traditional weather modeling to more fundamental turbulence studies \citep{clark2020synchronization,hasegawa2023continuous}.

%In fact, large-scale flow data can be sparse, as is typically irregularly distributed weather stations. While this is sufficient for climate or weather modelling which are boundary or initial value problems, the behaviour of such an approach is not clear when the small-scale simulation has to conform to a known full large-scale field.

Other methods employed in weather modelling, similar to spectral nudging, include super-parameterization techniques \citep{grabowski1999crcp,grabowski2004improved}, in which a simulation with a finer resolution is embedded in a large-scale calculation and forced volumetrically using mean velocities and gradients obtained from the latter. As in spectral nudging, the goal is not to achieve DNS-level small-scale solution fidelity, but rather enable coupling between various subgrid parameterizations and observe their effects on the large scales. Furthermore, such methods have been demonstrated only for specific boundary conditions and they only ensure that the small-scale fields match the large-scale solution on average. As in the case of other embedding methods described above, super-parameterization approaches cannot ensure a high-level of accuracy for the small scales or provide exact closures for an LES.

%Another method to reconstruct a turbulent flow field by nudging a simulation using known large-scale data (obtained using a measured field or another full-scale DNS) has also been proposed by . The nudging is applied only at a few selected points in the domain, thereby resulting in a small-scale forcing scheme, instead of large-scale forcing. However, this method is designed to be a data assimilation technique for a machine learning based reconstruction of a flow and not as a method for solving the governing equations. As such, this method assumes the availability of high-fidelity data. Moreover, the capability of this method to produce small-scale statistics of turbulence has not been demonstrated.

Finally, we note that a method for performing a resolved simulation, albeit for a short duration to obtain turbulence statistics, using initial conditions from the LES data has been proposed by \citet{yeung2020advancing}. It requires, however, a fully resolved simulation to be set up in the entire LES domain, which can be computationally prohibitive for the problems of engineering relevance, even if performed for a short duration. %Additionally, unlike that technique, the method developed in this work incorporates forcing in the smaller simulation and is not restricted to statistically stationary equilibrium flows or to spectral space solvers.

%To the authors' knowledge, there is no efficient and universally applicable method that can act as a large-scale forcing model in a high-fidelity simulation capturing all essential known and unknown features of the external flow. Additionally, no techniques exist for developing generalized LES closure models without prior knowledge of the nature of the flow. Considering these limitations, there are three main goals of the present study-
%Considering the lack of a versatile forcing strategy, there are three main goals of the present study-

\begin{figure*}
 \centerline{\includegraphics[width=0.6\textwidth]{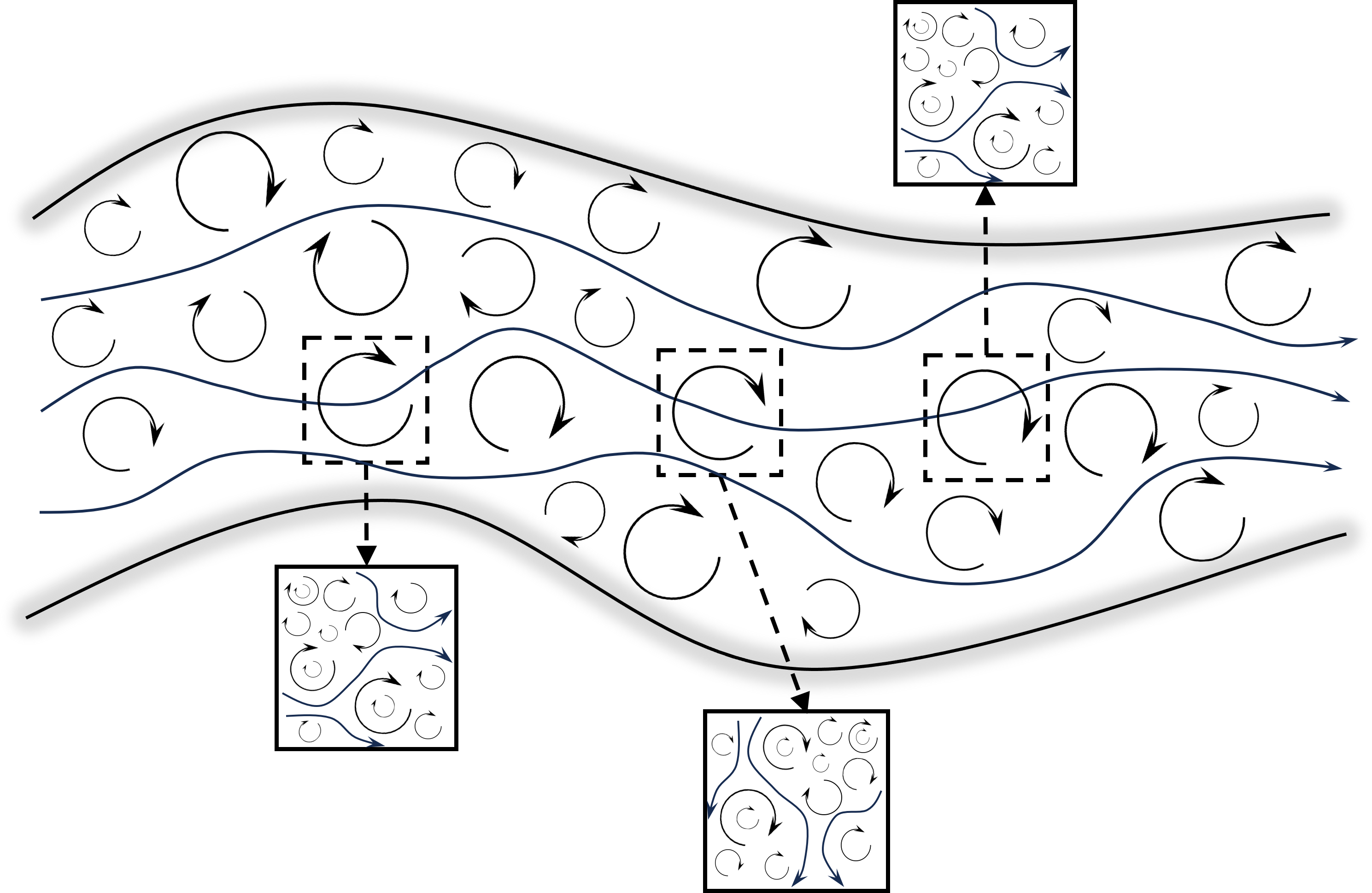}}
 \caption{L/SES concept. Large scales of a complex flow are captured by the low-cost, low-fidelity simulation. Detailed small-scale structure in the chosen regions of interest is then obtained using the L/SES method.}
\label{fig:LSESillustration}
\end{figure*}

Here we aim to combine the strengths, while addressing the limitations, of these two classes of turbulence forcing methods by formulating a general approach to perform fully resolved simulations of turbulence in a sub-region of interest of an arbitrary large-scale flow, as illustrated in Fig.~\ref{fig:LSESillustration}. In particular, the key three requirements of the proposed method are:
\begin{enumerate}
    \item It should be able to recover the actual instantaneous, as well as the time-averaged or statistically representative, small-scale structure in a particular region of the larger-scale flow. Such embedded calculation must provide the solution accuracy comparable to that of the DNS in terms of all turbulence metrics, from the global ones to the higher-order ones reflective of the nonlinear turbulent structure. Ability to recover instantaneous fields, although a stringent requirement, can be desirable in many practical situations, e.g., reacting flows, where instantaneous deviations of the velocity gradients from their mean may result in local rapid mixing of the reactants, auto-ignition, etc.
    \item No prior knowledge of the overall flow structure should be required, including assumptions of homogeneity, isotropy, equilibrium, or small-scale universality.
    \item In order to be useful in the studies of practical flows, proposed method should be formulated in physical space, and it should be applicable for arbitrary initial and boundary conditions, as well as system geometry.
\end{enumerate}
Ultimately, the goal is to develop a method, which will be able to achieve practically relevant turbulent Reynolds numbers in a fully resolved simulation beyond what is accessible in modern DNS.

Considering these objectives, we introduce a multi-fidelity simulation method, which starts with a lower-fidelity, time-dependent calculation of the flow of interest, in which a smaller region is identified based on its dynamical, physical, or other properties for high-fidelity modeling. The key requirement for this under-resolved calculation is that it captures the large-scale flow structure with the desired accuracy. Fidelity of the small scales in this calculation is not essential, so long as the inaccuracies in the small-scale solution do not degrade the accuracy of the large-scale flow. While in this work we primarily focus on the LES for such large-scale calculations, other lower fidelity alternatives, e.g., unsteady RANS, can also be considered in the future as they can further reduce the cost of the proposed method, thus increasing its practical utility. Flow data is extracted by low-pass filtering the velocity field in the sub-region of interest in the LES, and it is used to force the companion high-fidelity Small Eddy Simulation (SES), which resolves all small scales and uses the LES data as an effective model for the large scales.

Resulting Large/Small Eddy Simulation approach, or L/SES, can be viewed as a generalization of the second class of forcing methods, which aim to inject the actual large-scale flow structure into the high-fidelity simulation. At the same time, as will be shown in the discussion below, proposed method differs in several key aspects from the approaches discussed above. In particular, L/SES employs both boundary and volumetric forcing, which we find to be critical to achieve high solution accuracy. Furthermore, in contrast to the spectral nudging approaches, L/SES is formulated in physical space for complex domains with non-periodic boundaries, where Fourier decomposition would not be practical. As a result, L/SES relies on explicit filtering to separate different spectral components and thus to provide volumetric forcing formulated in physical space. Finally, in contrast to the traditional data assimilation approaches, in particular nudging techniques, which aim to inject the true reference flow field (obtained in experiments, observations, or another high-fidelity simulation) in order to achieve synchronization between such true data and the obtained solution, L/SES sets effectively an opposite goal. It crucially relies on a low-fidelity reference solution. Thus, we aim to demonstrate that using such low-fidelity large-scale forcing data, it is possible to recover the small-scale flow field with high accuracy.

Below we describe the L/SES approach, and we validate it for steady and unsteady HIT in an {\it a priori} sense, where the large-scale data is provided by filtered fields of a fully resolved large-scale DNS, rather than LES. Detailed {\it a posteriori} tests for a wide range of Re will be presented in the follow-up paper. We consider a range of metrics, including global kinetic energy content, time-averaged and time-dependent kinetic energy spectra, probability density functions (PDF) of the small-scale quantities, such as enstrophy and dissipation, as well as subgrid-scale dissipation, which is directly relevant for the LES closures.

We also contrast the L/SES with the two major types of turbulent forcing methods discussed above. First, we describe a generalization of the first class of turbulence forcing approaches, namely traditional volumetric forcing, in which the time-dependent energy injection rate is directly determined from the sub-region of interest in an LES instead of being prescribed as a free parameter. The small-scale calculation, which is intended to represent that LES sub-region, is then forced using the standard stochastic spectral forcing \citep{eswaran1988examination,Lemaster2009,poludnenko2010interaction}. While this method can capture some aspects of the time-varying evolution of the LES, the large-scale flow complexity is reduced to a single scalar parameter, similar to the traditional spectral forcing approach.  Second, we also compare the L/SES flow solution with the SMR/AMR-based approach similar to those describe above \citep{macart2021embedded,sirignano2023dynamic,towery2020scaling,christopher2022high}, in which a high-resolution region is directly embedded in a low-fidelity simulation using static mesh refinement.

The remainder of the paper is organized as follows. Details of the L/SES method are described in \S~\ref{sec:lses}. Alternative approaches, including the generalization of the stochastic spectral forcing and the SMR-based method, are summarized in \S~\ref{sec:altmethods}. The numerical method as well as the details of the DNS, which provides the forcing data, and the associated SES are given in \S~\ref{sec:num3}. Comparison of various turbulence metrics for all three methods is presented in \S~\ref{sec:Results}. We end with a discussion of the L/SES method, its implications for turbulence studies, limitations, and open questions, in \S~\ref{sec:discussion}, along with the concluding remarks.

\section{Large/small eddy simulation approach}
\label{sec:lses}

%--------------------------------------------------------------------------------------------
\subsection{Method description}
\label{ssec:method}

The key objective of the proposed L/SES approach is to utilize in the companion high-fidelity SES calculation the full large-scale flow-field information available in the LES. In other words, large scales in the SES should match the large-scale LES flow both at the boundaries and in the entire volume of the SES domain to ensure seamless transfer of the flow structure from the LES to the SES.

To achieve this, the SES solution is continuously adjusted to replace the flow structure produced by the SES on scales larger than a chosen filter scale, $\Delta$, with the known `true' large-scale LES flow field. A low-pass filtering operation is used to separate the large-scale from the small-scale components both in the LES and the SES. Mathematically, the $i^{\textrm{th}}$ component of the velocity perturbations introduced in the SES is
\begin{equation}
    \delta u_i = \tilde{u}_{i,LES}- \tilde{u}_{i,SES},
\label{eq:deltau}
\end{equation}
where $\tilde{u}_{i,LES}$ and $\tilde{u}_{i,SES}$ are the LES and SES velocity fields, respectively, filtered at the scale $\Delta$. Such velocity perturbations are then applied after each time step, giving the new SES velocity field at the time step $n+1$
\begin{equation}
    u_{i,SES}^{n+1}=u_{i,SES}^{n}+\delta u_i= (u_{i,SES}^{n}- \tilde{u}_{i,SES}^{n}) + \tilde{u}_{i,LES}^{n}.
\end{equation}
This operation removes the large-scale component of the SES flow field ($\tilde{u}_{i,SES}^{n}$ term) and replaces it with the large-scale velocities from the LES. Thus, the resulting field is a superposition of small scales produced by the SES (term in the parenthesis) and the large scales of the LES. We emphasize that low-pass filtering ensures that all LES scales $> \Delta$, including ones larger than the SES domain size $L_{SES}$, are involved in driving the SES solution. Formally, this formulation is equivalent to the traditional nudging techniques \citep{clark2020synchronization} with an important distinction, which concerns the reference solution. While in the data assimilation approaches it represents true data, typically obtained in experiments or observations, in L/SES it is a low-fidelity solution, in which only the large-scale component can be assumed to be accurate.

Here we are considering weakly compressible flows with the turbulent Mach number $Ma_t \sim 0.05$. Therefore, forcing the velocity field is sufficient. At the same time, for compressible flows, momentum should be perturbed instead through a similar procedure
\begin{equation}
    \delta (\rho u_i) = \overline{(\rho u_i)}_{LES}- \overline{(\rho u_i)}_{SES},
\label{eq:deltamom}
\end{equation}
where $\overline{(\rho u_i)} = \bar{\rho}\tilde{u_i}$, and $\overline{(...)}$ denotes Reynolds filtering.

%This method of forcing turbulence is analogous to the method suggested by \citet{von2000spectral}, who applied perturbations to a given variable $\Psi$ in a small-scale model through spectral nudging as
%\begin{equation}
%    \delta \Psi = \sum_{j=-J_a}^{J_a} \eta_j [\alpha_j^a(t)-\alpha_j^m(t)] \exp\Big(i\frac{jx}{L_x}\Big).
%\end{equation}
%This equation is shown for 1D, where $x$ is the coordinate, $L_x$ is the domain length, $\alpha_j^m$ and $\alpha_j^a$ represent the Fourier coefficients of wavenumber $j$ in the small-scale and large-scale models respectively, $\eta_j$ is the confidence in the accuracy of the large-scale model at wavenumber $j$, $J_a$ gives the range of wavenumbers in the large-scale model, and $i=\sqrt{-1}$. Introducing such perturbation removes the large scales produced in the small-scale model in the spectral range $[-J_a, J_a]$ and replaces them with the corresponding data from the large-scale model. In our case, however, we are interested in a method applicable in physical space in the domains with an arbitrarily complex geometry and non-periodic boundaries, where Fourier decomposition would not be practical. Hence instead, the separation of different spectral components is achieved by filtering. Finally, in L/SES, all scales in LES larger than the filter scale are considered accurate, and therefore we do not employ any scale-dependent confidence factor, $\eta_j$. 

In the proposed approach, there are three key parameters, which need to be chosen appropriately:
\begin{itemize}
\item \textit{Filter scale, $\Delta$:} In the explicitly filtered LES, $\Delta$ is set by definition. In the implicitly filtered ones, $\Delta$ needs to be chosen sufficiently larger than the LES grid size, $dx_{LES}$, to ensure that all scales $>\Delta$, which will be passed to the SES, are minimally affected by any grid effects and the associated numerical dissipation. As a result, specific choice of $\Delta$ would depend on the details of the numerical solver employed. For the finite-volume, Godunov-type solver used here, which is based on the constrained transport upwind (CTU) scheme with the piecewise-parabolic (PPM) spatial reconstruction \citep{colella1990multidimensional, saltzman1994unsplit,colella1984piecewise} (also see \S~\ref{ssec:DNS} below), we find $\Delta \sim 16dx_{LES}$ is an appropriate choice.
\item \textit{SES domain size, $L_{SES}$:} Criteria for choosing $L_{SES}$ are two-fold. On one hand, $L_{SES}$ must be close to $\Delta$ to minimize the computation cost since scales $> \Delta$ are replicated from the LES. On the other, some overlap between $\Delta$ and the largest SES scales is necessary to ensure that the SES field is consistent with the LES and the overall method is stable. We find that the choice of $L_{SES}=2\Delta$ provides optimal results while minimizing the range of scales covered both in LES and SES.
%At the same time, this choice allows a sufficient number of small scales for the L/SES to develop its own dynamics. Data from LES at scales larger than $\Delta$ is considered accurate and will be used to drive the L/SES, which in turn produces more accurate results for scales smaller than $\Delta$ in the specific LES sub-region. 
%since the method involves filtering L/SES fields which gives higher errors and computational overheads at scales close to $L_{L/SES}$
\item \textit{SES grid resolution, $dx_{SES}$.} Since the goal of the SES is to resolve fully all dynamically important flow scales, the choice of $dx_{SES}$ would typically be governed by the physics of the flow of interest. In the case of non-reactive, homogeneous, isotropic turbulence, which we consider in the tests described below, natural choice is $dx_{SES} = \eta/2$, where $\eta$ is the Kolmogorov scale of the flow. Such resolution was shown in prior studies to be sufficient to capture turbulence structure accurately near the dissipative scale \citep{donzis2010,jagannathan2016,yeung2018}. The caveat is that in the unsteady, non-equilibrium turbulent flows, such as considered in one of the tests described below, $\eta$ can vary in space and time, and thus the choice of $dx_{SES}$ must either be made conservatively or the time-varying SES resolution must be implemented.
\end{itemize}
The best performance of the method is expected when both $\Delta$ and $L_{SES}$ are $\gg dx_{SES}$, i.e., in the flows with sufficiently large Re$_{\lambda}$. In particular, in the SES, sufficient separation must exist between the dissipative and forcing scales to allow small scales to develop proper physical dynamics, which is not simply an imprint of the large-scale forcing by the LES flow field. Furthermore, when $dx_{SES}$ is close to $dx_{LES}$, LES can be viewed just as an under-resolved DNS, and the combined cost of L/SES will be comparable to that of a DNS.

Finally, we note that the approach described here implements one-way coupling between the large and small scales. This places constraints on the types of flows, which can be studied using the L/SES. In particular, large scales must not be controlled by the small-scale dynamics, more specifically, there should not be a pronounced up-scale energy transfer from the small scales \citep{towery2016spectral,o2017cross,towery2014spectral}. Extension of the L/SES method to a two-way coupled approach, in which SES effectively constrains the LES closures, is the subject of future work.

%--------------------------------------------------------------------------------------------
\subsubsection{Filter choice}
\label{sssec:filter}

\begin{figure}
  \centerline{\includegraphics[width=0.45\textwidth]{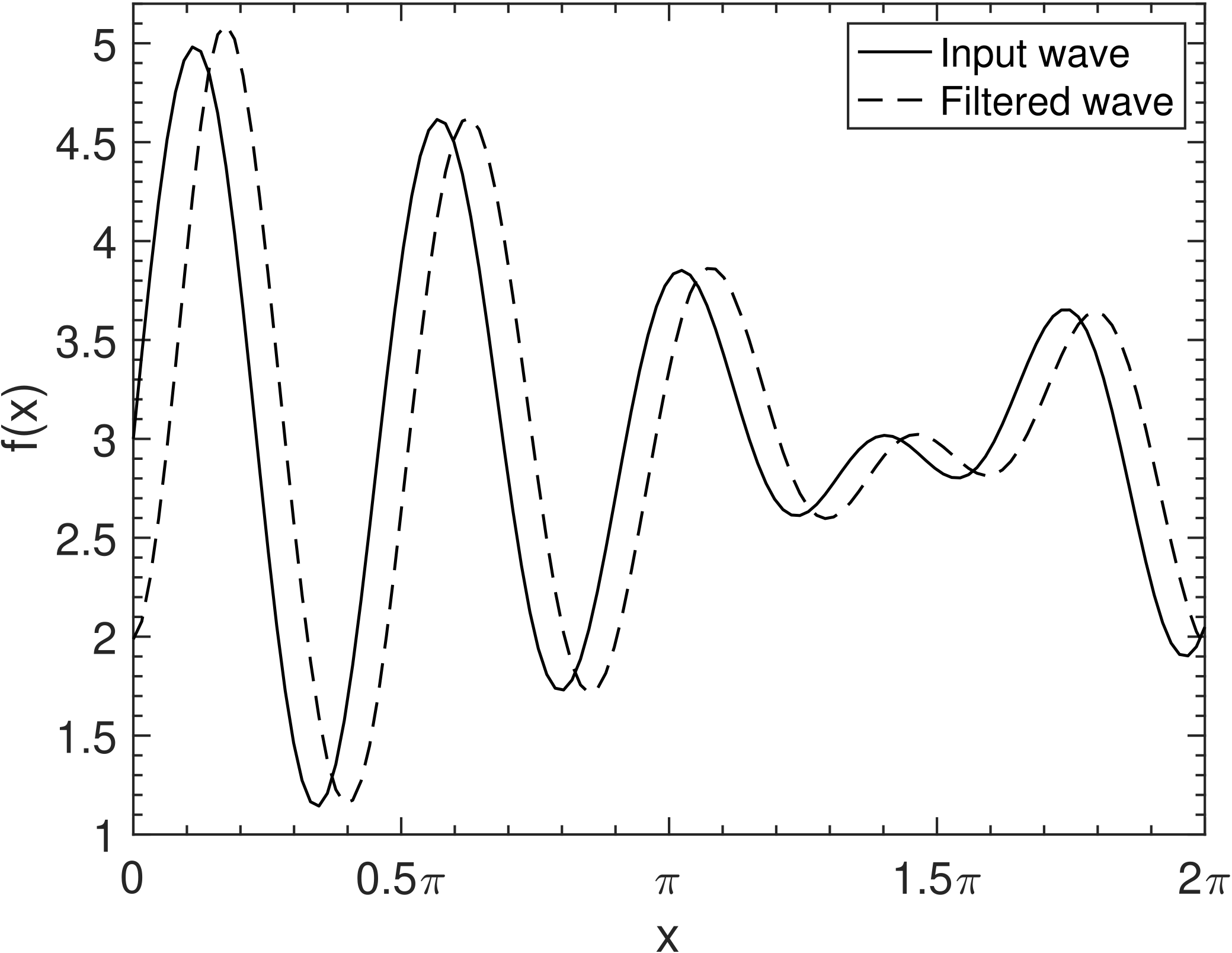}}
  \caption{Phase shift caused by the 4$^{\mathrm{th}}$-order Chebyshev filter.}
\label{fig:phaseshift}
\end{figure}

One of the key steps in the L/SES approach is the low-pass filtering, which isolates the large-scale flow structure in the LES and SES. An appropriate choice of such filter is critical both for the accuracy and stability of the overall method. In particular, the filter must exhibit the following four characteristics:
\begin{enumerate}
\item \textit{Spectral sharpness:} The filter must provide a spectral response as sharp as possible. 
\item \textit{Spatial locality:} The filter should also be local in space or, more specifically, it should have compact support. A non-local filter, such as the Fourier-transform-based spectral filter, when applied on a non-periodic domain would produce aliasing errors, which would affect the small scales disproportionately as they have inherently lower energy. Furthermore, a local filter is computationally more efficient.
\item \textit{Zero phase shift:} The filter should not produce a phase shift in the filtered variables. The LES and SES velocity fields are comprised of a wide range of wavenumbers. Filters with a non-zero phase response would cause motions at different scales, i.e., with different frequencies, to separate and create sharp gradients, thus destabilizing the overall method. An example of such dispersion is shown in Fig.~\ref{fig:phaseshift} for the 4$^\mathrm{th}$-order Chebyshev type-II filter \citep{daniels1974approximation}. In particular, shown is a low-frequency one-dimensional (1D) sinusoidal input wave (solid line), which lies in the pass band of the filter, similar to an LES flow field, along with its filtered output (dashed line). Since the input signal does not have any high-frequency component, the filtered output should coincide with the input. Instead, the phase-shift caused by the filter results in a non-zero velocity perturbation $\delta u$, calculated by the subtraction of these two waves. Such non-zero $\delta u$, if injected into an SES calculation, would destabilize it breaking the solution.
\item \textit{Commutativity:} Finally, an additional property to consider while choosing an appropriate filter is its commutativity with the derivative operators. Although this is not strictly necessary for the L/SES, this property is desirable if the same filter were to be used also as an explicit LES filter.
\end{enumerate}

%Secondly, as we wish to use the method in conjunction with an LES, we would ideally like the filter to commute with the derivative, although this is not strictly necessitated by the L/SES

Fourier-transform-based spectral filter would naturally provide the sharpest spectral response. At the same time, we seek a general method, which should be practical in non-periodic domains, ideally with an arbitrary geometry. While spectral filter can be applied directly in physical space, it is non-local and in a non-periodic domain it would produce a phase shift, which destabilizes the solution.
%While the spectral cutoff filter can be applied by transforming it to the physical space, it is non-local and thus it would it introduce errors in the non-periodic domains.% Fourier transform is non-local, truncated, and has negative weights in the domain, thus providing an inappropriate spectral response and destabilizing the method.
Therefore, here we adopt a higher-order differential filter of the form
\begin{equation}
    \phi=\bar{\phi}+(-1)^{n/2}a^n\nabla^n\bar{\phi},
\label{eq:Germano2}
\end{equation}
where $n$ is the filter order, $\phi$ is a variable being filtered, $\bar{\phi}$ represents the filtered value, $\nabla$ is the usual differential operator, and $a=\Delta/\sqrt{40}$ is the filter parameter chosen such that the filter transfer function becomes $0.5$ at the cutoff width $\Delta$.\citep{bose2011explicitly} The 2$^{\mathrm{nd}}$-order version of this filter, i.e., for $n=2$, was proposed by \citet{germano1986differential} for LES. This filter, although implicitly defined and computationally expensive, is chosen because it is local in space, and it gives both monotonic and non-dispersive frequency response. Furthermore, we use density-weighted Favre filtering $\tilde{u}_i=\overline{\rho u_i}/\bar{\rho}$. This, along with eq.~(\ref{eq:deltamom}), allows this method to be applicable also in the compressible flow regimes.

The transfer function of the above filter is
\begin{equation}
    \hat{f}(k)=\frac{1}{1+a^nk^n},
\label{eq:tf_Germano}    
\end{equation}
which can be found using the Laplace transform of eq.~(\ref{eq:Germano2}). Here $k$ is the wavenumber, and $a$ is defined in eq.~(\ref{eq:Germano2}) above.

\begin{figure*}
  \centerline{\includegraphics[width=1.0\textwidth]{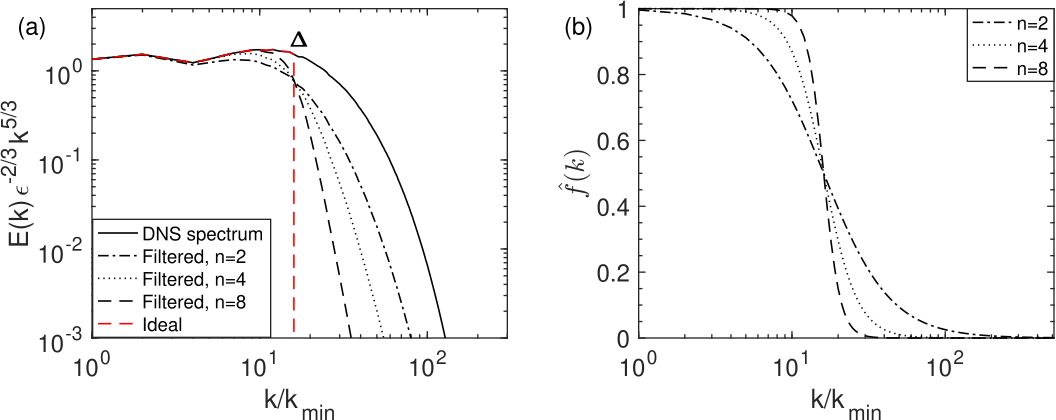}}
  \caption{$\textit{(a)}$ Spectral leakage in the differential filter at various filter orders. $\textit{(b)}$ Transfer functions of the differential filters of orders $n=2$, $4$, and $8$ given by eq.~(\ref{eq:Germano2}).}
\label{fig:leakage}
\end{figure*}

Figure~\ref{fig:leakage}(a) shows the compensated kinetic energy spectrum (solid line) in the DNS of HIT turbulence, described below, with the spectra of the same velocity field filtered using the 2$^{\textrm{nd}}$-, 4$^{\textrm{th}}$-, and 8$^{\textrm{th}}$-order differential filter given by eq.~(\ref{eq:Germano2}). Dashed red line shows the result for an ideal spectrally sharp filter. Figure~\ref{fig:leakage}(b) shows the corresponding transfer functions for these three differential filters. For all filter orders, not all small scales below the filter width $\Delta$ are removed, while some scales $> \Delta$ are suppressed. Crucially, such spectral leakage decreases considerably from the 2$^{\textrm{nd}}$- to 8$^{\textrm{th}}$-order filter as the transfer function becomes progressively narrower in the wavenumber space.

Such non-ideal spectral response of the filter introduces errors in the velocity or momentum perturbations injected into the SES calculation (eqs.~\ref{eq:deltau} or \ref{eq:deltamom}). In particular, eq.~(\ref{eq:deltau}) can be re-written for the actual injected velocity perturbations as
\begin{eqnarray}
    \delta u_i & = &
    \Big\{(\widetilde{u^*_{i}})_{LES} - (\widetilde{u_{i}^{LS}})_{LES} + (\widetilde{u_{i}^{SS}})_{LES}\Big\} \nonumber \\ & - &
    \Big\{(\widetilde{u^*_{i}})_{SES} - (\widetilde{u_{i}^{LS}})_{SES} + (\widetilde{u_{i}^{SS}})_{SES}\Big\} \\
    & \approx & 
    (\widetilde{u^*_{i}})_{LES} - (\widetilde{u^*_{i}})_{SES} - (\widetilde{u_{i}^{SS}})_{SES}.
\label{eq:real perturbation}
\end{eqnarray}
    %&=& (u_{i,LS,act}^{DNS} - u_{i,LS,act}^{L/SES}) +(\delta_{LS}^{L/SES}-\delta_{LS}^{DNS})+ (\delta_{SS}^{DNS}-\delta_{SS}^{L/SES})  \label{eq:termwise perturbation}
Here, $\widetilde{u^*_{i}}$ represents the velocities filtered with an ideal spectrally sharp filter, while $\widetilde{u_{i}^{LS}}$ and $\widetilde{u_{i}^{SS}}$ represent the suppression of large scales $> \Delta$ and enhancement of small scales $< \Delta$, respectively, due to the spectral leakage in LES and SES (Fig.~\ref{fig:leakage}). Since in the LES small scales $< \Delta$ are either completely absent, or they are suppressed due to the lack of resolution, the term $(\widetilde{u_{i}^{SS}})_{LES}$ can be viewed as small. On the other hand, since large scales $\gtrsim \Delta$ are close in LES and SES by design, terms $(\widetilde{u_{i}^{LS}})_{LES}$ and $(\widetilde{u_{i}^{LS}})_{SES}$ approximately cancel each other. Therefore, spectral leakage will result in a systematic error $(\widetilde{u_{i}^{SS}})_{SES}$ in the velocity perturbations, which will effectively act to modify the small scales in the SES degrading (or in some cases, destabilizing) the solution. Hence, there is a need to select the filter carefully. 

%superscripts LS and SS represent the true large- and small-scale velocities, respectively. For an ideal filter, both terms in the second bracket of eq. \ref{eq:real perturbation} will be filtered out and vanish. Simultaneously, such a filter will also obtain the true large scales perfectly (i.e. $ \widetilde{u_{i}^{LS}}=u_{i}^{LS}=\tilde{u}_i$ for all fields), thus recovering eq. \ref{eq:deltau}. A filter with spectral leakage however, will result in some loss of small-scale information from the L/SES as well as contamination from the large L/SES scales while computing $\delta u$. Likewise, it will also result in the loss of some large-scale information that needs to be extracted from the LES. This can be a source of inaccuracies and could potentially destabilize the simulation. The non-ideal spectral response might not cause problems for a priori simulations (forced from a filtered DNS field instead of an LES) because the spectral leakage from filtering the DNS could get compensated by a similar leakage from the filtering of the L/SES fields. However for a posteriori simulations, this would not be the case due to the absence of higher wavenumbers in the LES field. This would cause the small-scale spectral leakage from the L/SES to remain unbalanced, and a systematic error in the forcing term, due to a non-zero term in the second bracket of eq. \ref{eq:real perturbation}.

In the numerical tests discussed below, we use the 8$^{\textrm{th}}$-order differential filter. While higher-order filters can be constructed, they would require a significantly larger stencil while only marginally improving the spectral response. Thus, the choice of $n=8$ was found to be optimal. Resulting filter satisfies other requirements outlined above. In particular, it is spatially local as it can be calculated for each cell using a finite width stencil. And it has zero phase response, as the imaginary part of the transfer function in eq.~(\ref{eq:tf_Germano}) is zero.

Since solving the implicit eq.~(\ref{eq:Germano2}) in three dimensions (3D) is computationally expensive, we implement the filter in a directionally split way with the data being filtered in 1D successively along all three directions, as suggested by \citet{edoh2016discrete}. In 1D, the resulting diagonally dominant sparse system of linear equations is solved directly using LU decomposition. A 2$^{\mathrm{nd}}$-order, central, finite difference operator is used to discretize eq.~(\ref{eq:Germano2}).

%--------------------------------------------------------------------------------------------
\subsubsection{SES boundary conditions}
\label{sssec:sponge}

\begin{figure}
\includegraphics[width=0.4\textwidth]{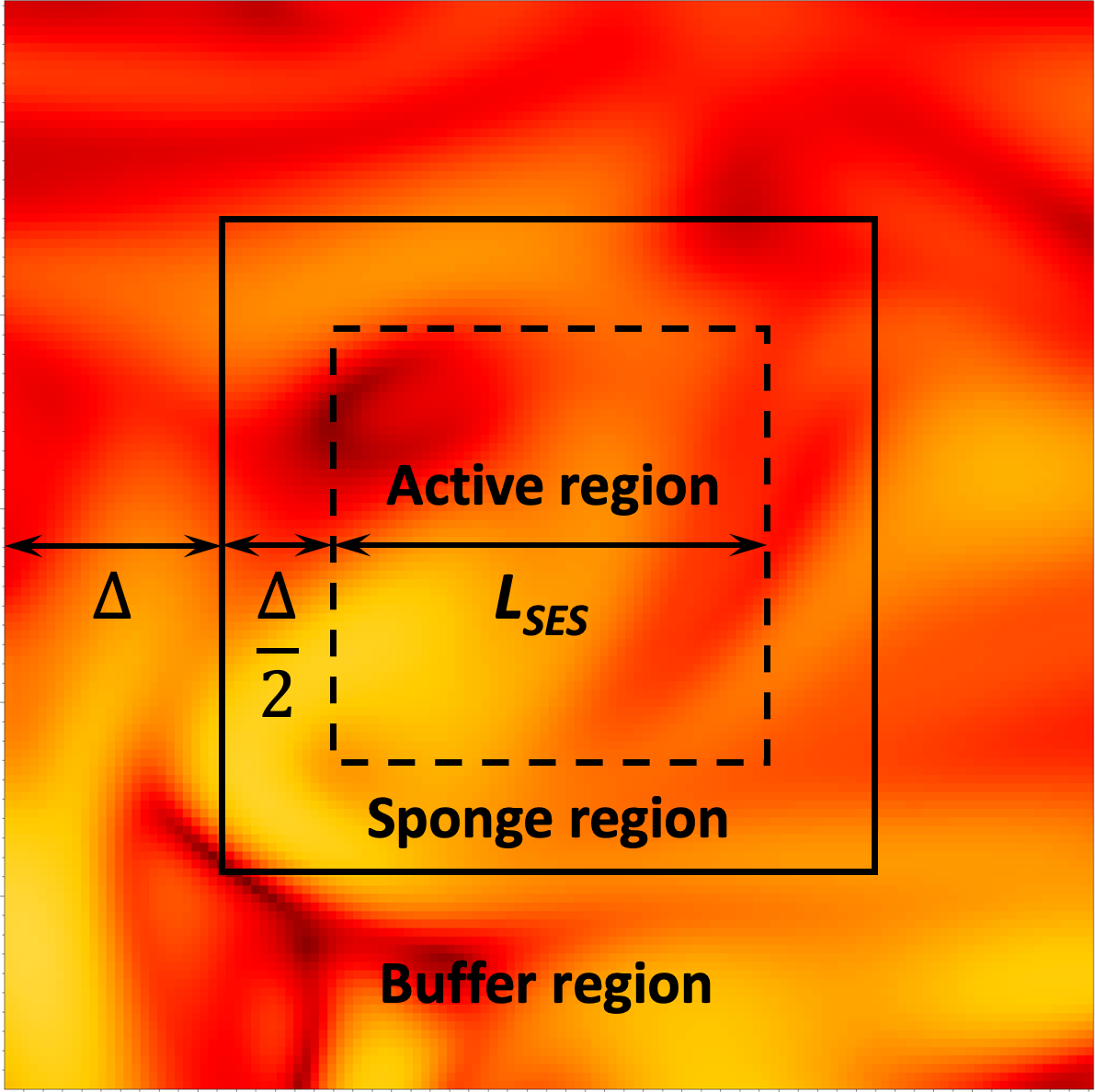}
  \caption{SES domain showing the sponge, buffer, and active regions.}
\label{fig:SESdomain}
\end{figure}

The L/SES approach can be viewed as the volumetric forcing since velocity (or momentum) perturbations are applied throughout the entire SES domain. At the same time, since the SES represents a sub-region of the LES, its domain is not periodic. This raises the question of the appropriate boundary conditions treatment in the  SES.

SES boundary conditions for all conserved variables are set directly using filtered LES data to ensure that at all times SES solution is fully consistent with the filtered LES field at the boundaries. To achieve this, SES solution is gradually relaxed to the filtered LES field using a sponge region near all boundaries. Such filtered LES field is constructed from the filtered LES density, pressure, and momenta, $\bar{\rho}$, $\bar{P}$, and $\overline{\rho u_i}$. Thus, the SES domain, which is evolved in the calculation, includes the sponge region and is thus larger than the active region of interest (see Fig.~\ref{fig:SESdomain}). As a result, since the filtered LES solution is injected throughout the SES volume, and it is also imposed at the SES domain boundaries, the L/SES approach can be viewed as a combination of forcing both at the physical and spectral boundaries.

\begin{figure*}
\includegraphics[width=1.0\textwidth]{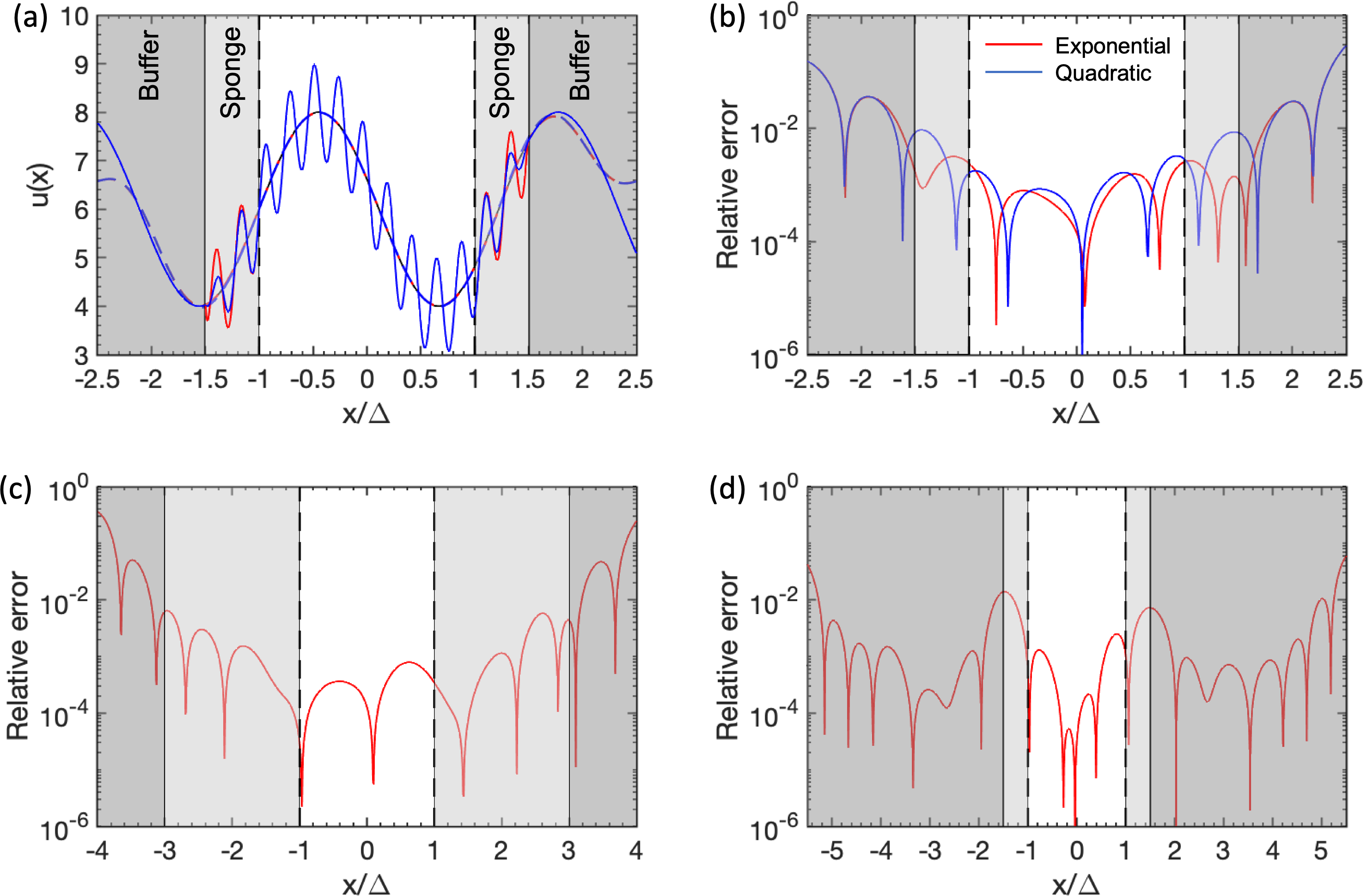}
\caption{Comparative tests of the sponge function type as well as the width of the sponge and buffer regions. a) Synthetic input signal with a high-frequency component in the active region damped in the sponge region using exponential (solid red line, eq.~\ref{eq_sponge}) and quadratic (solid blue line, eq.~\ref{eq_sponge_quadratic}) sponge functions to the reference low-frequency signal. Filtered signal is shown with the dashed lines. b) Relative error between the filtered output signal and the reference low-frequency signal for the exponential (red line) and quadratic (blue line) sponge functions. Sponge region (shaded light gray) width is $\Delta /2$ and buffer region (shaded dark gray) width is $\Delta$. c) Effect of the sponge region with $4\times$ larger width $2\Delta$ on the relative error of filtering. d) Effect of the buffer region with $4\times$ larger width $4\Delta$ on the relative error of filtering. See text for further details.}
\label{fig:sponge_tests}
\end{figure*}

The sponge relaxation function used in the tests described here has the exponential form
\begin{equation}
    f=\exp(\frac{1-j}{M})(\bar{\phi}_{LES} - \phi_{SES}),
\label{eq_sponge}
\end{equation}
where $j$ is the index of a cell from the nearest sponge region boundary (solid black line in Fig.~\ref{fig:SESdomain}), $\phi_{SES}$ is any conserved variable in the SES (note that the total energy is not filtered directly, but obtained from the filtered pressure and momenta), $\bar{\phi}_{LES}$ is the filtered LES value of the same conserved variable at that location, and $M$ is the sponge layer parameter, which is taken to be equal to $3$ following \citet{marbaix2003lateral}. This value of $M$ was found to give the optimal results. Resulting quantity $f$ is then added to all conserved variables in the sponge region modifying each variable as $\phi'=\phi+f$. Other forms of the sponge relaxation function can be employed, for instance, based on the quadratic polynomials \citep{marbaix2003lateral}
\begin{equation}
f=(x/D_s)^2(\bar{\phi}_{LES}-\phi_{SES}),
%\phi=(1-(x/D_s)^2)\phi_{SES}+(x/D_s)^2\bar{\phi}_{LES},
\label{eq_sponge_quadratic}
\end{equation}
where $x$ is the distance of a cell in the sponge to the nearest active region (dashed black line in Fig.~\ref{fig:SESdomain}) and $D_s$ is the width of the sponge region.

The sponge region alone, however, does not address another important aspect related to the boundary conditions. Note that the filter is given by the inhomogeneous Helmholtz equation (\ref{eq:Germano2}), the unique solution of which formally requires a boundary condition at infinity. Prescribing such a boundary condition is not possible on a finite discrete mesh. Therefore, the boundary will always introduce an error, unless the boundary is periodic, which is not the case in general since the SES flow field is non-periodic.

In principle, the non-periodic boundary can be treated by reducing the filter width near the sponge region boundaries. Such approach, however, is not ideal as in this case the filtered SES field will include smaller scales near the outer boundary of the sponge region. This will lead to systematic errors in the forcing term and a mismatch between the SES and the LES solutions at the boundary. Resulting errors can advect into the active region and destabilize the overall SES solution.

\begin{figure*}
\includegraphics[width=0.8\textwidth]{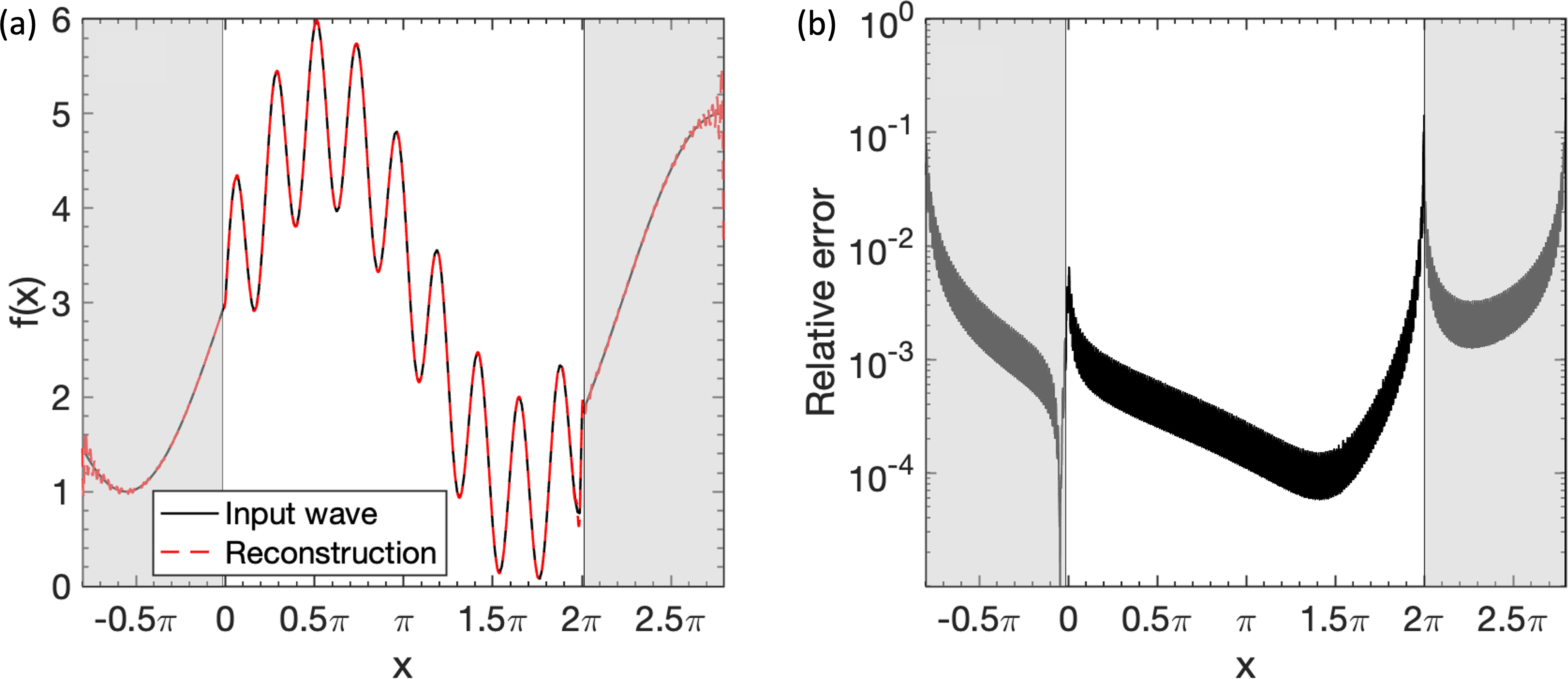}
\caption{(a) Spectral (trigonometric) spatial interpolation of a non-periodic, bi-modal 1D signal (cf. Fig.~\ref{fig:sponge_tests}). (b) Relative error of the interpolation. The non-shaded area represents the entire SES domain (including sponge and buffer regions), while the shaded areas represent additional regions included only in the interpolation.}
\label{fig:interpolation}
\end{figure*}

To address this difficulty, an additional buffer region is created outside of the sponge region (cf. Fig.~\ref{fig:SESdomain}). Such buffer is filled at every time step directly with the LES filtered data, as described above. Periodic boundary condition is assumed at the outer boundary of the buffer. This eliminates the need to reduce the filter width, and the entire domain, which includes active, sponge, and buffer regions, can be filtered with a constant filter width. Furthermore, buffer region also pushes the domain boundary further away from the active region, minimizing the errors introduced into it.
%The filtered data obtained this way is not affected by any domain boundary inaccuracies in the actual L/SES region, and can be used as $\tilde{u}_{i,L/SES}$ in eq. \ref{eq:deltau}.
%The filtered data in the buffer region itself is discarded.
In particular, since even the extended SES domain with the buffer is non-periodic, using periodic boundary condition, as well as any other boundary condition not at infinity, will create errors in the filtered result. However, if the buffer region is sufficiently large, these errors primarily affect the buffer, where they are discarded. Furthermore, although the appended data is taken from a large-scale simulation, which does not contain a small-scale component, this does not affect the filtering, as the filter is low-pass in nature and it is aimed at removing the higher wavenumbers. Finally, Navier-Stokes equations are not solved in the buffer, and thus the additional computational cost of including the buffer is modest amounting only to the extra cost of filtering in that region as well as the additional memory cost.
 
%To address this difficulty, an additional buffer region is created outside of the sponge region (cf. Fig.~\ref{fig:SESdomain}). Such buffer is filled at every time step directly with the LES filtered data, as described above, and the entire extended field is filtered with a constant filter width. Periodic boundary condition is assumed at the outer boundary of the buffer. After filtering, the result in the buffer region is discarded and the buffer is repopulated with LES data from the next instant. The filtered data so obtained is unaffected by any domain boundary inaccuracies in the active + sponge region and eliminates the need to reduce the filter width near the edges. Since even the extended data is non-periodic, the periodicity assumption causes errors in the filtered result. However, if the buffer is large, these errors are confined to it, and are discarded without affecting the sponge+active region. Although the appended data is taken from a large-scale simulation which does not contain high wavenumbers, it does not affect the filtering, as the filter is low-pass in nature and is aimed at removing the higher wavenumbers. Finally, Navier-Stokes equations are not solved in the buffer, and thus the incremental computational cost of including the buffer is modest amounting only to the extra cost of filtering in that region as well as the additional memory cost.

Figure~\ref{fig:sponge_tests}a shows comparison of the effect of the two sponge functions (eqs. ~\ref{eq_sponge} and \ref{eq_sponge_quadratic}) on the accuracy of the filtering of a synthetic 1D signal. The input signal (solid red and blue lines in panel a) is comprised of the low-frequency component, analogous to the LES field, which is given by the function $f_{low}(x)=6+2\sin(0.9x)$ and which extends through the entire domain. This input signal also has a high-frequency component, analogous to the SES field, given by the function $f_{high}(x)=\sin (8.8x)$. This high-frequency component is present in the central active region, and it is damped in the sponge region (shaded light gray areas) using either exponential (solid red line, eq.~\ref{eq_sponge}) or quadratic (solid blue line, eq.~\ref{eq_sponge_quadratic}) functions. Only the low-frequency component is present in the buffer region (shaded dark gray areas). Note that both signal components are non-periodic in the domain being considered. The result of filtering the input signal damped using the two sponge functions is shown with the dashed red and blue lines, which are indistinguishable from the low-frequency reference signal in the active and sponge regions. Filtering successfully preserves the low-frequency component and removes the high-frequency component in the central region, with deviations from $f_{low}(x)$ occurring only in the buffer, in which they become pronounced near the outer boundaries.

More quantitative comparison is given in Fig.~\ref{fig:sponge_tests}b, which shows the relative error between the filtered output signal and the reference low-frequency input signal $f_{low}(x)$. Both sponge functions give very similar results, and thus we choose to employ the exponential sponge function for all the tests discussed below.

Figure~\ref{fig:sponge_tests} also demonstrates the effect of the sponge and buffer region sizes. In particular, in panels (a) and (b), the sponge and buffer region widths are $\Delta/2$ and $\Delta$, respectively. The resulting relative error is $\sim 10^{-3}$ in the active region and it gradually increases through the sponge and buffer regions exceeding $10\%$ near the boundaries. Increasing the sponge region size by a factor of four to $2\Delta$ (Fig.~\ref{fig:sponge_tests}c) somewhat reduces the relative error in the active region to below $10^{-3}$, while the error in the buffer region remains approximately the same. In contrast, increasing the buffer region size by a factor of four to $4\Delta$ (Fig.~\ref{fig:sponge_tests}d) does not change the relative error in the active and sponge regions, though error in the buffer region decreases by about an order of magnitude. Decreasing the sponge and buffer region sizes below $\Delta/2$ and $\Delta$, respectively, increases filtering errors substantially. 

%This is due to the fact that when a filter is applied at the scale $\Delta$, this scale sets the filter kernel size. Therefore by imposing periodic boundary condition at the domain edges, cells located up to a distance $\Delta$ away from the boundary are primdirectly affected and must be discarded from the solution. As a result, making the buffer larger than $\Delta$ does not affect the accuracy of the filtered signal in the active and sponge regions, as can be seen in Fig.~\ref{fig:sponge_tests}d. On the other hand, the sponge is used to make the SES data consistent with the LES and it is not directly linked to forcing. However, this parameter is indirectly affected by filtering as the SES data is filtered for calculating the perturbations. Moreover, specifying sponge length in terms of $\Delta$ is desirable because such a method would automatically set the sponge size based on one parameter without the need for additional calibrations for different flow conditions.

Therefore, sponge and buffer region sizes shown in Figs.~\ref{fig:sponge_tests}a,b, namely $\Delta/2$ and $\Delta$, respectively, represent the optimal balance between the resulting error and the associated computational and memory costs. Such sizes are used in all tests discussed below, and they are found to provide good overall stability and accuracy of the SES. %Comparative analysis of different sponge functions, as well as different sponge and buffer region sizes, using the actual L/SES calculations rather than the synthetic data, will be presented in a companion paper.

\subsubsection{Spatial and temporal interpolation}
\label{sssec:interp}

As discussed in \S~\ref{ssec:method}, velocity or momentum perturbations in eqs.~(\ref{eq:deltau}) or (\ref{eq:deltamom}) are applied in SES at every timestep and in every cell. Since SES by construction has finer spatial and temporal resolution than LES, the filtered LES forcing field must be interpolated both in space and time. Traditional interpolants, such as trilinear or cubic spline, introduce spurious oscillations in the spatially interpolated velocity field at higher wavenumbers. Such oscillations were found to be detrimental to the SES accuracy as they are retained in the forcing field throughout the simulation, and they affect the very scales that we seek to capture.

If the large-scale data, which is used to drive the SES calculations, are periodic, as in the DNS  used in the {\it a priori} tests described below, spectral interpolation represents the most accurate approach. In particular, the Fourier transform of the filtered large-scale field is first obtained. Resulting field, which lacks scales smaller than the filter scale $\Delta$, is then padded with zeros at the higher wavenumbers present in the SES, and the inverse Fourier transform is performed. Resulting spatially interpolated velocity field is ensured to contain no small-scale component.
%The resulting field is not necessarily real-valued, however only magnitude of each velocity component is needed. Since the magnitude can only be non-negative, a constant must be added to the initial data before taking FT such that it becomes positive everywhere, and the constant is subsequently subtracted after the interpolation. This step only adds a zero-frequency signal over the data and remains unaffected by the FT.

The entire procedure is equivalent to the trigonometric interpolation
%, as the data is known at equidistant cells
\citep{campuzano2022trigonometric}. %The advantage of the trigonometric interpolation is that it can be performed directly in physical space. 
Although this procedure is applied here via Fourier transforms, it can be performed directly in physical space \citep{zygmund2002trigonometric}. While ideally it would require periodic data, it can also be applied to non-periodic LES domains. The original spectral content would be preserved by the operations and mild distortions of the data would occur near the domain edges. The distortions can be mitigated by adding a few extra points to the SES domain size (including the buffer), performing the interpolation on this extended grid, and later discarding the excess interpolated data outside the SES domain. Figure~\ref{fig:interpolation} illustrates the accuracy of this procedure applied to a non-periodic synthetic 1D data represented by the same bi-modal sine wave as in Fig.~\ref{fig:sponge_tests}. The ratio of the resolution (or step size) between the coarser input and the finer output signals is $8$. Setting the number of extra points on each side of the domain boundary to be twice the ratio of the grid spacing between the coarse and fine data proves to be sufficient to eliminate the distortions. Figure~\ref{fig:interpolation}b shows that the relative error of the interpolation is the largest in this added region (shaded gray area) and near the boundaries of the central region, which in the SES would lie in the buffer region, and hence would not affect the SES solution. Otherwise, the relative error is $\lesssim 10^{-3}$, which is comparable to the filtering errors shown in Fig.~\ref{fig:sponge_tests}.

Finally, the filtered and spatially interpolated LES flow field also has to be interpolated in time to provide forcing data at each finer SES time step. For the present tests, linear interpolation proved to be sufficient. While ideally, such interpolation would be done between each successive LES time steps, it may be too costly computationally to filter and interpolate LES data so frequently especially in compressible solvers, in which the time step is constrained by the sound speed rather than the flow velocity. We find that a much larger time interval for the forcing LES data is sufficient
\begin{equation}
    \Delta t_{forcing} \lesssim \frac{\Delta}{U_{rms}},
\label{eq:dtLES}
\end{equation}
where $\Delta$ again is the filter scale, and $U_{rms}$ is the r.m.s. velocity in the region of interest. This effectively represents a characteristic filter-scale crossing time by the flow, which means that the changes in the large-scale flow structure are minimal on this timescale. As a result, snapshots of the flow-field taken at this time interval will be close to each other, and thus one can interpolate between them in time with a minimal loss of accuracy.

In the case of an explicit compressible solver, such as used in this study, the time step of the LES calculation is constrained by the Courant-Friedrichs-Lewy condition
\begin{equation}
\Delta t_{LES} \approx \textrm{CFL} \frac{\Delta x_{LES}}{(U_{rms} + c_s)},
\end{equation}
where $c_s$ is the sound speed and \textrm{CFL} $\sim O(1)$ is the Courant-Friedrichs-Lewy number. Then
\begin{equation}
\frac{\Delta t_{forcing}}{\Delta t_{LES}} \lesssim \frac{1}{\textrm{CFL}}\frac{\Delta}{\Delta x_{LES}}\Big(1+\frac{1}{M_{rms}}\Big).
\end{equation}
Here $M_{rms} = U_{rms}/c_s$. For the low-Mach-number turbulence with $M_{rms} < 0.1$, and assuming the ratio of the filter scale to the LES cell size $\Delta/\Delta x_{LES} \gtrsim 10$, this relation suggests that LES data for forcing the SES can be collected every $\gtrsim 100$ time steps or even less frequently. As a result, the cost of LES data filtering and interpolation becomes negligible compared to the overall computational cost of the LES calculation. This consideration applies both to the SES performed synchronously with the LES (`online' mode) or in isolation from it (`offline' mode). In the tests described below, SES are performed offline with the LES data stored with the time interval $\Delta t_{LES}$, and filtering and interpolation done as a post-processing step.

%In the present work, L/SES is done in an `offline' way, i.e. the LES is performed beforehand and L/SES is not used to provide small-scale information back to the LES. The offline approach is chosen for simplicity, and to demonstrate the performance of the method in isolation. A two-way coupled LES-L/SES remains a subject of future work. Since LES is not done simultaneously, its data to be used for forcing is stored at finite intervals and is not available at every timestep. Hence A similar time interpolation would also be required for a coupled `online' L/SES, since the time step taken by the higher resolution L/SES will be smaller than the LES.
%can give reference for linear timestep

%--------------------------------------------------------------------------------------------
\subsubsection{Mean flow}
\label{sssec:meanflow}

The filtered LES field in the active region of the SES may have a non-zero mean velocity. Even in the case of the HIT, mean velocity over a chosen SES sub-region is non-zero instantaneously and of the order of $U_{rms}$. If such mean velocity becomes large, the residence time of the large-scale flow injected into the SES may not be long enough to allow the corresponding small-scale dynamics to develop. For instance, for the HIT, such mean residence time of the large-scale flow can be estimated as 
\begin{equation}
    \tau_{mf}=\frac{L_{SES}}{U_{rms}},
\label{eq:taumf}
\end{equation}
where $L_{SES}$ is the SES domain size. This can be compared with the characteristic turbulent eddy turnover timescale in the SES domain
\begin{equation}
    \tau_{ed,SES}=\frac{L_{SES}}{u_{\Delta}},
\label{eq:taued}
\end{equation}
where $u_{\Delta}$ is the characteristic velocity at the filter scale, $\Delta$. The ratio of these two timescales would be
\begin{equation}
    \frac{\tau_{mf}}{\tau_{ed,SES}} = \frac{u_{\Delta}}{U_{rms}} \sim \Big(\frac{\Delta}{L_{LES}}\Big)^{1/3},
\end{equation}
assuming Kolmogorov scaling, where $L_{LES}$ is the LES domain size. In high-Re$_{\lambda}$ flows, which are the primary target of the proposed approach, $\Delta \ll L_{LES}$, which could lead to very short residence times of the mean large-scale flow relative to the eddy turnover time in the SES, $\tau_{mf} \ll \tau_{ed,SES}$. In such cases, the mean velocity could be subtracted from the filtered LES data before calculating $\delta u_i$, i.e., eq.~(\ref{eq:deltau}) can be modified as
\begin{equation}
    \delta u_i=(\tilde{u}_{i,LES}-\overline{\tilde{u}_{i,LES}})-\tilde{u}_{i,{SES}},
\label{eq:deltau_mean}
\end{equation}
where $\overline{\tilde{u}_{i,LES}}$ is the average filtered LES velocity in the SES domain. This can be interpreted as removing the zero-wavenumber component of the forcing LES velocity field, or alternatively it can be viewed as an SES domain, which is co-moving with the local LES flow in the Lagrangian framework.

In the tests discussed below, subtracting the mean velocity from the forcing flow did not affect the results since $\tau_{mf}/\tau_{ed,SES} \sim O(1)$, and all SES simulations presented below do not use this modification. Detailed analysis of the L/SES method performance in flows with large mean velocities is a subject of future work. 
%As the L/SES-M2 domain is a small sub-region of the LES domain, the filtered velocity field used for driving
%Since the present work only considers HIT flows in periodic LES simulations, the time scales of the mean flow were comparable to the eddy turnover timescales of the L/SES, given by

%--------------------------------------------------------------------------------------------
\subsubsection{Summary of the L/SES algorithm}
\label{sssec:summary}

Finally, we summarize the sequence of algorithmic steps in the proposed L/SES approach described above:
\begin{enumerate}
    \item Set up the large-scale simulation (LES), which resolves the large scales of the flow of interest, but does not capture the smallest scales of turbulence.
    \item Select a sub-region of interest with size $\gtrsim 2\Delta$ within the LES domain for the SES calculation, where $\Delta$ is the filter scale chosen such that all scales $> \Delta$ are minimally affected by the lack of LES resolution or the subgrid-scale model uncertainties.
    \item Set up a fully resolved SES grid in this sub-region. Also allocate an additional region around the SES domain for the sponge and buffer zones of size $\Delta$/2 and $\Delta$, respectively.
    \item Advance the LES calculation. Low-pass filter the LES fields at the filter scale $\Delta$ using eq.~(\ref{eq:Germano2}).
    \item Perform spectral (or trigonometric) interpolation of the filtered LES data onto the SES grid, including the sponge and buffer regions, and a linear interpolation in time between the successive LES data snapshots.
    \item Update the SES buffer with the interpolated LES data to specify the boundary conditions.
    \item Inject the filtered and interpolated LES data into the SES simulation (only the active and sponge regions) at every time step using eq.~(\ref{eq:deltau}) or (\ref{eq:deltamom}).
    \item Relax the SES solution to the LES field in the sponge region, according to eq.~(\ref{eq_sponge}).
    \item Advance the SES solution in time, and then repeat either from step 4 (`online' approach) or step 6 (`offline' approach).
\end{enumerate}

%\begin{figure}
%\includegraphics[width=12cm]{Old_figures/Flowchart2.png}
%\caption{Sequence of steps for performing L/SES.}
%\label{fig:flowchart}
%\end{figure}

%This procedure is pictorially depicted in fig. \ref{fig:flowchart}. For L/SES-M1, only the first two steps are used, and the filtered data is used to calculate an energy injection rate using eqs. \ref{eq_uDeltaAk_M1}-\ref{eq_eps_M1}. This quantity is then used used to drive turbulence in a periodic domain in a conventional way. All the information presented above can be used to set up the L/SES simulations for an HIT flow. The numerical method used for all the simulations in the present work is described in the next section.

%--------------------------------------------------------------------------------------------
\section{Alternative approaches}
\label{sec:altmethods}

A question can be asked whether other simpler approaches can achieve a similar goal as the L/SES method of embedding a high-fidelity local simulation in a lower-fidelity global model. Here we discuss two potential alternatives: one, which is similar to the traditional spectral forcing techniques, and one, which is based on the static (or adaptive) mesh refinement. In contrast to the L/SES, which combines both volumetric and boundary driving of the embedded simulation, approaches discussed here implement only one or the other type of driving. These two alternative methods also provide insights into the role of these two types of forcing and if both of them are essential to achieve high fidelity of the resulting solution.

%--------------------------------------------------------------------------------------------
\subsection{Embedding of a high-resolution calculation through a scalar energy injection rate}
\label{ssec:lses-eps}

%Traditional methods of turbulence driving aim to inject energy into the simulation domain at a prescribed rate, $\epsilon$ \citep{eswaran1988examination,seror2001radiated,witkowska1997numerical,lundgren2003linearly}. In a steady-state turbulence, this energy injection balances energy dissipation on small scales and allows an equilibrium cascade to establish. At the same time, a question could be asked if such single scalar quantity $\epsilon$ is sufficient to allow SES to develop the flow, which would accurately reproduce the physically correct structure of the flow in a sub-region of a larger system of interest. Furthermore, for a flow that is not in equilibrium, this rate of energy injection may be time-dependent and it may differ from the energy dissipation rate. In this section, we present a method where this energy injection rate, $\epsilon$, is determined on-the-fly from the LES using just the filtered fields of the large-scale flow, without making any assumptions of equilibrium. This method can be viewed as a generalization of the traditional volumetric forcing approaches, including spectral \citep{eswaran1988examination} and linear \citep{lundgren2003linearly,rosales2005linear,petersen2010forcing} forcing.

Traditional methods of turbulence driving\citep{eswaran1988examination,seror2001radiated,witkowska1997numerical,lundgren2003linearly} aim to inject energy into the simulation domain at a prescribed rate, $\epsilon$. In a steady-state turbulence, this energy injection balances energy dissipation on small scales and allows an equilibrium cascade to establish. At the same time, for an arbitary sub-region of interest in a larger system with an arbitrary flow, such rate of energy injection may differ from the global $\epsilon$ and it may not be immediately known. Furthermore, for a flow that is not in equilibrium, this rate of energy injection may be time-dependent and it may differ from the energy dissipation rate. In this section, we present a method where this energy injection rate into the chosen SES sub-region is determined on-the-fly from the LES using just the filtered fields of the large-scale flow, without making any assumptions of equilibrium. Resulting single scalar quantity, $\epsilon$, is then used to force turbulence in the SES domain in a traditional way. This method can be viewed as a generalization of the traditional volumetric forcing approaches, including spectral \citep{eswaran1988examination} and linear \citep{lundgren2003linearly,rosales2005linear,petersen2010forcing} forcing.

The starting point is the same as in the L/SES approach described above. We choose a sub-region of interest of the LES calculation to be represented by an SES calculation with the domain size $L_{SES}$, as well as the scale, $\Delta_{\epsilon}$, at which energy would be injected. Considerations for the choice of $\Delta_{\epsilon}$ are similar to the choice of the filter scale, $\Delta$, in the L/SES above. This scale should be contained within the SES, i.e., $\Delta_{\epsilon} \leq L_{SES}$. For an explicitly filtered LES, $\Delta_{\epsilon}$ could be chosen equal to the scale of explicit filtering, whereas for an implicitly filtered LES, it could be any scale that is minimally affected by numerical dissipation or SGS modelling of the LES. Here, we choose $\Delta_{\epsilon}$ to be the largest scale available in the SES, i.e., $\Delta_{\epsilon}=L_{SES}$.
%, so that $\epsilon$ estimated at this scale can be used to drive turbulence in the SES using one of the standard volumetric forcing method at only the smallest wavenumber in the L/SES, following \citet{poludnenko2010interaction}.
%$\Delta$ is the scale at which LES and L/SES are stitched. Scales larger than $\Delta$ can be considered to be accurately simulated by LES, and scales smaller than $\Delta$ are to be captured by the L/SES

%For estimating $\epsilon$, we also consider two other scales, one larger ($\Delta_1$) and one smaller ($\Delta_2$) than $\Delta$ (fig. \ref{fig:epscalc_show}(a)). We assume that the slope of the spectral kinetic energy density with respect to wavenumber $E(k)$ on a logarithmic scale is a constant (-$\alpha$) between $\Delta_1$ and $\Delta_2$ and takes the following form
%\begin{equation} \label{eq:Ek_equation}
%    E(k_\Delta)=Ak_\Delta^{-\alpha}=C_k\epsilon^{2/3}k_\Delta^{-\alpha}
%\end{equation}
%check this equation
%where $k_\Delta=2\pi/\Delta$ is the wavenumber associated with scale $\Delta$ and $C_k$ is the Kolmogorov constant whose value lies in the range 1.2 to 2 for HIT \citep{aspden2009analysis}, and $C_k=1.5$ was chosen for this work following \citet{poludnenko2010interaction}. Both $A$ and $\alpha$ are unknown and time dependent in the general case. The assumption of the form of $E(k)$, although not perfect, is still reasonable since the two scales can be chosen as close to $\Delta$ as possible. For the present work, a choice of $\Delta_1=2\Delta$ and $\Delta_2=\Delta/2$ appeared to work best. 

\begin{figure}
\includegraphics[width=0.45\textwidth]{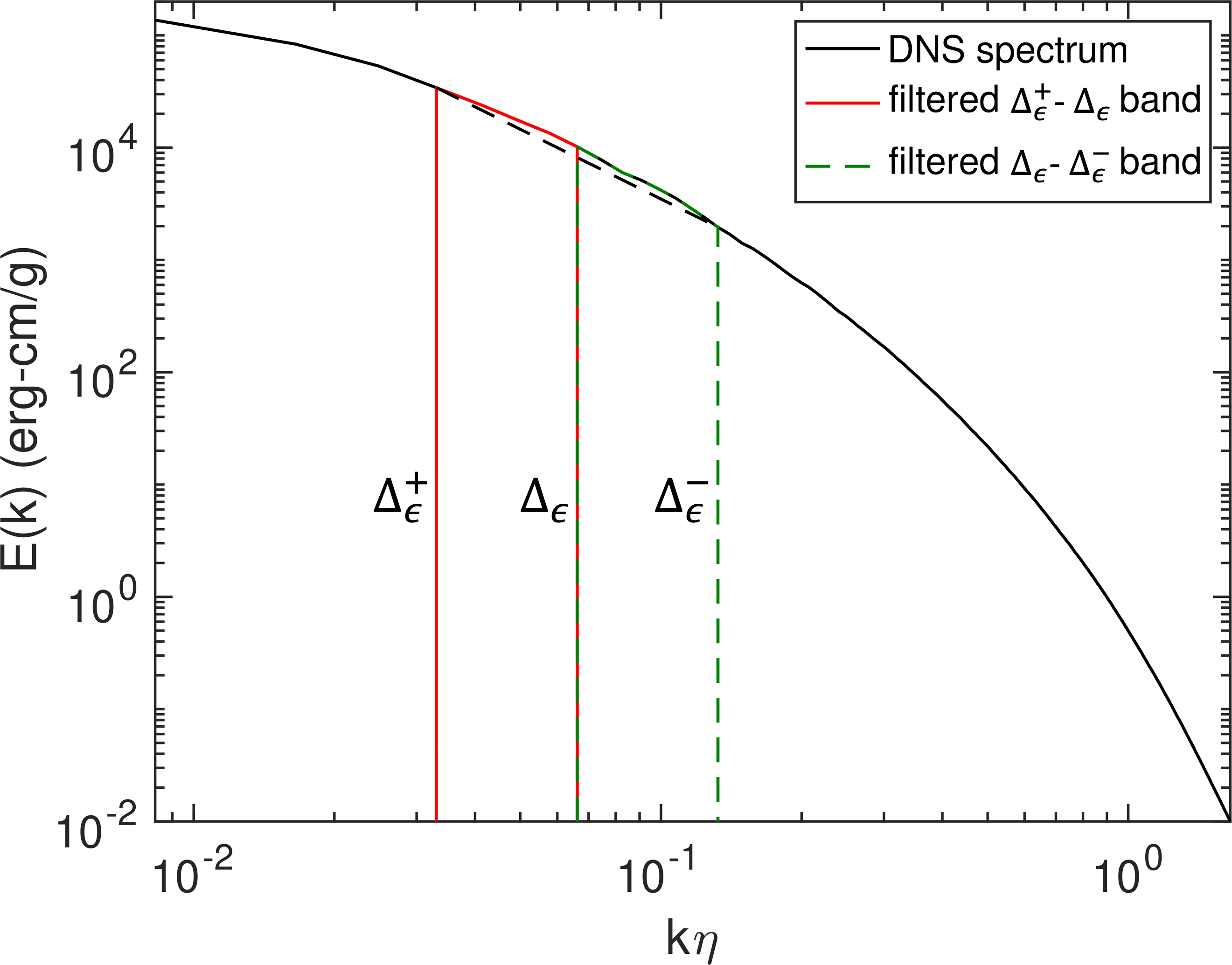}
\caption{Generic HIT spectrum illustrating the scale $\Delta_{\epsilon}$ and the two bands used for the calculation of $\epsilon$.}
\label{fig:LSESeps_bands}
\end{figure}

For equilibrium turbulence, the specific kinetic energy spectral density takes the form $E(k)=C\epsilon^{2/3}k^{-5/3}$, where $k$ is the wavenumber and $C$ is a constant. For a more general case of a non-equilibrium flow, the rate of energy transfer between different scales may be wavenumber-dependent, i.e., $\epsilon=\epsilon_0k^{x}$. Then, spectral kinetic energy density at the scale $\Delta_{\epsilon}$ can be written as
\begin{equation}
    E(k_{\Delta_{\epsilon}})=C\epsilon_0^{2/3} k_{\Delta_{\epsilon}}^{2x/3-5/3}=Ak_{\Delta_{\epsilon}}^{-\alpha},
 \label{eq:Ek_equation}
\end{equation}
where $k_{\Delta_{\epsilon}}=2\pi/\Delta_{\epsilon}$. Both $A$ and $\alpha$ are unknown and time-dependent in the general case.

In order to determine $\epsilon$ at the scale $\Delta_{\epsilon}$, we consider two additional scales, one larger, $\Delta^+_{\epsilon}$, and one smaller, $\Delta^-_{\epsilon}$, than $\Delta_{\epsilon}$ (Fig. \ref{fig:LSESeps_bands}). We assume that the slope of $E(k)$ on a logarithmic scale between $\Delta^+_{\epsilon}$ and $\Delta^-_{\epsilon}$ equals a constant $-\alpha$. This approximation of the local form of $E(k)$ is reasonable since the two scales $\Delta^+_{\epsilon}$ and $\Delta^-_{\epsilon}$ can be chosen arbitrarily close to $\Delta_{\epsilon}$. We found that $\Delta^+_{\epsilon}=2\Delta_{\epsilon}$ and $\Delta^-_{\epsilon}=\Delta_{\epsilon}/2$ gave the best results.

From the original LES field, we find the band-pass filtered velocity field in the two bands $\Delta^+_{\epsilon}$ to $\Delta_{\epsilon}$ and $\Delta_{\epsilon}$ to $\Delta^-_{\epsilon}$, $u_{\Delta^+_{\epsilon}}$ and $u_{\Delta^-_{\epsilon}}$, respectively (Fig.~\ref{fig:LSESeps_bands}). Band-pass filtering is achieved by applying a low-pass filter at each of the two scales and subtracting one result from the other. The filtering procedure is the same as described in \S~\ref{sssec:filter}. Next, we calculate the r.m.s. of the resulting two band-pass filtered velocity fields $u_{\Delta^+_{\epsilon}}$ and $u_{\Delta^-_{\epsilon}}$. The r.m.s. is only taken over the sub-region corresponding to the SES domain.

The total specific kinetic energy in the two band-pass filtered velocity fields can then be related to $E(k_{\Delta_{\epsilon}})$ (eq.~\ref{eq:Ek_equation})
\begin{eqnarray}
    \frac{u_{\Delta^+_{\epsilon}}^{2}}{2} = \int_{k_{\Delta^+_{\epsilon}}}^{k_{\Delta_{\epsilon}}} Ak^{-\alpha} dk & = & \frac{A}{\alpha-1} (k^{-\alpha+1}_{\Delta^+_{\epsilon}} - k^{-\alpha+1}_{\Delta_\epsilon}),
    \label{eq_uDeltaAk_M1} \\
    \frac{u_{\Delta^-_{\epsilon}}^{2}}{2}  =  \int_{k_{\Delta_{\epsilon}}}^{k_{\Delta^-_{\epsilon}}} Ak^{-\alpha} dk & = & \frac{A}{\alpha-1} (k^{-\alpha+1}_{\Delta_{\epsilon}} - k^{-\alpha+1}_{\Delta^-_{\epsilon}}).
    \label{eq:uDeltaAk_M2}
\end{eqnarray}
From these two equations, two unknowns $\alpha$ and $A$ in eq.~(\ref{eq:Ek_equation}) can be found
\begin{eqnarray}
    \frac{u_{\Delta^+_{\epsilon}}^{2}}{u_{\Delta^-_{\epsilon}}^{2}} & = & \frac{k_{\Delta^+_{\epsilon}}^{-\alpha+1} - k_{\Delta_{\epsilon}}^{-\alpha+1}}{k_{\Delta_{\epsilon}}^{-\alpha+1} - k_{\Delta^-_{\epsilon}}^{-\alpha+1}} \implies \alpha = \alpha (u_{\Delta^+_{\epsilon}},u_{\Delta^-_{\epsilon}}),
\label{eq:A_alpha1}\\
    A & = & C\epsilon_0^{2/3} = \frac{u_{\Delta^-_{\epsilon}}^{2}}{2} \frac{\alpha-1}{(k^{-\alpha+1}_{\Delta^+_{\epsilon}} - k^{-\alpha+1}_{\Delta^-_{\epsilon}})}.
\label{eq:A_alpha2}
\end{eqnarray}
Equation~(\ref{eq:A_alpha1}) can be solved numerically to determine $\alpha$. Substituting these constants into eq.~(\ref{eq:Ek_equation}), we finally determine the rate of energy injection in the SES domain
\begin{equation}
    \epsilon = \epsilon_0k^x = \Big(\frac{A}{C}\Big)^{3/2} k_{\Delta_{\epsilon}}^{(5-3\alpha)/2}.
\label{eq:eps}
\end{equation}
The constant $C \approx 1.5$ for HIT \citep{wang1996examination,zhou1993interacting}.
%This method extracts the rate of energy transfer to the small scales at a chosen wavenumber $k_\Delta$ in any given flow field.
Once $\epsilon$ is determined, it can subsequently be applied to force the SES simulation using any conventional volumetric forcing approach. Particular approach used in the tests below is described in \S~\ref{sssec:forcing}. We hereafter refer to this method as L/SES-$\epsilon$.

L/SES-$\epsilon$ can be viewed as an extension of the traditional turbulence driving methods, in which $\epsilon$ actually represents the rate of the inter-scale kinetic energy transfer in a realistic flow instead of being an arbitrary parameter. Such energy transfer rate can be obtained directly without any assumptions regarding the nature of the flow, or the existence of equilibrium. As such, this method can be viewed as a test of the effectiveness of the traditional volumetric forcing methods in representing the effect of a large-scale ambient flow. Although proposed formulation extracts just a single scalar $\epsilon$ at the scale $\Delta_{\epsilon}$, this method can also be extended to implement
%a broadband forcing scheme by adjusting the scale and the band of filtering. A
anisotropic forcing by computing directional components of $\epsilon$ from the kinetic energy carried by the corresponding velocity components. At the same time, the key aspect is that in contrast to the L/SES approach, which injects the entire large-scale flow field into the SES, in the L/SES-$\epsilon$ approach, such large-scale flow is reduced to just a single scalar $\epsilon$, which may not be sufficient to fully recover the corresponding structure of the small scales.
%Therefore another driving approach is proposed.
%The fully resolved small scale simulation can also be used to calculate LES closure terms directly within the specific sub-region, without the need for modelling. Statistics of turbulence from the small scale simulation can be compared with statistics from the corresponding region of a fully resolved reference large scale simulation, which can demonstrate how well a large scale flow is described through the parameter $\epsilon$. 

%--------------------------------------------------------------------------------------------
\subsubsection{Spectral forcing with a given energy injection rate}
\label{sssec:forcing}

A number of forcing strategies have been developed in the past\citep{eswaran1988examination,petersen2010forcing} for injecting kinetic energy volumetrically into the turbulent flow with a prescribed rate, $\epsilon$ (also see \S~\ref{sec:Introduction}). Here we describe a spectral forcing approach \citep{Lemaster2009,poludnenko2010interaction}, which can inject energy into the flow at a particular scale or over a range of scales. This forcing method is used both in the tests of L/SES-$\epsilon$, as well as in the DNS described below.

First, we define velocity perturbations in spectral space as an independent realization of a Gaussian random field
\begin{equation}
    \delta \hat{u}_i''(\pmb{k})=\sqrt{-2\ln\phi_1} \cos(2\pi\phi_2)e^{i2\pi\phi_3},
\end{equation}
where $\phi_1$, $\phi_2$, and $\phi_3$ are random variables uniformly distributed in the interval $(0,1)$, $\pmb{k}$ is the wavenumber vector, $u_i$ is the $i^{th}$ component of the velocity, and $\hat{u}$ represents a variable in the Fourier space.

Next, the function $\delta \mathcal{E}(k)$, which defines the spectral distribution of the energy injection rate, is prescribed as
\begin{equation}
    \delta \mathcal{E}(k) = 1 \quad  \forall \quad \frac{|k_i|}{2\pi/L} = 0 , 1; \quad 0 \quad \mathrm{otherwise.}
\label{eq:Einj}
\end{equation}
Here $k \equiv ||\pmb{k}||$ is the magnitude of the wavenumber. This particular form provides energy injection at the largest scale of the simulation domain, although in general the function $\delta \mathcal{E}(k)$ can be arbitrarily complex, and it can support broadband energy injection. This energy injection spectrum is applied to the velocity perturbation field as
\begin{equation}
    \delta \hat{u}_i'(\pmb{k}) = \frac{\sqrt{\mathcal{E}(k)}}{k} \delta \hat{u}_i''(\pmb{k}).
\end{equation}
 
To ensure that the perturbations are solenoidal in physical space, an orthogonal projection is taken
\begin{equation}
    \delta \hat{u}_i(\pmb{k})= \Big(\delta_{ij}-\frac{k_ik_j}{k}\Big)\delta \hat{u}_j'(\pmb{k}),
\end{equation}
where $\delta_{ij}$ is the Kronecker delta. Resulting perturbations are then transformed to the physical space and normalized to provide the prescribed energy injection rate $\epsilon$. The total momentum of the perturbations is subtracted to ensure that no net momentum is introduced. Finally, resulting velocity perturbation field is directly added to the velocity field at each time step
\begin{equation}
    u_i^{n+1}=u_i^n+\delta u_i.
\label{eq:vperturb}
\end{equation}

%--------------------------------------------------------------------------------------------
\subsection{Embedding of a high-resolution region through mesh refinement}
\label{ssec:lses-smr}

\begin{figure}
\includegraphics[width=0.45\textwidth]{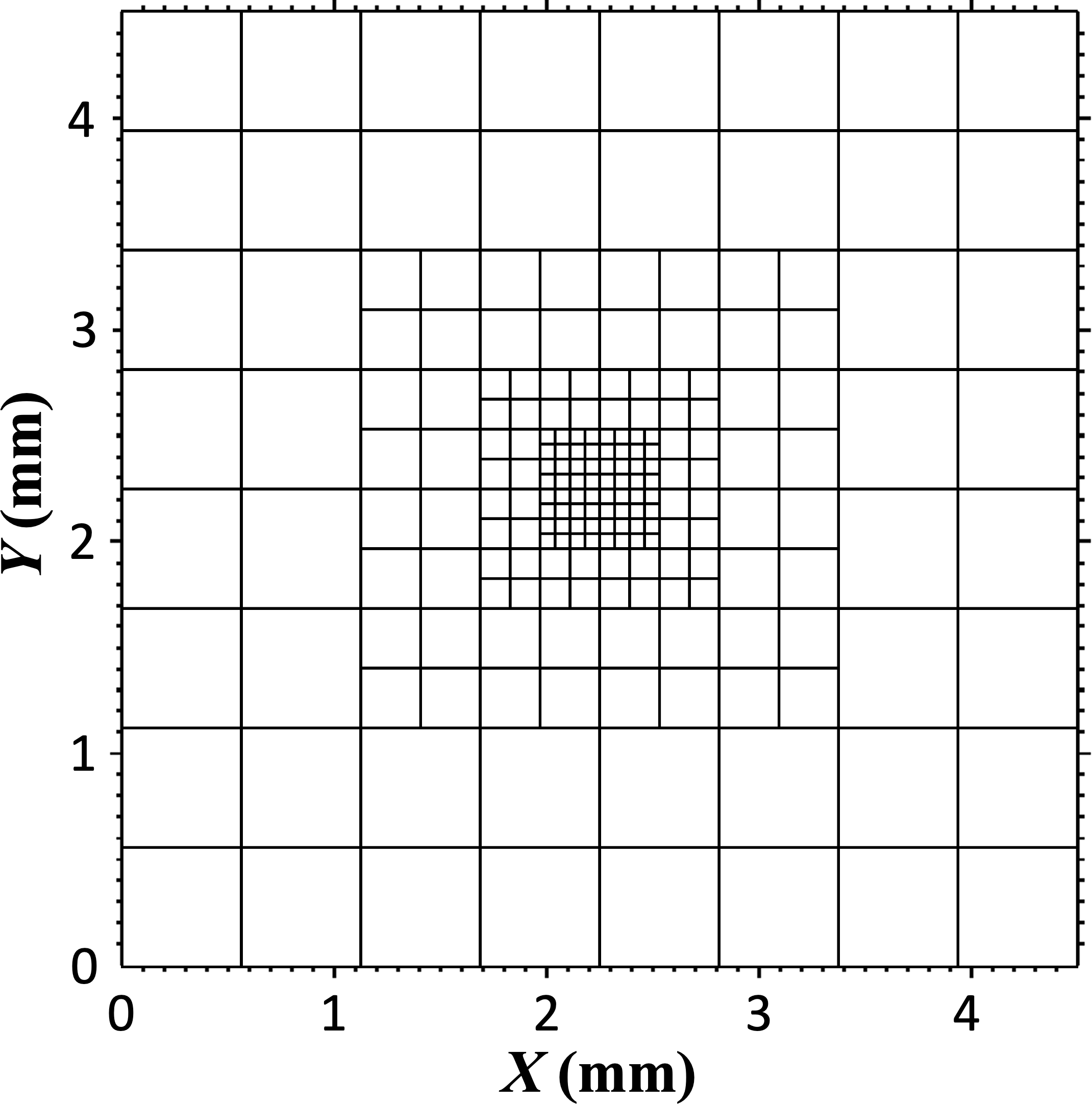}
\caption{Grid structure in the midplane of the SMR calculations. Each block represents $16^3$ computational cells.}
\label{fig:Meshblock}
\end{figure}

Both L/SES and L/SES-$\epsilon$ methods described above require two separate simulations LES and SES, running sequentially or concurrently, as well as explicit filtering of the velocity fields. Therefore, a question can be asked whether these approaches can be replaced with a simple mesh refinement, which can be embedded inside an LES sub-region of interest and which would not require any elaborate forcing mechanisms \citep{macart2021embedded,sirignano2023dynamic,towery2020scaling,christopher2022high}. In this approach, such sub-region is statically or adaptively refined to resolve the full range of small scales. Figure~\ref{fig:Meshblock} shows such statically refined mesh in a periodic cube containing HIT. Each block represents $16^3$ cells. The central high-resolution region is surrounded by progressively coarser cells to limit the refinement factor to $2$ between adjacent mesh blocks. This proper nesting requirement is typical for the modern mesh refinement algorithms \citep{colella2021chombo, zhang2016boxlib, henry2023pelec}, and it is in agreement with the prior analysis by others \citep{towery2020scaling,christopher2022high}.

Static (SMR) and adaptive (AMR) mesh refinement are well established computational techniques, which have been used over the years in a wide range of problems from engineering \citep{vanella2014adaptive,hay2007adaptive,ganesh2009r} to astrophysics \citep{plewa2005adaptive}. Traditionally, proper use of SMR or AMR assumes that the entire flow feature of interest, such as a shock wave, a flame, or a shear layer are fully resolved. This ensures that such features do not cross the coarse-fine grid boundaries, which are prone to introduce errors in the solution. This particularly concerns regions with turbulence, which typically are expected to be captured with uniform resolution. Therefore, proposal to refine only a sub-region of a turbulent flow, as in Fig.~\ref{fig:Meshblock}, is contrary to the established practices for the use of SMR/AMR. Nevertheless, we investigate here whether such selective refinement of a turbulent flow can yield positive results, which could obviate the need for more complex methods, such as L/SES or L/SES-$\epsilon$.

Finally, in contrast to the L/SES-$\epsilon$ strategy described above, which implements purely volumetric forcing, this approach that we refer to as L/SES-SMR, effectively represents only boundary forcing. The large-scale flow structures enter the high-resolution region through the coarse-fine boundaries. Thus, the detailed treatment of such boundaries is central to this method. Any artifacts associated with flux interpolation at the coarse-fine boundaries, or any other aspects of the numerical algorithm, will be advected into the region of interest, where they can affect the small-scale flow structure that has to develop there. Detailed tests of both the L/SES-$\epsilon$ and L/SES-SMR approaches, and their comparison with the performance of the L/SES method, are described below.

\section{Numerical model and simulation setup}
\label{sec:num3}

\begin{table}
  \begin{center}
\def~{\hphantom{0}}
  \begin{tabular}{lcc}
\hline
  & \quad \quad e-HIT DNS \quad \quad \quad & ne-HIT DNS \\
  \hline
 Domain width, $L_{DNS}$ (cm)                       & 0.45                  & 0.9                    \\
 Grid size                                          & $1024^3$              & $1024^3$               \\
 Energy injection rate, $\epsilon$ (erg/g s)        & 5.29$\times 10^9$     & 3.43$\times 10^9$       \\
 Density, $\rho$ (g/cm$^3$)                         & 1.10$\times 10^{-3}$  & 8.73$\times 10^{-4}$   \\
 %$Re_\lambda$                                       & 156                   & 266                    \\
 %$\eta/dx$                                          & 2.05                  & 1.35                   \\
 Kinematic viscosity, $\nu$ (cm$^2$/s)              & 0.15                  & 0.19                   \\
\hline
\end{tabular}
\caption{Parameters of the DNS.}
\label{DNS_param}
\end{center}
\end{table}
% Cell size, dx ($cm$) & 4.395e-4 & 4.395e-4   \\

All methods described above solve the Navier-Stokes equations
\begin{eqnarray}
    \frac{\partial \rho}{\partial t} + \frac{\partial \rho u_j}{\partial x_j} & = & 0,
    \label{DNS equations1} \\
    \frac{\partial \rho u_i}{\partial t} + \frac{\partial \rho u_i u_j}{\partial x_j} & = & -\frac{\partial p}{\partial x_i} + \frac{\partial \sigma_{ij}}{\partial x_j}+f_i,
    \label{DNS equations2}
\end{eqnarray}
\begin{equation}
\begin{split}
    \frac{\partial \rho E}{\partial t} + \frac{\partial (\rho E+p) u_j}{\partial x_j} = &\frac{\partial }{\partial x_j} \Big(\kappa\frac{ \partial T}{\partial x_j}\Big) + \frac{\partial u_i \sigma_{ij}}{\partial x_j} \\ &+ u_jf_j, \label{DNS equations3}
\end{split}
\end{equation}
%+f_i  +u_jf_j 
where $u_i$ is the fluid velocity, $\rho$ is the density, $p = (\gamma-1)\rho e$ is the pressure, $E=e+ u_ju_j/2$ is the total energy per unit mass, $e$ is the internal energy per unit mass, $\gamma$ is the specific heat ratio set to a constant value of $1.197$ in all tests described below, $\sigma_{ij}=2\mu S_{ij}-2/3\mu S_{kk}\delta_{ij}$ is the viscous stress tensor with the strain rate $S_{ij}= \frac{1}{2}(\frac{\partial u_i}{\partial x_j} + \frac{\partial u_j}{\partial x_i})$, $\delta_{ij}$ is the Kronecker delta, $\mu=\rho \nu$ is the coefficient of viscosity, $\nu$ is the kinematic viscosity, $T$ is the temperature, and $\kappa=6.234*10^{3}$ erg/(s cm K) is the thermal conductivity.

Additional terms $f_i$ and $u_i f_i$ represent external forcing, which takes on different meaning depending on the calculation. In particular, forcing described by these terms in the DNS and in the SES-$\epsilon$ tests is detailed in \S~\ref{sssec:forcing} above. In the SES, the forcing procedure is described in \S~\ref{sec:lses}. Finally, the details of forcing in the SES-SMR tests is given in \S~\ref{ssec:SMR} below.

All simulations, with the exception of the SES-SMR, are performed using a fully compressible, massively parallel, numerical solver \texttt{Athena-RFX} \citep{poludnenko2010interaction}, a reacting-flow extension of the code \texttt{Athena} \citep{stone2008athena}. The code uses a higher-order, fully conservative, Godunov-type method for integration based on the unsplit corner transport upwind (CTU) algorithm \citep{colella1990multidimensional, saltzman1994unsplit}. The integration scheme uses piecewise parabolic method (PPM) for spatial reconstruction \citep{colella1984piecewise}, along with an approximate nonlinear Harten–Lax–van Leer contact (HLLC) Riemann solver. The overall scheme is 3$^\mathrm{rd}$-order accurate in space and 2$^\mathrm{nd}$-order accurate in time. Further details of the algorithm and its implementation can be found in \citet{gardiner2008unsplit} and \citet{stone2008athena}, while the detailed tests of the code for modeling the HIT are described in \citet{hamlington2011interactions,hamlington2012intermittency}. The code has been extensively used in a wide variety of reacting and non-reacting flow studies \citep{jozefik2024modeling, xu2024modeling, poludnenko2010interaction, hamlington2011interactions, hamlington2012intermittency, poludnenko2015pulsating, poludnenko2019unified} including detonations \citep{kozak2020weno, dammati2021numerical, kozak2019novel}. All calculations performed with \texttt{Athena-RFX} use a uniform, Cartesian grid.

The SES-SMR calculations were performed with the code \texttt{Athena-RFX++}, a reacting flow extension of the code \texttt{Athena++}, which implements both static and adaptive mesh refinement on a block-structured grid \citep{Stone2020}. This code also implements a range of higher-order, Godunov-type integrators, including the van Leer predictor-corrector scheme used in this work \citep{stone2009simple}. Spatial reconstruction and the Riemann solver are the same as in \texttt{Athena-RFX}, giving the same order of accuracy of the overall algorithm. Resulting turbulent flow fields obtained with both codes were verified to be statistically identical.
%Additionally, as the central refined region is well-resolved, the choice of the algorithm would have little effect on the results. Forcing for SMR can be done in multiple ways and further tests are presented in the appendix.
 
%Finally, LES simulations are also performed for the entire domain for both cases using equation set \ref{LES equations}. We drop the unclosed terms and solve for the filtered fields on the LES grid directly, without using an explicit model. This implicit LES method is shown to work well for HIT where the grid provides the necessary dissipation \citep{Aspdenetal:2008}. 

%Test conditions Simulation setup
%--------------------------------------------------------------------------------------------
\subsection{Direct numerical simulations for the {\it a priori} tests}
\label{ssec:DNS}

In the present work, we restrict ourselves to the {\it a priori} tests of the method, in which no LES are directly performed. Instead, a fully resolved DNS is explicitly filtered at an appropriate scale (in the inertial range) to mimic the LES field and obtain the forcing data for the SES. The {\it a priori} approach is chosen here for two reasons. First, we seek to demonstrate that the L/SES method can work in principle for physically correct large-scale flows not affected by the numerics or the subgrid-scale models. Second, DNS also provides the reference small-scale flow structure for direct comparison with that obtained in the SES. This is in contrast with the {\it a posteriori} tests based on the LES, for which results can be compared with the DNS only statistically. Such statistical comparison is not sufficient in the case of unsteady flows. Detailed {\it a posteriori} tests for a range of Re$_{\lambda}$ values will be presented in the follow-up paper.

%We consider two types of fully resolved DNS, both performed in a large domain with size $1,024^3$. This DNS serves as a benchmark for comparing the small-scale statistics from the L/SES. The DNS is also used for a priori tests of the L/SES methods, where instead of an LES, the large-scale driving data is taken from DNS fields filtered explicitly at some scale to resemble LES data.
%We will first test the L/SES methods in an a priori way, where the large scale driving data is taken from a fully resolved large scale DNS, instead of an LES. The a priori tests are done for a preliminary evaluation of the L/SES method, as well to test its theoretical limit of accuracy. For these tests, the DNS will be filtered explicitly at a scale of $8 dx_{DNS}$ to resemble LES data. The DNS will also be treated as a reference to compare the performance of the L/SES methods.

Two reference DNS of the HIT are performed by solving eqs. (\ref{DNS equations1}) - (\ref{DNS equations3}) in a cubic, triply periodic domain on a $1024^3$ grid. Parameters of both DNS are given in Table~\ref{DNS_param}. The first DNS considers driven equilibrium HIT in a steady state and is hereafter referred to as the e-HIT DNS. Turbulence in this simulation is stirred at the scale of the domain width, $L_{DNS}$, using the approach described in \S~\ref{sssec:forcing}. The calculation is initialized with uniform density $\rho=1.104 \times 10^{-3}$ g/cm$^3$ and atmospheric pressure. The initial velocity field in the domain is prescribed with the kinetic energy spectrum given by the power-law of the form $E(k)\propto k^{-5/3}$. To reach an equilibrium, the flow is allowed to develop for two large-scale eddy turnover times, $\tau_{ed}$,
\begin{equation}
    \tau_{ed}=\frac{L_{DNS}}{U_{L,DNS}}.
\end{equation}
Here $U_{L,DNS}$ is the velocity at the scale $L_{DNS}$. We note that $\tau_{ed}$ is approximately four times larger than the integral-scale eddy turnover time \citep{poludnenko2010interaction}. After the initial equilibration period, the e-HIT DNS is continued for another $2 \tau_{ed}$ to provide forcing data and collect the statistics for evaluating the performance of the companion SES.  

\begin{figure*}
\includegraphics[width=0.9\textwidth]{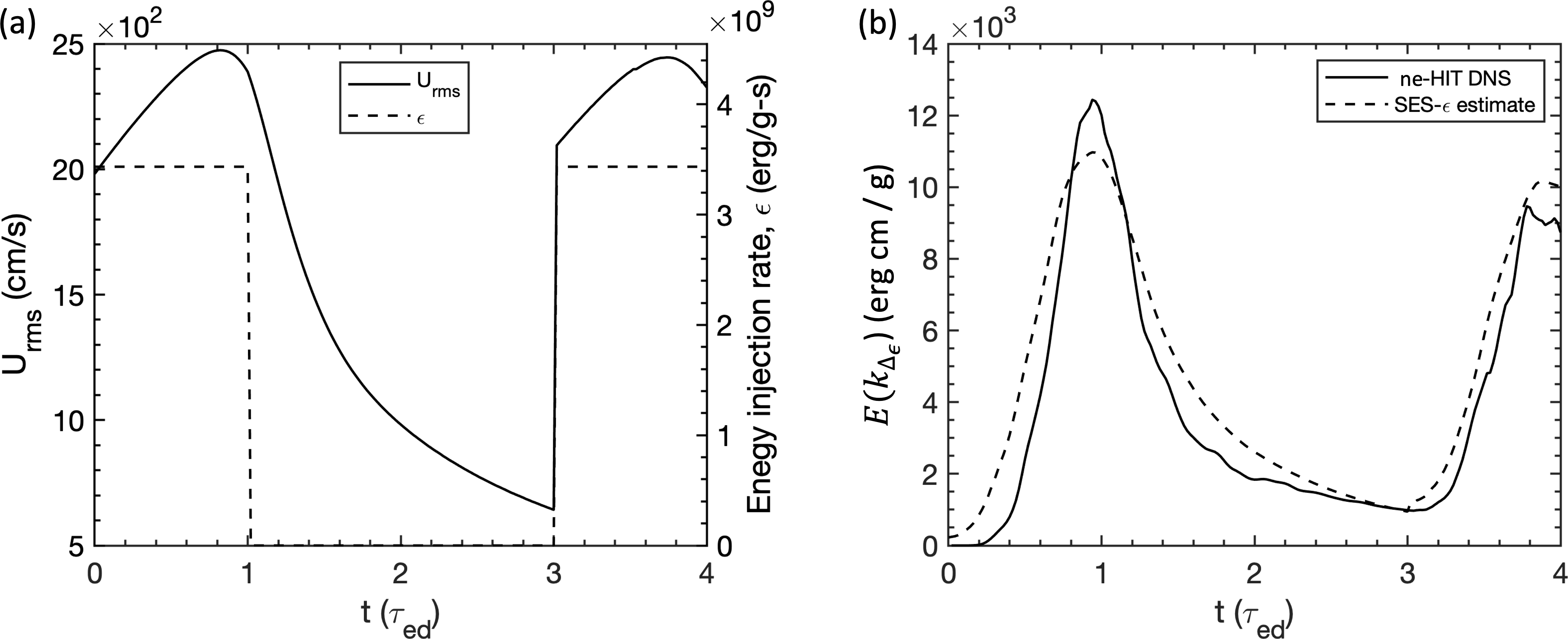}
\caption{a) The $U_{rms}$ and the energy injection rate $\epsilon$ as a function of time in the ne-HIT DNS. b) Spectral kinetic energy density at the scale $\Delta_{\epsilon}$ in the ne-HIT DNS obtained directly from the spectrum (solid line) and using the SES-$\epsilon$ approach (\S~\ref{ssec:lses-eps}) (dashed line).}
\label{fig:Unst_Urms}
\end{figure*}

Since we seek to devise a general method, which is agnostic to the underlying flow conditions, the e-HIT DNS is supplemented with the second reference DNS that we use to study the time-dependent performance of the proposed L/SES approach in turbulent flows out of equilibrium. This unsteady DNS, hereafter referred to as ne-HIT, is initialized in the same way as the e-HIT DNS with the parameters given in Table~\ref{DNS_param}. For the first $\tau_{ed}$, the simulation is forced with a constant $\epsilon$ using the same method as in the e-HIT DNS. Driving is then stopped, and turbulence is allowed to decay for the next $2\tau_{ed}$. Note that during this time, the actual $\tau_{ed}$ would be changing, however hereafter we use the equilibrium steady-state value of $\tau_{ed}$ to specify time intervals both in the ne-HIT DNS and the companion SES. At $3\tau_{ed}$, driving is restarted by injecting at the largest scale the amount of energy needed to increase the r.m.s. velocity in the domain, $U_{rms}$, to the value close to that at $t = 0$. After that, forcing is continued for another $\tau_{ed}$ with the original energy injection rate $\epsilon$. This progression is illustrated in Fig.~\ref{fig:Unst_Urms}a, which shows both $\epsilon$ and $U_{rms}$ in the ne-HIT DNS as a function of time. As a result, ne-HIT DNS consists of periods of turbulence growth, decay, and sudden increase of energy injection creating strong turbulent non-equilibrium. Thus, this case allows us to test the L/SES approach under more complex turbulent flow conditions compared to the more idealized, steady, equilibrium HIT.

In both DNS calculations, grid resolution was chosen to provide at least two cells per Kolmogorov scale (see Table~\ref{tab:Urms}). Such resolution was previously found sufficient to ensure high fidelity of the DNS solution \citep{donzis2010,jagannathan2016,yeung2018}. Resulting turbulence in both DNS has a very low degree of compressibility with the r.m.s. Mach number $\approx 0.05$.

Finally, we note that these two calculations do not represent DNS of some realistic flow due to the presence of artificial large-scale forcing. For the purposes of this work, however, we use these two calculations to provide: i) the canonical, well established reference flow with a sufficiently high Re$_{\lambda}$, ii) a realistic large-scale turbulent flow of sufficient complexity for SES forcing, and iii) a fully resolved reference small-scale flow for direct comparison with the SES.

\begin{table*}
\begin{center}
\def~{\hphantom{0}}
\begin{tabular}{lcccc}
\hline
& \quad \quad e-HIT SES \quad \quad         & e-HIT SES-$\epsilon$ \quad \quad  & ne-HIT SES \quad \quad  & ne-HIT SES-$\epsilon$   \\
\hline
$L_{SES}$ (cm)      & 0.05625      & 0.05625   & 0.1125       & 0.1125            \\
Sponge size (cm)    & 0.0140625    & ---       & 0.028125     & ---            \\
Grid size           & $(128+64)^3$  & $128^3$   & $(128+64)^3$  & $128^3$           \\
%$\eta/dx$           & 2.049         & 2.049     & 1.348         & 1.348             \\
\hline
\end{tabular}
\caption{Parameters of the SES and SES-$\epsilon$. For the SES, grid size in parenthesis includes both the active and sponge regions.}
\label{SES_param}
\end{center}
\end{table*}
%7.81e6 (L/SES-M1 estimate)

%--------------------------------------------------------------------------------------------
\subsection{SES and SES-$\epsilon$ tests}
\label{ssec:SES}

Companion SES and SES-$\epsilon$ tests are set up and driven with the flow fields obtained in the e-HIT and ne-HIT DNS using methods described in \S~\ref{sec:lses} and \ref{ssec:lses-eps}. All SES tests were performed in the `offline' mode, i.e., separately from the DNS, using large-scale forcing data recorded and stored at discrete time intervals $\Delta t_{forcing} \approx 0.36\Delta/U_{rms}$ for the e-HIT DNS and $\approx 0.64\Delta/U_{rms}$ for the ne-HIT DNS in agreement with the criterion given in eq.~(\ref{eq:dtLES}) (see \S~\ref{sssec:interp}).
%This interval needs to be chosen carefully. The time scale associated with the filter scale is $t_\Delta=\Delta/U_{rms}$. As the forcing data is filtered at this scale, $t_\Delta$ becomes the smallest timescale of the larger driving eddies, and any driving interval smaller than $t_\Delta$ would sufficiently capture the large-scale effects. For the present work, 
%, which were $1.07 \times 10^{-5} s$ and $1.96 \times 10^{-5} s$ for E-HIT and NE-HIT respectively. 
%For the E-HIT case, this interval for the first $2 \tau_{ed}$ was chosen to be higher ($0.06 \tau_{ed}$) as this was the turbulence development period where statistics were not considered. Both these intervals conform to the $t_\Delta$ requirement and ensure that the L/SES remains correlated with the forcing field while also getting ample time to develop its own dynamics.

For the e-HIT SES-$\epsilon$ test, $\epsilon$ obtained from the e-HIT DNS was time-averaged and the resulting value was used as a constant energy injection rate for the entire calculation. This time-averaged value was found to be $3.92\times 10^9$ erg/g$\cdot$s vs. the actual $\epsilon = 5.29\times 10^9$ erg/g$\cdot$s in the e-HIT DNS (Table~\ref{DNS_param}), or $26\%$ lower.

For the ne-HIT SES-$\epsilon$, Fig.~\ref{fig:Unst_Urms}b shows the time-dependent spectral density of the specific kinetic energy at the filter scale $\Delta_{\epsilon} = L_{SES}$, or $E(k_{\Delta_{\epsilon}})$. Values obtained directly from the kinetic energy spectra at this scale in the ne-HIT DNS are shown with a solid black line, while the dashed line shows values calculated using eq.~(\ref{eq:Ek_equation}) with $A$ and $\alpha$ found following the procedure in \S~\ref{ssec:lses-eps} (eqs.~\ref{eq:A_alpha1}, ~\ref{eq:A_alpha2}). These $A$ and $\alpha$ are then used to calculate $\epsilon$ based on eq.~(\ref{eq:eps}). Close agreement of the $E(k_{\Delta_{\epsilon}})$ obtained using the SES-$\epsilon$ approach with the exact DNS values demonstrates that both $A$ and $\alpha$, and thus $\epsilon$ derived from them and used to force the SES-$\epsilon$ calculation, accurately capture the complex flow evolution in the ne-HIT DNS.
%Minor differences observed are due to the fact that the $\epsilon$ estimate is obtained for a specific sub-region which would only match the DNS on average.

%For all other SESes, data is supplied as a function of time.

The central cubic sub-region of the DNS domain was chosen as the region of interest for the SES tests. In both SES cases, the filter scale $\Delta = 64dx_{DNS}$, and the size of the active region of the domain is set to $L_{SES} = 2\Delta$ as discussed in \S~\ref{ssec:method}. For SES-$\epsilon$, $\Delta_{\epsilon} = L_{SES} = 128dx_{DNS}$. Both the grid size and $L_{SES}$ for the SES and SES-$\epsilon$ tests are given in Table~\ref{SES_param}. The choice of the location of the sub-region is arbitrary and any other sub-region could be considered instead. Grid cell size in the SES and SES-$\epsilon$ is equal to that in the corresponding DNS.

Initial conditions in the SES were set directly based on the full flow field in the central region of the corresponding DNS at $t = 0$. Boundary conditions were also set directly from the DNS field as described in \S~\ref{sssec:sponge}. In contrast, SES-$\epsilon$ tests were run with periodic boundaries. They were initialized with the same constant density and pressure as the corresponding DNS. Initial velocity field was set with the same spectral kinetic energy distribution of the corresponding DNS.
\subsection{SES-SMR tests}
\label{ssec:SMR}

The L/SES-SMR approach does not admit {\it a priori} testing since most of the flow in the domain is by design unresolved. Nevertheless, we include the tests of this method since it is considerably simpler algorithmically than the L/SES and L/SES-$\epsilon$ approaches, as it can be implemented effectively using any existing SMR/AMR code. Thus it is important to understand whether it can potentially provide a more cost-effective alternative to L/SES and L/SES-$\epsilon$.

Tests of the L/SES-SMR (\S~\ref{ssec:lses-smr}) consider only equilibrium, steady-state HIT in a 3D periodic cube, and thus they are similar in setup to the e-HIT DNS. In particular, SES-SMR tests have the same domain size and other parameters of the e-HIT DNS listed in Table~\ref{DNS_param}. The only difference is the non-uniform computational grid shown in Fig.~\ref{fig:Meshblock}. The central high-resolution region has the same physical size $L_{SES}$, grid size equal to $128^3$, and the computational cell size as the SES (excluding sponge regions) and SES-$\epsilon$ tests. Outside this central region, resolution gradually decreases by a factor of two over the next 3 levels of refinement. The coarsest region has the resolution equivalent to that of the $128^3$ grid. Note that even though the coarse regions of these simulations are not well-resolved, eqs. (\ref{DNS equations1}) - (\ref{DNS equations3}) are directly evolved in the non-uniform grid without using any closure models.

Turbulence is forced with the prescribed rate $\epsilon$ at the scale of the domain width using the same spectral forcing approach outlined in \S~\ref{sssec:forcing} as in the e-HIT DNS. At the same time, presence of a non-uniform mesh prevents the forcing method from being applied in its standard form. Instead, we consider three modifications of forcing.

In the first one, velocity perturbation field $\delta u_i$ is first formed in exactly the same way as in the e-HIT DNS on a uniform $1,024^3$ mesh spanning the entire computational domain (Fig.~\ref{fig:Meshblock}). Subsequently, $\delta u_i$ are applied as is in the fine central region. In the coarser outer cells, $\delta u_i$ are averaged over the corresponding finer cells of the source uniform mesh before being applied to the velocity field using eq.~(\ref{eq:vperturb}).
%This is the most straightforward way of forcing the L/SES-SMR simulation, and it serves as an equivalent case to directly contrast the L/SES results.
We refer to this test as SES-SMR1.

This approach, however, is somewhat artificial as it presumes that the effect of large scales is known in the central region of interest, and thus it injects the large-scale flow information into the central region volumetrically, and not just through the region boundaries. In a practical LES simulation, no such large-scale forcing would be present as the large-scale flow dynamics would directly emerge in the process of flow evolution. Therefore, another test, SES-SMR2, is performed, in which exactly the same velocity perturbation field $\delta u_i$ is applied everywhere except for the most-refined central region, i.e., in that region $\delta u_i$ are set to zero. This approach produces the same flow field outside the central region, where it maintains the same rate of kinetic energy injection per unit volume as in SES-SMR1, however the total amount of injected energy in the domain is reduced. 

Finally, the third test, SES-SMR3, is performed, in which similar to SES-SMR2 no energy is injected in the most refined region. Instead, volumetric energy injection rate in the outer regions is increased to ensure that the total rate of energy injection in the domain is the same as in the first SES-SMR1 case. Together, these three cases span the range of possibilities for forcing an embedded SMR simulation, and thus they allow us to elucidate the difference in the solution accuracy achievable through SMR vs. L/SES.

%Comparisons are also made with the SMR simulation to contrast the behaviour of this method versus a mesh refinement techniques. Note that the performance of SMR can only be observed when a simulation contains both well-resolved and under-resolved regions, hence this test is not done for a priori analysis.

%All three SMR cases are also performed in Athena++ with the same grid as shown in fig. \ref{fig:Meshblock}.

%\begin{figure}
%  \centerline{\includegraphics[width=14cm]{pdf_aposteriori_dda.eps}}
%  \caption{Time averaged PDFs for normalized $\textit{(a)}$ enstrophy ($\Omega$) and $\textit{(b)}$ dissipation rate ($\epsilon$) for all three SMR cases compared with LES and E-HIT DNS.}
%\label{fig:pdf_SMR}
%\end{figure}

%--------------------------------------------------------------------------------------------
\section{Results}
\label{sec:Results}

% $5.35 \times 10^{-4} s$, which represents the final

\begin{figure*}[htpb]
\includegraphics[width=\textwidth]{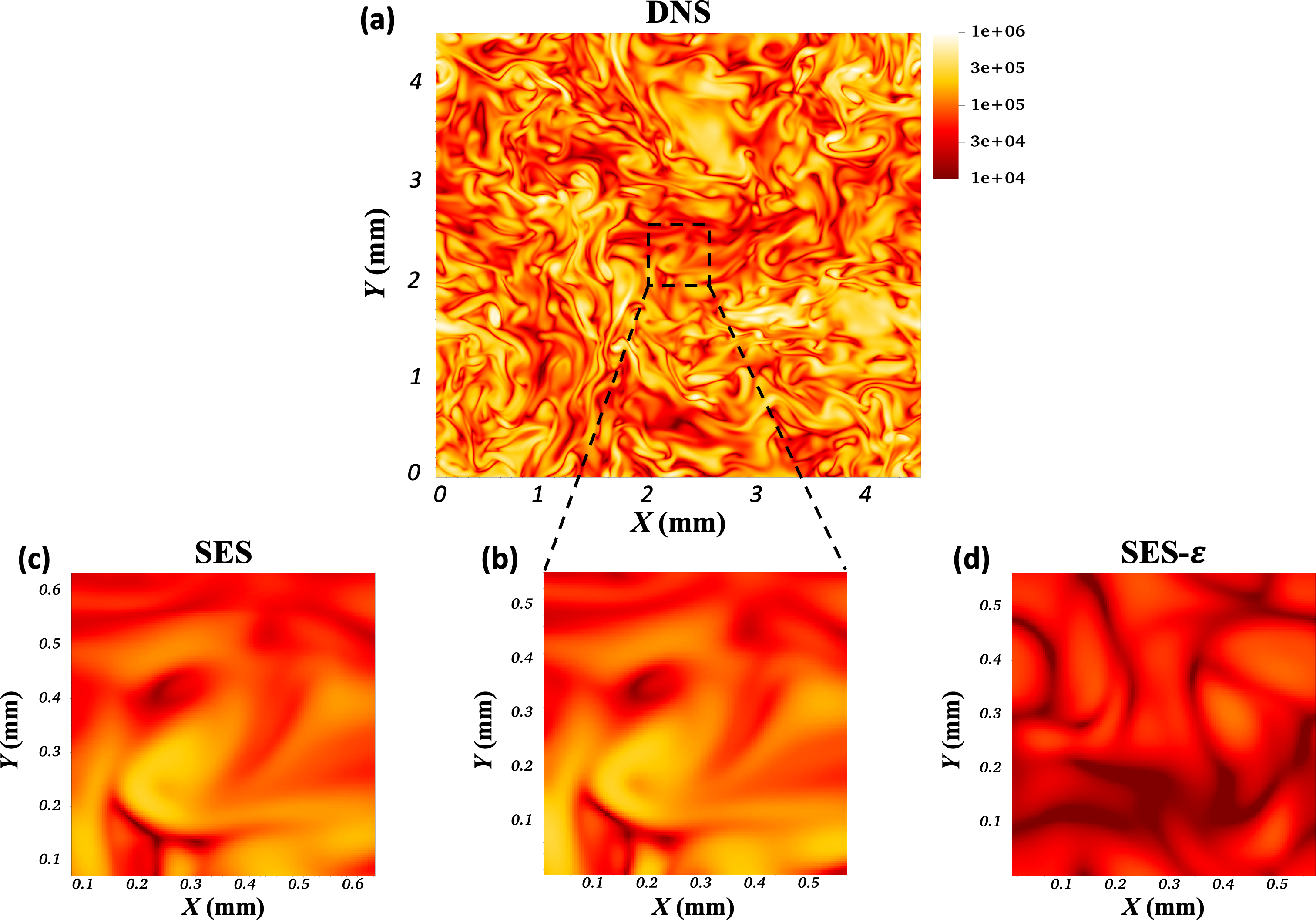}% Images in 100% size
\caption{Vorticity magnitude field at a representative time instant in the e-HIT simulations. (a) Full DNS, (b) magnified view of the central sub-region of the DNS, (c) SES, and (d) SES-$\epsilon$.}
\label{fig:LSES}
\end{figure*}

Figure~\ref{fig:LSES}a shows the instantaneous flow structure, in particular vorticity magnitude in the midplane of the computational domain, in the fully developed e-HIT DNS. Central region chosen for the SES is marked with a dashed black line. Magnified view of this central region is given in Fig.~\ref{fig:LSES}b. Corresponding flow fields in the SES and SES-$\epsilon$ at the same time instant are shown in Figs.~\ref{fig:LSES}c and \ref{fig:LSES}d, respectively. Forcing of the two SES was performed using the instantaneous DNS velocity fields, as described in \S\S~\ref{sec:lses} and \ref{sec:altmethods}.

Qualitatively, the flow in the SES calculation appears to be virtually indistinguishable from the corresponding region in the full DNS. At the same time, in the SES-$\epsilon$ calculation, not only is the flow structure completely different, but more importantly the vorticity magnitude is considerably lower. We quantify this observation below using $U_{rms}$. We do not show here the instantaneous flow structure in the SES-SMR calculations since they are not correlated with the DNS.

%, as well as the kinetic energy spectrum. Next, we examine how accurately the proposed methods can capture turbulence intermittency, in particular the statistics of enstrophy and dissipation rate. Finally, we consider the ability of SES, SES-$\epsilon$, and SES-SMR to provide accurate SGS closure terms relevant to the LES. More specifically, we compare the SGS dissipation at various scales. %These terms are described in detail in Appendix~\ref{appA}.
%Additional comparison of the closures in the energy conservation equation can be found in Appendix~\ref{app:EnergyClosures}.

%--------------------------------------------------------------------------------------------
\subsection{Global characteristics of turbulence}
\label{ssec:Ekin}

\begin{figure}
\includegraphics[width=0.45\textwidth]{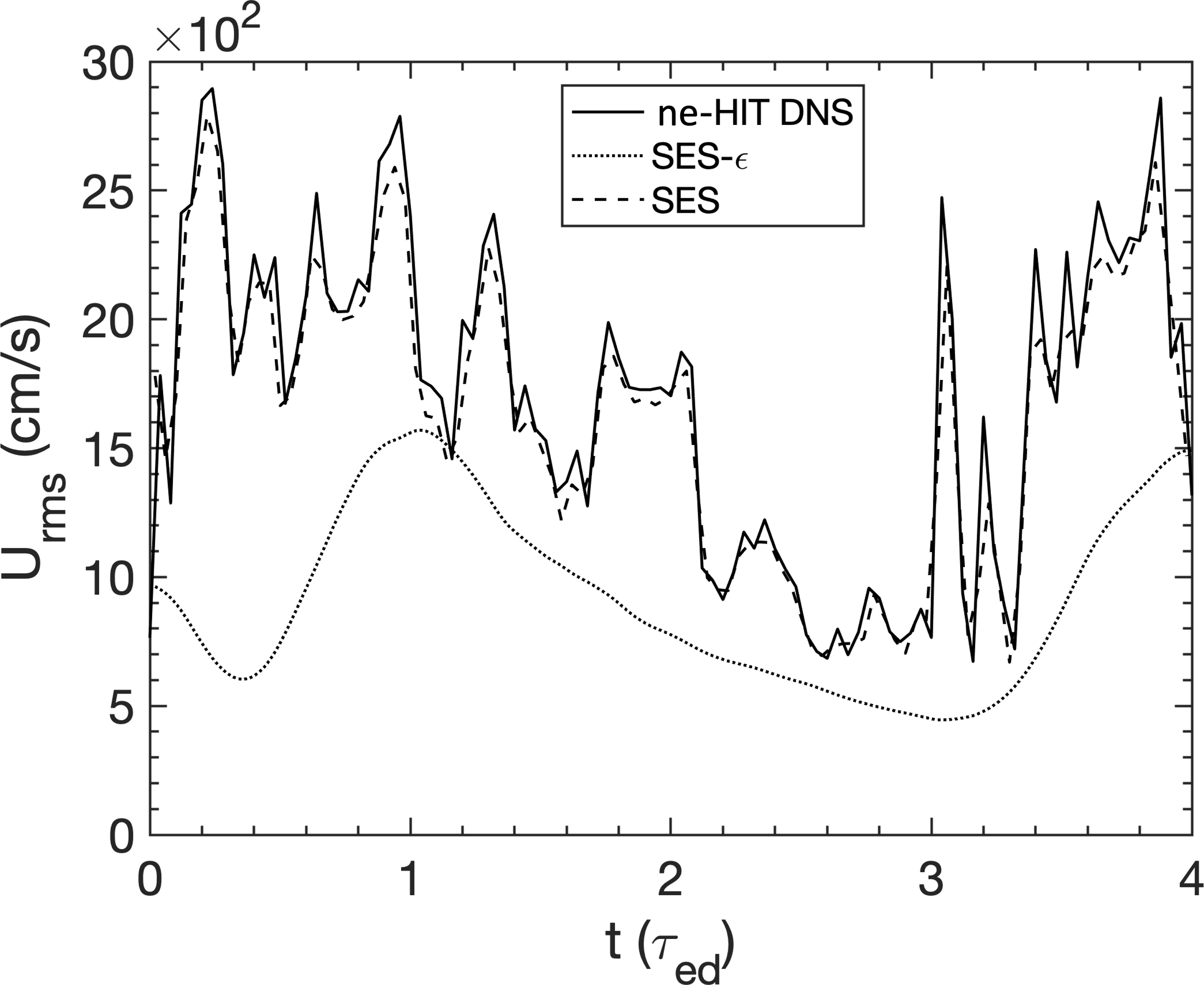}
  \caption{Evolution of the r.m.s. velocity in the ne-HIT DNS, SES, and SES-$\epsilon$. DNS values are considered over the central region only.}
\label{fig:UrmsUnsteady}
\end{figure}

While the qualitative comparison of the flow structure suggests that the SES calculation, unlike SES-$\epsilon$, is able to capture in detail the small-scale flow in the region of interest in the DNS, it is important to consider more quantitative metrics of the solution accuracy. First, we compare global turbulence characteristics, such as kinetic energy in the domain in terms of $U_{rms}$, Taylor scale $\lambda$ and the associated Reynolds number Re$_{\lambda}$, and the average Kolmogorov scale, $\eta$. For the equilibrium turbulence calculations, all statistics are averaged in time over $267$ samples during the final $2 \tau_{ed}$. For ne-HIT, statistics are considered with respect to time with all data being collected every $9.8 \times 10^{-6}$ s, or $0.02 \tau_{ed}$, from the start of the simulation.

Key turbulence properties obtained in all e-HIT SES tests are listed in Table~\ref{tab:Urms}, along with the values in the corresponding central region of the e-HIT DNS for comparison. In all calculations, turbulence is well resolved \citep{donzis2010,jagannathan2016,yeung2018} with the Kolmogorov scale
\begin{equation}
\eta = \Big(\frac{\nu^3}{\langle\epsilon'\rangle}\Big)^{1/4} \approx 2\Delta x.
\end{equation}
Here, $\Delta x$ is the computational cell size, $\epsilon'=2\nu S_{ij}S_{ij}$ is the dissipation (different from the energy injection rate $\epsilon$), $S_{ij}$ is the strain rate, and $\langle \epsilon'\rangle$ indicates averaging over space and time.

\begin{table}
  \begin{center}
\def~{\hphantom{0}}
  \begin{tabular}{lcccc}
        \hline
        & \quad $\overline{U}_{rms}$ (cm/s) \quad & \quad $Re_\lambda$ \quad \quad & \quad $\eta/\Delta x$ \quad \quad & \quad $\lambda/\eta$ \quad \quad \\ [3pt]
        \hline
       DNS              & 1797.3  & 145.6 & 2.04 & 23.75 \\
       SES              & 1681.5  & 114.2 & 1.91 & 20.75 \\
       SES-$\epsilon$   & 850.44  & 51.6  & 2.02 & 11.13 \\
       SES-SMR1         & 1454.1  & 88.3  & 1.84 & 18.72 \\
       SES-SMR2         & 1323.6  & 59.6  & 1.98 & 15.36 \\
       SES-SMR3         & 1422.2  & 78.7  & 1.85 & 17.24 \\
       \hline
  \end{tabular}
  \caption{Turbulence properties in the DNS and SES calculations.}
  \label{tab:Urms}
  \end{center}
\end{table}

The total amount of kinetic energy represented by the time-averaged $\overline{U}_{rms}$ shows a wide spread among different tests. SES gives by far the best agreement with the DNS, with $\overline{U}_{rms}$ being within $6.4\%$ of the DNS value. In contrast, in the SES-$\epsilon$, $\overline{U}_{rms}$ is more than a factor of two smaller. Finally, for all three SES-SMR tests (see \S~\ref{ssec:SMR}), the agreement is worse than in SES but substantially better than for SES-$\epsilon$ with the deficit relative to DNS ranging from $19\%$ to $26\%$. Note that such large deficit in total kinetic energy is observed regardless of whether the volumetric forcing is applied in the central high-resolution region (SES-SMR1 test), and also even when the total rate of kinetic energy injection in the domain is the same as in the DNS (SES-SMR1 and SES-SMR3 tests).

Similar trends are also observed for the turbulent Taylor scale
\begin{equation}
\lambda = \frac{\langle u_ju_j \rangle^{1/2}}{\langle(\partial u_i/\partial x_i)^2\rangle^{1/2}},
\label{eq:lambda}
\end{equation}
and the associated Taylor-scale Reynolds number
\begin{equation}
\textrm{Re}_{\lambda} = \frac{U_l\lambda}{\nu},
\label{eq:Rel}
\end{equation}
in the e-HIT calculations shown in Table~\ref{tab:Urms}. Here, $\langle ... \rangle$ again indicates space and time averaging, $U_l = U_{rms}/\sqrt 3$ is the integral velocity, and kinematic viscosity $\nu$ is given in Table~\ref{DNS_param}. In particular, while in the SES, Re$_{\lambda}$ is within $22\%$ of the DNS value, it is almost a factor of 3 too low in SES-$\epsilon$, and between $40\%$ to more than a factor of 2 too low for the SES-SMR tests. Similar level of agreement is also seen for the ratio $\lambda/\eta$.
%This shows that in terms of these global turbulence metrics, SES gives by far the best agreement with the DNS, with SES-$\epsilon$ failing to capture both the $U_{\rms}$ and $\lambda$ in the flow.

Figure~\ref{fig:UrmsUnsteady} shows $U_{rms}$ in the unsteady ne-HIT DNS, SES, and SES-$\epsilon$. In the SES, $U_{rms}$ (dashed line) tracks the DNS (solid line) very closely through all stages of the flow evolution including both forced equilibrium and non-equilibrium turbulence, as well as decaying turbulence. 

In contrast, while $U_{rms}$ in SES-$\epsilon$ (dotted line in Fig.~\ref{fig:UrmsUnsteady}) is generally correlated with the ne-HIT DNS on larger timescales and it captures all the transient phases of the flow, the actual values of $U_{rms}$ are again substantially lower. Thus, L/SES-$\epsilon$, which relies only on the large-scale energy transfer rate in the cascade, significantly underpredicts kinetic energy in the region of interest both in the e-HIT and ne-HIT, thereby also leading to much lower Reynolds numbers compared to the host flow (cf. Table~\ref{tab:Urms}).

This disagreement cannot be attributed solely to the under-prediction of $\epsilon$ by the SES-$\epsilon$ procedure (\S~\ref{ssec:lses-eps}). For instance, it was stated in \S~\ref{ssec:SES} that $\epsilon$ in the e-HIT SES-$\epsilon$ is $\approx 26\%$ lower than in the corresponding DNS. This, however, would reduce $U_{rms}$ by only $10\%$ based on the Kolmogorov scaling, and thus this difference in $\epsilon$ would not explain the fact that $U_{rms}$ in SES-$\epsilon$ is twice lower than in the e-HIT DNS. Instead, the cause of much lower kinetic energy in the SES-$\epsilon$ is the fact that in this approach, turbulent velocities at the largest scale $L_{SES}$ in the domain are comparable to the velocities at the same scale in the DNS, but much smaller than the velocities at the largest scale of the DNS $L_{DNS} \gg L_{SES}$. As a result, kinetic energy in the domain, and thus $\lambda$ and Re$_{\lambda}$, are significantly under-predicted compared to the DNS. This is confirmed if we compare the ratio of $U_{rms}$ in the SES-$\epsilon$ and DNS with the inertial range scaling law \citep{pope2001turbulent}
\begin{equation}
    \frac{U_{rms,SES-\epsilon}}{U_{rms,DNS}}=0.47 \approx \Big(\frac{L_{SES}}{L_{DNS}}\Big)^{1/3}=0.50.
\end{equation}

%Thus in SES-$\epsilon$, $U_{rms}$ is indeed representative of the characteristic velocities in the DNS at the scale $L_{SES}$ rather than $L_{DNS}$

%For equilibrium SESes forced by higher Reynolds number LES simulations, no reference DNS $U_{rms}$ is available. However $U_{rms}$ for these SESes are found to be comparable to the corresponding LES value, the potential of this method to accurately represent the dynamics of a large Reynolds number flow.
%$U_{rms}$ from L/SES-M1 for both simulations is much lower than the DNS, as is also reflected by a much lower Taylor scale Reynold's number (table \ref{Eqm_param}), defined as $Re_\lambda=u_0\lambda/\nu$ where $u_0$ is the integral velocity and $\lambda$ is the Taylor length scale. This indicates that reducing the external field to a single scalar $\epsilon$ and using it for energy injection at the largest scale is inadequate. However the general trend of non-equilibrium turbulence is qualitatively observed, and $U_{rms}$ agreement is not too bad in the turbulence decay phase.

%This is despite the fact that the energy injection rate determined using the approach described in \S~\ref{ssec:lses-eps} in the SES-$\epsilon$ calculation was very close to that in the DNS, namely $???$ vs. $5.85\times 10^6$ erg/cm$^3$ s. 

In contrast, L/SES does not suffer from this deficiency. While not capturing the large scales explicitly, the SES domain nevertheless contains the velocities associated with the large-scale motions since they are directly introduced into the SES domain via the forcing procedure. As a result, the L/SES method can generate turbulence with global characteristics very close to a full-scale simulation both in the steady-state and unsteady turbulence despite having a much smaller dynamical range of scales available in the domain.
%As a result, SES is able to recover the correct kinetic energy in the DNS.

%--------------------------------------------------------------------------------------------
\subsection{Kinetic energy spectra}
\label{ssec:spectra}

Next, we compare the spectral distribution of kinetic energy density in all calculations. In the periodic DNS and SES-$\epsilon$ domains, 1D kinetic energy spectrum is calculated in the usual way. For the DNS, spectrum is taken over the entire domain to show the energy content at all wavenumbers.

At the same time, since SES uses a non-periodic domain, while SES-SMR has a non-periodic, high-resolution central region of interest, care must be taken obtaining the corresponding turbulent spectra. To minimize the aliasing errors, we use a Hann window function of the form
\begin{equation}
    w=cos^2\big(\frac{\pi x}{L_x}\big) cos^2\big(\frac{\pi y}{L_y}\big) cos^2\big(\frac{\pi z}{L_z}\big).
\label{eq:HannWindow}
\end{equation}
Here, $x$, $y$, $z$ are the cell-center coordinates relative to the center of the domain, and $L_x=L_y=L_z=L_{SES}$.

\begin{figure*}
\includegraphics[width=0.48\textwidth]{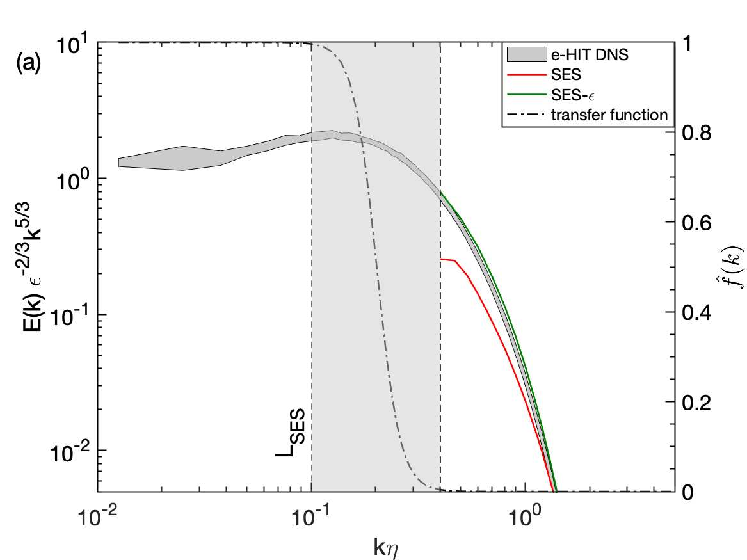}
\includegraphics[width=0.48\textwidth]{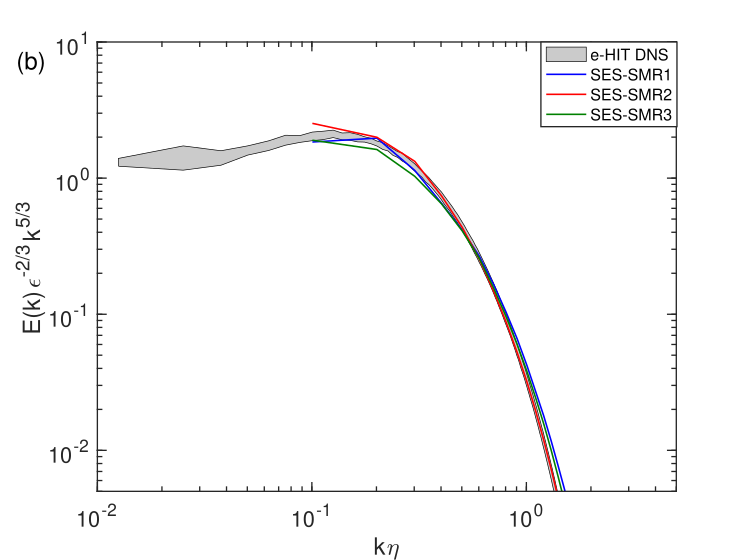}
  \caption{Time-averaged kinetic energy spectra in the e-HIT SES and SES-$\epsilon$ calculations (panel a) and the e-HIT SES-SMR tests (panel b). 
  %The $t=1 \tau_{ed}$: DNS driving is disabled, $t=2 \tau_{ed}$: middle of the time interval when turbulence is decaying, $t=3 \tau_{ed}$: DNS driving is restarted, and $t=4 \tau_{ed}$: end of the simulation.
  Both panels show the range of spectra variability in the e-HIT DNS (gray band). All spectra are compensated by the Kolmogorov scaling. In panel (a), scales affected by filtering are marked as a vertical gray region, and SES spectra are not shown in this range. Transfer function of the filter is shown on the right axis of panel (a).}
\label{fig:spectra_eq}
\end{figure*}

Resulting spectrum of the windowed field needs to be appropriately re-scaled to compensate for the reduction in the energy caused by the windowing operation and thus to ensure that the windowed field has the same energy as the original field. The exact correction factor is signal dependent. Since the velocity-field signal can be considered stationary in space due to the flow homogeneity, the correction factor can be estimated as the inverse of the r.m.s. of the window function, $w_{rms}$. For the Hann window in 1D,
\begin{equation}
    w_{rms}=\Big(\frac{1}{L_x}\int_{-L_x/2}^{L_x/2} \cos^4\big(\frac{\pi x}{L_x}\big) dx\Big)^{1/2}=\frac{3}{8}.
\end{equation}
Thus, the kinetic energy spectral density of the windowed velocity field must be scaled by the factor $((8/3)^3)^2$.

Finally, another correction must be introduced in the SES and SES-SMR spectra to account for the fact that velocities associated with scales larger than $L_{SES}$ are present in the SES domain. Such large-scale motions cannot be properly represented in the spectrum, which does not contain scales $> L_{SES}$. At the same time, energy associated with such motions is distributed among the available scales in the SES spectrum since the total kinetic energy in the SES domain and in the spectrum must be equal to satisfy the Parseval's theorem (also cf. $U_{rms}$ in Table~\ref{tab:Urms}). This leads to a higher spectral kinetic energy density in the SES. To correct for this, another scaling factor is applied, which represents the ratio of the energy on small scales $< L_{SES}$ to the total specific kinetic energy in the SES 
\begin{equation}
    \frac{E^{SS}}{E}=\frac{U_{rms}^2-\tilde{U}_{rms}^2 }{U_{rms}^2}.
\label{spec_scale_fac}
\end{equation}
Here, $U_{rms}$ and $\tilde{U}_{rms}$ are the r.m.s. of the raw and filtered velocity field respectively, with filtering applied at the scale of the SES active domain size $L_{SES}$, and $SS$ represents scales smaller than this cutoff. Therefore, finally, all SES and SES-SMR spectra are normalized by the factor $ (E^{SS}/E)/w_{RMS}^6$.

\begin{figure}[ht]
\includegraphics[width=0.45\textwidth]{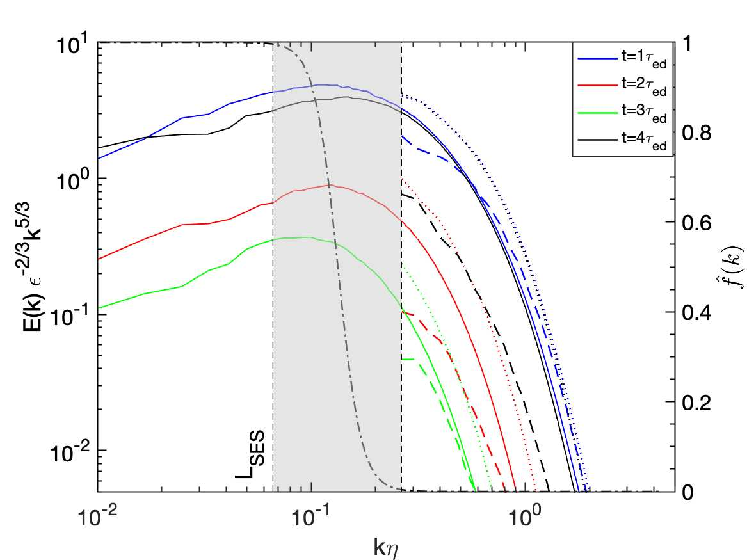}
  \caption{Kinetic energy spectra at a few representative time instants (cf. Fig.~\ref{fig:Unst_Urms}) in the ne-HIT SES and SES-$\epsilon$ calculations (see text for further details). Solid lines: ne-HIT DNS, dashed lines: SES, dotted lines: SES-$\epsilon$. All spectra are compensated by the Kolmogorov scaling. Scales affected by filtering are marked as a vertical gray region, and SES spectra are not shown in this range. Transfer function of the filter (dash-dot line) is shown on the right axis.}
\label{fig:spectra_neq}
\end{figure}

Figure~\ref{fig:spectra_eq}a shows the time-averaged, compensated energy spectra in the e-HIT SES and SES-$\epsilon$, along with the range of variability of the spectra in the DNS. SES-$\epsilon$ spectra agree closely with the DNS, which is expected since SES-$\epsilon$ obtains $\epsilon$ directly from the DNS, and it is forced in the same way as the DNS. For the SES, agreement with the DNS is slightly worse, in particular at scales close to the filter scale, $\Delta$. This is the result of these scales being affected by the large-scale flow-field injection. At sufficiently high Re$_{\lambda}$, the impact on small scales that one is interested in should be minimal. Note that limited spectral sharpness of the filter affects scales close to the filter size $\Delta$ (see \S~\ref{sssec:filter}). Therefore, scales in the range [$\Delta/2, 2\Delta$], represented by the vertical gray region, were removed from the SES spectra in both panels. Similar comparison of the time-averaged spectra for the SES-SMR tests, is shown in Fig.~\ref{fig:spectra_eq}b. Close agreement with the DNS, similar to that for SES-$\epsilon$, can be seen.

\begin{figure*}
\includegraphics[width=0.9\textwidth]{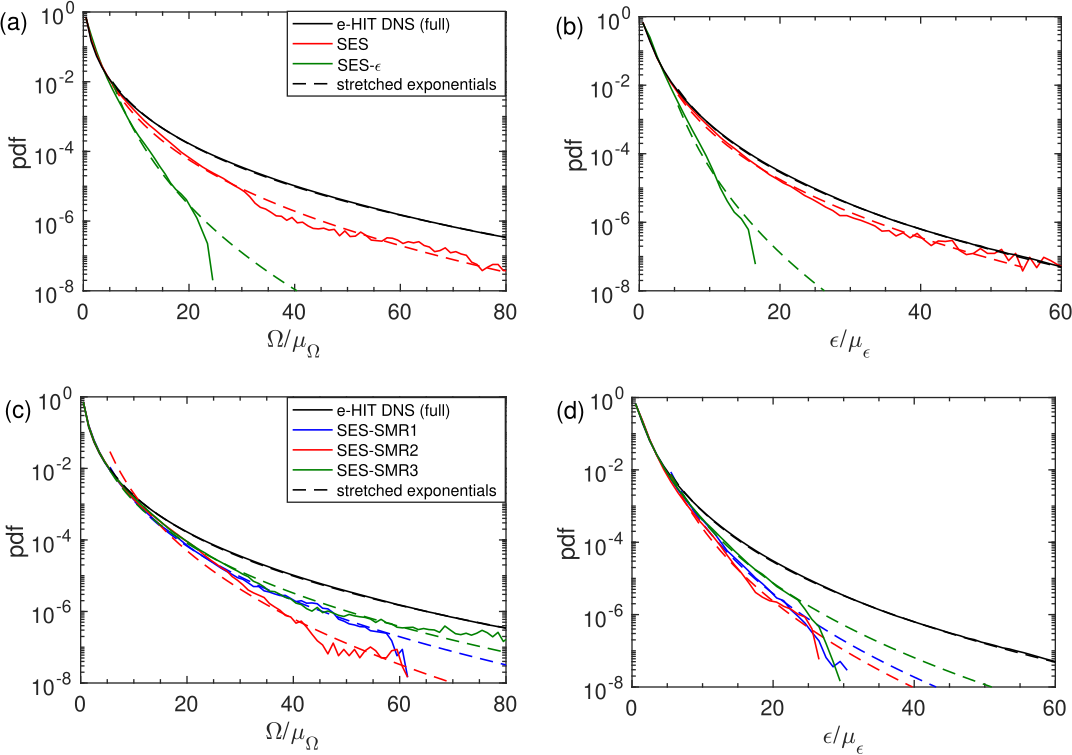}
  \caption{Time-averaged PDF of enstrophy, $\Omega$ (a, c), and dissipation rate, $\epsilon$ (b, d), for the e-HIT DNS (full domain), SES and SES-$\epsilon$ (a, b), and SES-SMR (c, d). Dashed lines show the stretched exponential fits.}
\label{fig:pdf_st}
\end{figure*}

Evolution of spectra in the unsteady ne-HIT case is illustrated in Fig.~\ref{fig:spectra_neq}, which shows the instantaneous compensated spectra in the SES and SES-$\epsilon$ calculations at several representative time instants. These correspond to the end of the forcing stage when turbulence becomes fully developed ($t = \tau_{ed}$), middle of the decaying turbulence phase ($t = 2\tau_{ed}$), time when forcing is restarted ($t = 3\tau_{ed}$), and end of the simulation when turbulence again re-equilibrated ($t = 4\tau_{ed}$) (cf.~Fig.~\ref{fig:Unst_Urms}).  Again, similar to Fig.~\ref{fig:spectra_eq}, scales affected by the filter are excluded. 
For the SES-$\epsilon$ (dotted lines), agreement with the DNS (solid lines) is worse than in the e-HIT case, however, the SES-$\epsilon$ is able to follow the overall spectral evolution over different stages. For the SES (dashed lines), agreement with the DNS is comparable to or better than for the SES-$\epsilon$, with the exception of $t = 4\tau_{ed}$, where the SES spectrum deviates more strongly from the DNS. %The full evolution of the SES and SES-$\epsilon$ spectra with time in the ne-HIT case can be found in the animations provided in the Supplementary Materials.

%However other higher order metrics are needed to quantify the accuracy of the method. Also spectrum test is not clean for L/SES-M2 and has many approximations.

% \begin{figure}
%   \centerline{\includegraphics[width=10cm]{spectrum_unst.png}}
%   \caption{Spectrum comparison at a few selected time instants for NE-HIT. $t=1 \tau_{ed}$ represents the time just before the DNS driving is disabled, $t=2 \tau_{ed}$ represents a time where the turbulence has decayed for $1 \tau_{ed}$, $t=3 \tau_{ed}$ is the instant just before the driving is restarted, and $t=4 \tau_{ed}$ is the end of the simulation where the turbulence has developed for $1 \tau_{ed}$}
% \label{fig:spectrum_unst}
% \end{figure}

%--------------------------------------------------------------------------------------------
\subsection{Small-scale intermittency}
\label{ssec:intermittency}

While turbulent velocities in the HIT are Gaussian \citep{donzis2008dissipation}, velocity gradients and the associated small-scale quantities, such as enstrophy, $\Omega=\omega_i\omega_i/2$, where $\pmb{\omega}=\pmb{\nabla}\times\pmb{u}$ is the vorticity, and dissipation rate, $\epsilon=2\nu S_{ij}S_{ij}$, where $S_{ij}$ is the strain rate, exhibit large local deviations from the mean with their PDF being strongly non-Gaussian \citep{donzis2008dissipation}. Such small-scale intermittency is one of the central characteristics of the nonlinear dynamics of a fully developed turbulence \citep{tsinober2009informal,chen1997inertial,zeff2003measuring}. Thus, the accuracy with which it can be captured is a sensitive test of the overall quality of the solution. 
%Turbulence models have been developed to explicitly incorporate these effects through an intermittency transport equation \citep{cho1992ak}. However previous studies on intermittency from LES were restricted to intermittencies in first-order variables such as velocity or passive scalar \citep{gilliland2012external}, but not higher order quantities like enstrophy. Capturing small-scale intermittency in higher-order quantities is relevant for example in reacting flows, where it can affect flame extinction and re-ignition, and its statistical properties are also used in PDF-based combustion models.
In particular, a question could be asked whether an SES calculation with a much smaller range of available scales compared to the DNS, and thus also a much smaller spatial ensemble size, can represent accurately the statistics of the velocity-gradient-based quantities.
%Enstrophy is defined as  and the dissipation rate is given by  defined previously. These quantities are important as they represent the rotation of turbulence and the local straining, respectively. 

For the e-HIT cases, Fig.~\ref{fig:pdf_st} shows the time-averaged PDF of the enstrophy and the dissipation rate normalized by their corresponding mean values. %For the DNS, in panels (a) and (b), PDF are shown both for the entire domain and for the central region, which corresponds to the SES. While for $\Omega$, these two PDF are very close, for dissipation central region exhibits stronger intermittency. 
%Intermittencies from the entire DNS region show a similar behaviour, due to the ergodic nature of turbulence \citep{galanti2004turbulence} and are also plotted for reference.
Panels (a) and (b) show that SES-$\epsilon$ significantly under-predicts the intermittency of both quantities due to the lower $U_{rms}$ and $\lambda$, and thus $Re_\lambda$. At the same time, in the SES, statistics of $\Omega$ and $\epsilon$ are much closer to those in the DNS. In particular, while for enstrophy, the SES PDF is somewhat lower than in the DNS, for dissipation both PDF are in close agreement. Observed discrepancy in the PDF of $\Omega$ between the DNS and SES is the result of the relatively low Re$_{\lambda}$ of the flow and thus a limited range of scales in the SES. Sufficient separation between the large scales, which are being forced, and the small scales, on which the statistics is collected, is required to minimize the impact of forcing on the latter. Analysis of the intermittency properties in the SES for a wide range of Re$_{\lambda}$ will be presented in the follow-up paper.

\begin{figure*}
\includegraphics[width=0.9\textwidth]{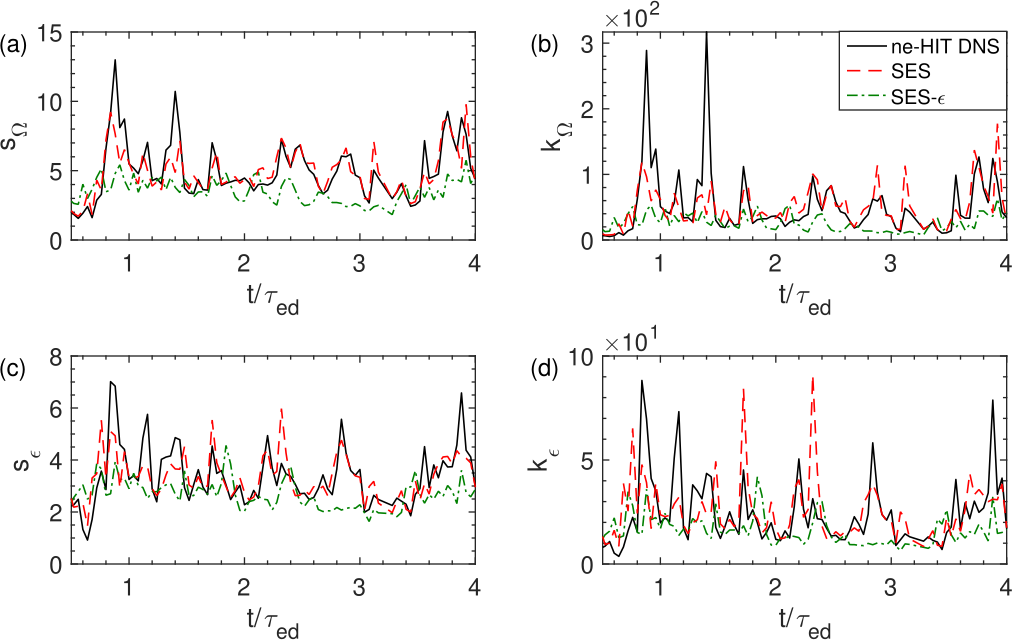}
  \caption{Skewness (a, c) and kurtosis (b, d) of the PDF of enstrophy, $\Omega$ (a, b), and dissipation rate, $\epsilon$ (c, d), in the ne-HIT cases as a function of time. DNS values are shown only for the central region of the domain.}
\label{fig:skew_kurt_unst}
\end{figure*}

\begin{table*}
\renewcommand{\arraystretch}{1.2}
  \begin{center}
\def~{\hphantom{0}}
  \begin{tabular}{p{0.1\textwidth}>{\centering}p{0.05\textwidth}>{\centering}p{0.05\textwidth}>{\centering\arraybackslash}p{0.05\textwidth}>{\centering\arraybackslash}p{0.05\textwidth}>{\centering\arraybackslash}p{0.05\textwidth}>{\centering\arraybackslash}p{0.05\textwidth}>{\centering\arraybackslash}p{0.05\textwidth}>{\centering\arraybackslash}p{0.05\textwidth}}%{lcccccccc}
        \hline
        & $\sigma_\Omega$ & $s_\Omega$ & $k_\Omega$ & $b_\Omega$ & $\sigma_\epsilon$ & $s_\epsilon$ & $k_\epsilon$ & $b_\epsilon$  \\[3pt]
        \hline
       DNS              & 1.89 & 8.13 & 182.20 & 7.04  & 1.26 & 4.86 & 59.75 & 9.55    \\
       SES              & 1.59 & 4.68 & 43.47  & 8.48  & 1.14 & 4.02 & 37.75 & 9.98    \\
       SES-$\epsilon$   & 1.20 & 3.29 & 22.36  & 13.15 & 0.91 & 2.28 & 11.36 & 16.98   \\
       SES-SMR1         & 1.57 & 4.34 & 35.14  & 8.73  & 1.09 & 3.13 & 19.71 & 13.28   \\
       SES-SMR2         & 1.58 & 4.26 & 33.15  & 10.84  & 1.02 & 2.87 & 17.12 & 13.77   \\
       SES-SMR3         & 1.61 & 4.81 & 48.30  & 8.02  & 1.08 & 3.15 & 19.61 & 11.87   \\
       \hline
  \end{tabular}
  \caption{Statistical parameters of the PDF of enstrophy and dissipation in the e-HIT calculations. See eqs.~(\ref{eq:sigma})-(\ref{eq:stretched_exp}) for the definitions.}
  \label{tab:skew_kurt}
  \end{center}
\end{table*}

For the e-HIT SES-SMR tests shown in Fig.~\ref{fig:pdf_st}c,d, PDF generally have worse agreement with the DNS exhibiting suppressed PDF tails at extreme values, which is consistent with the lower Re$_{\lambda}$ in these calculations. This disagreement is particularly pronounced in the PDF of dissipation. The only notable exception is the PDF of enstrophy in the SES-SMR3, which is comparable to the SES. Overall, SES-SMR are somewhat better able to capture turbulence intermittency than SES-$\epsilon$, albeit not to the degree of SES.

%This further demonstrates that information carried by intermediate scales in the range $(L_{SMR},L_{LES})$ is required for accurate small-scale behaviour and can only be transferred to a smaller simulation if all such scales are forced volumetrically. However, such information can be partially exchanged through the boundary interfaces, which leads to higher intermittency in SMR compared to a priori L/SES-M1 (table \ref{tab:skew_kurt_large}).

%Although the agreement is not perfect, the performance is much better than L/SES-M1, particularly for the most extreme events. 

To compare the obtained statistics more quantitatively, we examine the $2^{\mathrm{nd}}$-, $3^{\mathrm{rd}}$-, and $4^{\mathrm{th}}$-order central moments of both enstrophy and dissipation, i.e., standard deviation, skewness, and kurtosis, defined respectively as 
\begin{eqnarray}
    \sigma_X &=& \langle (X-\mu_X)^2 \rangle^{1/2}, \label{eq:sigma} \\
    s_X      &=& \langle (X-\mu_X)^3 \rangle/\sigma_X^3, \label{eq:skew} \\
    k_X      &=& \langle (X-\mu_X)^4 \rangle/\sigma_X^4, \label{eq:kurt} 
\end{eqnarray}
where $\mu_X$ is the mean of a given quantity $X$. Values of these three moments for both quantities are given in Table~\ref{tab:skew_kurt}. Obtained DNS values are in agreement with prior findings  \citep{donzis2008dissipation,hamlington2012intermittency}, and furthermore they are in much closer agreement with the SES than SES-SMR and especially SES-$\epsilon$.

\begin{figure*}
\includegraphics[width=0.9\textwidth]{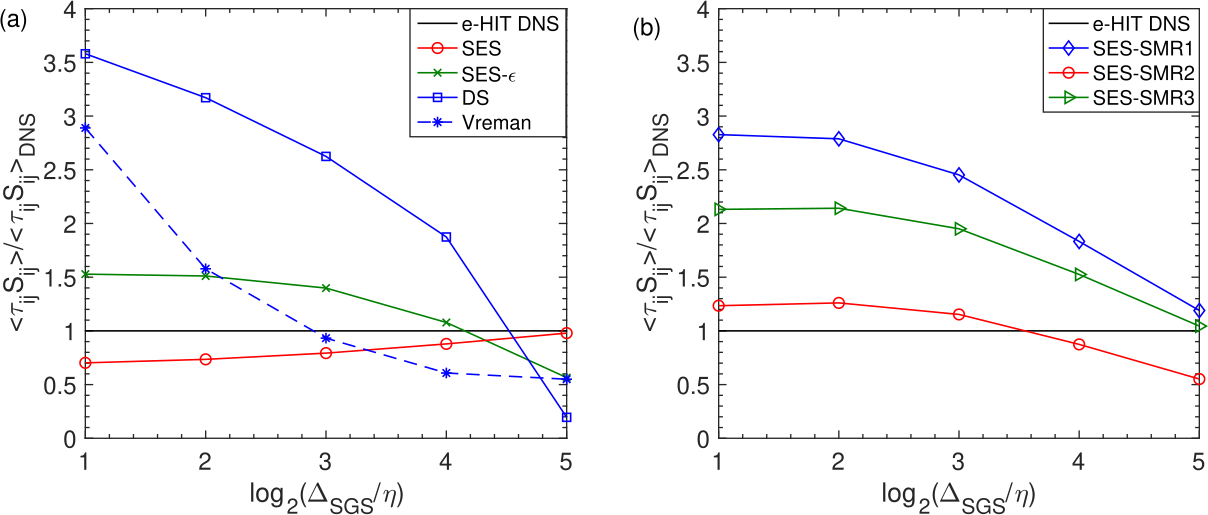}
  \caption{Time- and space-averaged mean sub-filter scale dissipation, $<\tau_{ij}S_{ij}>$, normalized with the DNS results as a function of the filter scale in the e-HIT SES and SES-$\epsilon$ (panel a) and in the SES-SMR (panel b). In panel (a), results of the DS and Vreman SGS models are shown for comparison.}
\label{fig:diss}
\end{figure*}

Another common way to characterize the PDF is by fitting them with the stretched exponential functions \citep{donzis2008dissipation}
\begin{equation}
    f(\psi) \sim exp(-b_\psi \psi^{1/4}),
\label{eq:stretched_exp}
\end{equation}
where $\psi$ is a normalized quantity $\Omega/\mu_\Omega$ or $\epsilon/\mu_\epsilon$, and $b_{\psi}$ is the parameter that gives the best fit of the observed PDF. Stretched exponential fits are typically obtained for values $\psi > 5$, and they are also shown in Fig.~\ref{fig:pdf_st} as dashed lines. Corresponding values of the parameter $b$ are listed in Table~\ref{tab:skew_kurt}. Again, similar to the moments, values of $b$ for the DNS are in agreement with \citet{donzis2008dissipation}, and they are closer to the SES than to the SES-SMR and SES-$\epsilon$. In particular, in contrast to the SES, the shapes of the PDF in SES-$\epsilon$ and SES-SMR do not agree well with the stretched exponential functions, especially at the extreme tails.

%The discrepancies between DNS and L/SES-M2 suggest that forcing turbulence using this method still does not result in a flow with $Re_\lambda$ equivalent to the DNS, however, it can be used to obtain flow fields that are characteristic of much higher Reynolds numbers compared to conventional driving methods. A more thorough quantification of intermittency can be done based on structure functions, however, it is not the primary focus of the present study and will be addressed in a future work.

% \begin{figure}
%   \centerline{\includegraphics[width=11cm]{pdfs_st.eps}}
%   \caption{Time averaged PDFs of $\textit{(a)}$ enstrophy ($\Omega$) and $\textit{(b)}$ dissipation rate ($\epsilon$) for E-HIT along with 1 standard deviation variability.}
% \label{fig:pdf_st}
% \end{figure}

PDF for the ne-HIT cases cannot be averaged over time. Instead, Fig.~\ref{fig:skew_kurt_unst} shows the evolution of skewness and kurtosis of the PDF of enstrophy and dissipation in time both in the DNS and in the SES and SES-$\epsilon$. Note that when considering the instantaneous statistics, only the central DNS region must be compared with the SES due to the large difference in the ensemble size between the central region and the full DNS domain. Once again, SES demonstrates closer agreement with DNS than SES-$\epsilon$. Skewness appears to be better correlated with DNS in time for the SES, compared to kurtosis. In contrast, SES-$\epsilon$ shows lower skewness and kurtosis due to the lower Re$_\lambda$. %The time evolution of the PDF for all the ne-HIT cases can be found in the animations provided in the Supplementary Materials.

%We have shown based on several metrics that the performance of L/SES-M2 is comparable to DNS in a priori tests.

%Since the performance of L/SES-M2 was better than L/SES-M1 in terms of intermittencies and SGS quantities in the a priori case, L/SES-M1 is not used for the a posteriori case.
% at two scales, and is used to drive two separate L/SES-M2 simulations.    Two L/SES simulations are then performed   .

%The use of sponge regions to relax the solution to the forcing field could help partially mitigate the problem, but inaccuracies in this region can advect and affect the core region of the simulation.  the overall filtering behaviour is not well-defined.

%The Germano filter was successful for the a priori tests because the forcing field was obtained directly from a DNS processed by the same filter. This could have helped partially mitigate some of the errors as the spectral leakage in the L/SES would be matched and compensated by the leakage from the filtered forcing field. However this would not be the case if the forcing fields are taken from an LES which does not have any high wavenumber component by design ($\widetilde{u_{i,SS}^{LES}}=0$), which may cause a non-zero $\widetilde{u_{i,SS}^{L/SES}}$ to remain unbalanced and remove excessive energy from small scales during L/SES forcing. Moreover we would like to develop a general method that could work with large scale fields filtered with any filter. 

%--------------------------------------------------------------------------------------------
\subsection{LES closures and SGS quantities}
\label{ssec:closures}

The last set of test metrics we describe concerns the accuracy of the proposed L/SES methods in capturing the LES closure terms. %Obtaining these quantities is essential to allow the use of the L/SES method for developing and validating new LES models.
A detailed description of the LES equations and the associated unclosed terms is provided in Appendix~\ref{appA}. Both in the reference DNS and in the SES calculations, all scales are fully resolved, and thus any unclosed terms of interest can be directly computed by filtering the flow fields.

\begin{figure*}[ht]
\includegraphics[width=0.9\textwidth]{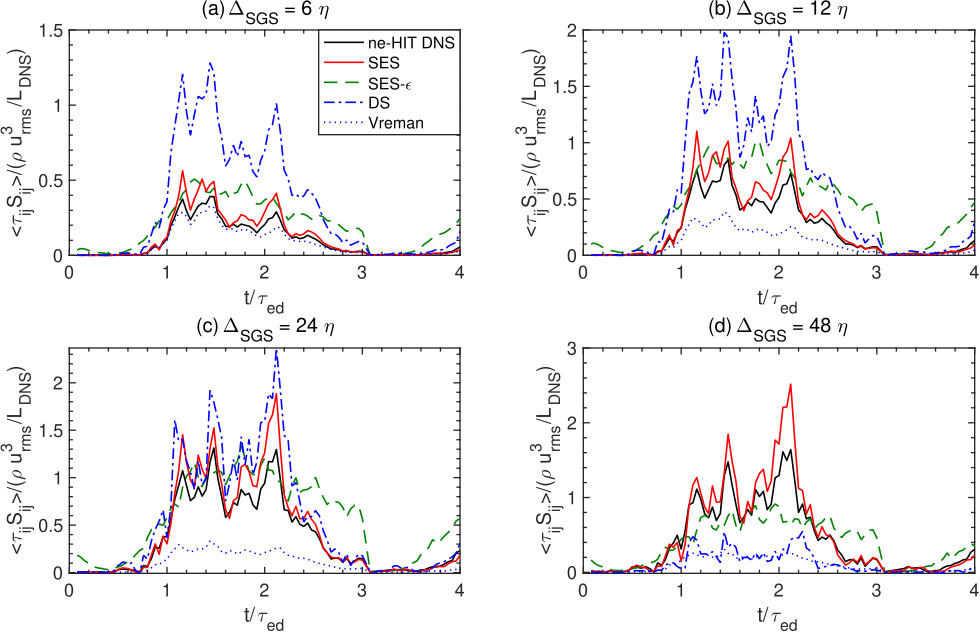}
  \caption{Mean sub-filter scale dissipation normalized by $U_{rms,DNS}^3/L_{DNS}$ as a function of normalized time at various scales for the ne-HIT case. Spatial averaging is performed only over the central region in the DNS.}
\label{fig:diss_NE-HIT}
\end{figure*}

In weakly compressible low-Mach-number flows, the dominant closure term is the Reynolds stress, $\tau_{ij}$ (eq.~\ref{eq:tauij}). In particular, since the role of the SGS terms is to describe the effect of small scales, which are primarily dissipative in nature, it is important for the SGS model to produce the correct SGS dissipation, which is given by the product of the Reynolds stress with the resolved strain rate tensor. %Obtaining accurate SGS dissipation is a necessary condition to recover the correct second-order moments of the velocities \citep{meneveau1994statistics}.
%In particular, one of the central unclosed LES terms is the SGS dissipation, given by the product of the Reynolds stress, $\tau_{ij}$ (eq.~\ref{eq:tauij}), with the resolved strain rate tensor. It is the dominant term for weakly compressible flows.
%\citet{meneveau1994statistics} showed that a necessary condition for obtaining correct second-order moments of velocity from an LES is for a closure model to produce the correct mean SGS . 

As LES closures may be needed at any scale depending on the LES resolution, we compare the sub-filter scale dissipation at several different scales, $\Delta_{SGS}$, by calculating it using the $8^{\mathrm{th}}$-order differential filter. For the e-HIT cases, it is temporally and spatially averaged and normalized by the corresponding DNS mean. Resulting values are shown in Fig.~\ref{fig:diss}a for the SES and SES-$\epsilon$, and in Fig.~\ref{fig:diss}b for the SES-SMR tests. To ensure that the same ensemble size is used in the comparison with the DNS values, the spatial average for the reference DNS is taken just over the central sub-region corresponding to the SES.

Obtained results are also compared in Fig.~\ref{fig:diss}a with the two widely used SGS models: the Dynamic Smagorinsky (DS) \citep{moin1991dynamic} and Vreman models \citep{vreman2004eddy}. For these models, we form a pseudo-LES field by explicitly filtering the resolved DNS data, and then we apply test filters directly on the resulting data. Further summary of the DS and Vreman model coefficients is provided in eqs.~(\ref{eq: DS closure1})-(\ref{eq: DS closure_last}) and eqs.~(\ref{eq: Vre closure1})-(\ref{eq: Vre closure_last}), respectively, in Appendix~\ref{appA}. We emphasize that our goal is not to analyze the fidelity of these two models, but rather we provide them here for reference.

Figure~\ref{fig:diss}a shows two opposite trends for the SES and SES-$\epsilon$. At the smallest filter scale close to the Kolmogorov scale, both SES and SES-$\epsilon$ are off the target DNS value, with SES being in better agreement with the DNS. However, at larger filter scales, SES values approach the DNS becoming virtually identical with the DNS at $\Delta_{SGS} = 32\eta$ or one half of the SES domain width, which is the filter scale of the SES calculation. This improvement in the accuracy of the sub-filter scale dissipation in the SES on larger scales right up to the scales approaching the inertial range is an important property, as closures are sought at the relevant LES filter scales, which are $\gg \eta$ in practical situations. In contrast, SES-$\epsilon$ values start to decrease with increasing $\Delta_{SGS}$ becoming approximately half of the DNS value at the largest filter scales. This suggests that SES-$\epsilon$ would not be able to provide reliable values of the SGS dissipation on large scales.

%All SES-SMR test show a very similar trend, which suffers from the same deficiency as the SES-$\epsilon$, namely accuracy of the predicted values of the SGS dissipation becomes progressively worse on larger scales with the error reaching $50\%$ on the largest scale $L_{SES}/2$. This indicates that simply refining the mesh would not allow one to obtain reliable SGS closures.

All SES-SMR cases show a trend similar to SES-$\epsilon$, though they tend to be much more dissipative. SES-SMR2 produces the closest agreement with DNS at the smaller scales, however its accuracy drops on larger scales reaching $\approx 50\%$ of the DNS value at the SES filter scale $L_{SES}/2$. In contrast, SES-SMR1 and SES-SMR3 have the worst disagreement with DNS on small scales, but they give better agreement at the largest scales. If LES closures are to be obtained using SMR at some particular scale, it is not clear based on these tests which method of forcing the SMR would be the most appropriate. In general, SMR results tend to disagree with the DNS at most of the scales. This indicates that simply refining the mesh would not allow one to obtain reliable SGS closures.

It is interesting to note that the DS and Vreman models exhibit much worse agreement with DNS than any of the SES tests. In particular, the disagreement is the largest at the smallest dissipative scales close to $\eta$. Higher dissipation of these SGS models is in accordance with previous findings \citep{li2006subgrid,fureby1997comparative}. At larger scales, predicted dissipation tends to decrease eventually crossing the DNS values, and at the largest scales, DS and Vreman models tend to under-predict the SGS dissipation.

Finally, for the unsteady ne-HIT, evolution of the spatially averaged sub-filter scale dissipation, which is non-dimensionalized by $\rho U_{rms,DNS}^3/L_{DNS}$, is shown in Fig.~\ref{fig:diss_NE-HIT} for several filter scales. Again, SES shows very close correlation with the DNS values throughout all stages of turbulent growth and decay. This demonstrates the general and flow-agnostic nature of the L/SES, which does not require any calibration for different types of flows. In contrast, SES-$\epsilon$, while exhibiting comparable values to DNS, is virtually uncorrelated with the DNS. Furthermore, both SES and SES-$\epsilon$ outperform the existing DS and Vreman models in this unsteady, non-equilibrium turbulent flow.

\section{Discussion and conclusions}
\label{sec:discussion}

We presented a description of a novel multi-fidelity Large/Small Eddy Simulation (L/SES) method for studying turbulent flows at high Reynolds numbers. The key aspect of the proposed approach is the coupling of a lower-fidelity, unresolved, time-dependent calculation, in particular Large-Eddy Simulation (LES), and a high-fidelity, fully resolved Small-Eddy Simulation (SES) of a small sub-region of interest of the LES. In the present formulation, such coupling between the LES and the SES is one-way and it is provided by continuous replacement of the large scales of the SES flow field with the low-pass filtered velocity field from the LES. Resulting L/SES approach is a combination of both volumetric and boundary forcing. We described all the algorithmic steps of the L/SES approach, including explicit filtering of the LES and SES fields, spatial and temporal interpolation of the large-scale LES data, and the boundary condition treatment. 

\begin{figure*}
\includegraphics[width=0.9\textwidth]{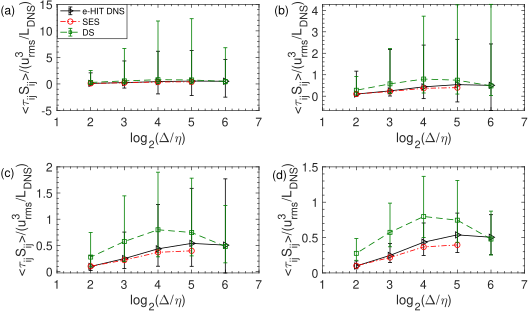}
  \caption{Time-averaged, normalized SGS dissipation from the entire e-HIT DNS domain at different scales, along with the corresponding DS and SES values. Error bars show the minimum and maximum values of $\langle\tau_{ij}S_{ij}\rangle$ in the sub-regions of different sizes, namely (a) $32^3$ cells, (b) $64^3$ cells, (c) $128^3$ cells, and (d) $256^3$ cells. See text for further details.}
\label{fig:diss_variab}
\end{figure*}

The L/SES can be viewed as a generalized method for forcing turbulence in resolved simulations, which are effectively embedded in a large-scale flow of arbitrary complexity. In this sense, SES is also a dual counterpart of the traditional LES. In the latter, large-scale structures are explicitly evolved, while the sub-filter scale dynamics is modeled with the appropriate closures. In contrast, in the SES, small-scale dynamics is explicitly resolved, while the dynamics of the large scales (above the filter scale) not captured in the SES is introduced using the realistic `large-scale model' provided by the companion LES calculation. As such, L/SES is qualitatively different from the traditional turbulence forcing methods, e.g., spectral or linear forcing, which are idealized in nature and are thus decoupled from any large-scale flow properties.

Proposed approach is applicable for an arbitrary geometry, as well as any initial and boundary conditions, as long as the filtering operation can be applied in the physical space. L/SES does not make assumptions of periodicity or turbulence equilibrium, and it is agnostic to any underlying flow features such as anisotropy or shear, which can be captured directly without additional modifications. % While L/SES-M1 only compares well with the DNS in terms of a few metrics, the performance of L/SES-M2 was found to be much better for all the tested quantities.
Thus, L/SES is potentially applicable to complex turbulent flows in practical physical and engineering systems. We demonstrated the method performance in the {\it a priori} tests of both steady and unsteady (including decaying) homogeneous, isotropic turbulence (HIT). Analysis of the method accuracy in more complex turbulent flows will be presented in future work.

The only implicit assumption made in the L/SES is that the large scales in the LES are sufficiently accurate. Furthermore, one-way coupling between the LES and the SES in the current formulation also implies that the large-scale dynamics is not critically controlled by the small scales. This is reasonable for the flows, which are dominated by the downward cascade of kinetic energy on average, such as HIT considered in the {\it a priori} tests described here. Although such flows would also contain regions where the upscale transfer, or backscatter, of kinetic energy is locally observed \citep{piomelli1991subgrid,leith1990stochastic}, the net effect of such backscatter on the large scales is limited. Furthermore, since the SES domain is not periodic, any transfer of kinetic energy to scales larger than $L_{SES}$ will be able to leave the domain, thus maintaining the accuracy of the small scales. %Such regions of upscale kinetic energy transfer would not be detrimental to the method if the large scales of the forcing simulation are accurate (i.e. for {\it a posteriori} cases, the LES model reasonably captures the effects of the sub-grid scales).%In fact, we show in a companion paper that in such flows, even implicit LES (ILES), which does not use any subgrid-scale closures and just relies on numerical dissipation, can still provide DNS-level fidelity of the L/SES solution for a wide range of Reynolds numbers. In particular, such accuracy is verified up to Re$_{\lambda} \approx 600$. 
At the same time, special treatment would be required in the flows with a strong kinetic energy backscatter or net up-scale transfer of kinetic energy. This is the case, for instance, in reacting flows \citep{towery2016spectral,o2017cross,towery2014spectral}, in which small scales can efficiently energize the large scales in the turbulent flame, and thus large-scale dynamics may not necessarily be determined independently of the small scales. In such cases, a two-way coupling between the SES and the parent LES is necessary, which is a subject of future work.

In the context of such two-way coupling, an important question is whether L/SES can be used to provide the closures for the LES calculation directly without the need for their modelling. Presented {\it a priori} tests demonstrate that SES can indeed give the correct SGS dissipation in the sub-region of interest, outperforming the classically used closure models, such as the Dynamic Smagorinsky (DS) and Vreman models \citep{moin1991dynamic,vreman2004eddy}. In particular, in addition to the close agreement of the mean $\langle\tau_{ij}S_{ij}\rangle$ in the equilibrium HIT, L/SES also achieves good temporal correlation with the DNS in the time-evolving, non-equilibrium HIT.%, as well as an agreement for the actual Reynolds stresses for both steady and unsteady turbulence. This shows the capability of the method to work in conditions when the  traditional assumptions do not hold / unknown regimes

The key aspect here, however, is that the SES domain is restricted to a pre-selected sub-region of the larger flow, and as a result, it can provide the SGS closures only in that particular region. At the same time, even in the ideal case of a steady-state HIT, the instantaneous values of the relevant quantities, such as the SGS dissipation, could differ significantly between various sub-regions \citep{Aoyamaetal:2005}. Such variability is illustrated in Fig.~\ref{fig:diss_variab}. In particular, black symbols show the normalized SGS dissipation at different scales in the e-HIT DNS. These values are calculated by explicitly filtering the fully-resolved DNS data, and subsequently averaging resulting values temporally and spatially over the entire $1024^3$ domain. Green symbols also show the average values of the SGS dissipation obtained using the DS model at the same scales. Next, we partition the entire DNS domain into smaller sub-regions with sizes ranging from $32^3$ cells to $256^3$ cells. For each particular sub-region size, the average SGS dissipation, $\langle\tau_{ij}S_{ij}\rangle$, is calculated in each sub-region both directly from the DNS data and using the DS model. Resulting minimum and maximum values among all sub-regions are shown with error bars in Fig.~\ref{fig:diss_variab}. For comparison, $\langle\tau_{ij}S_{ij}\rangle$ in the SES domain, which corresponds to the central $128^3$ sub-region of the DNS, is also shown with red symbols. As can be seen, there is significant variability in the mean SGS dissipation between different sub-regions of the flow at all scales and for all sub-region sizes. This variability can result in an order of magnitude deviation of $\langle\tau_{ij}S_{ij}\rangle$ from the global mean, particularly for smaller region sizes. An SGS model, such as DS, which is computed everywhere in the flow, also exhibits variability in different sub-regions, although agreement with the DNS, both in terms of the mean values, as well as the observed minimum/maximum range, can be quite poor.

In contrast, the SES captures the mean SGS dissipation in the DNS quite well. At the same time, since SES is only performed for the central region, by definition it cannot describe the variability of the dissipation throughout the entire flow. Therefore, a single SES calculation cannot provide SGS closures for a more extended region of the host large-scale flow, or for the entire LES calculation. This is not necessarily a shortcoming of the method, as it is inherent in the formulation itself, whereby the SES can only accurately capture the flow structure in the region, which was used to force it. 

At the same time, L/SES fundamentally differs from the traditional LES in that it provides closure terms directly from first principles, without any assumptions or approximations. Such ability to correctly recover sub-filter quantities can be useful to validate new LES models or verify their applicability in the poorly explored regimes. L/SES can also serve as a useful discovery tool to probe specific LES regions with higher fidelity. If the large-scale flow is non-homogeneous, multiple SES may be required to probe different sub-regions with sufficiently distinct small-scale dynamics. Therefore, L/SES can serve as a tool for the development of the SGS models, which can then be used to provide two-way coupling with the LES.
%deThe method can be extended to involve a two-way coupling in the future, where the driving LES itself takes feedback from the L/SES for obtaining the correct closure term, which could be useful in enabling LES simulations for extreme flow regimes.

It is important also to highlight the differences between the proposed L/SES approach and the existing data assimilation techniques including various nudging methods. For instance, \citet{clark2020synchronization} applied nudging in homogeneous and isotropic turbulence, with the reference data obtained from a DNS and injected either at isolated locations in physical space or at a set of spectral modes. In particular, the latter spectral case is analogous to the L/SES approach in that the reference data was Fourier filtered at low wavenumbers. At the same time, such methods differ from the L/SES in two key respects. First, from the fundamental standpoint, traditional data assimilation assumes that the reference data is a true solution obtained either through the experiments/observations or in a high-fidelity simulation \citep{clark2020synchronization}. As a result, the principal goal of such techniques is to reach synchronization between the reference field and the obtained solution. Achieving such synchronization for the entire spectral range is a non-trivial task, especially when only a subset of the wavenumbers are nudged. Indeed, as shown by \citet{clark2020synchronization}, the error between the reference field and the obtained solution is low only for the wavenumbers being nudged, in particular at large scales. This in turn raises the question of the accuracy of the solution on scales, on which such synchronization is not reached. In contrast, in the L/SES, the forcing field is obtained from a low-fidelity solution, in which by definition small-scale structure cannot be assumed to be accurate and is therefore discarded. L/SES does not require the availability of the true reference solution, which in many cases may be difficult or impossible to obtain, and thus, by construction, there is no goal to achieve the synchronization for the entire range of scales. Crucially, we show that even using a low-fidelity reference solution, the cost of obtaining which may be relatively low, it is possible to recover the small-scale flow structure very accurately. Second, from a practical standpoint, traditional spectral nudging techniques \citep{von2000spectral,clark2020synchronization} rely on spectral decomposition of the reference field and thus they cannot be directly applied in physical space. This limits their utility in the engineering contexts, which may not have periodic boundary conditions and simple geometries. Proposed L/SES approach does not have this limitation as it is entirely formulated in physical space, which requires various algorithmic steps described in \S~\ref{sec:lses} above. Moreover, for periodic cases such as HIT, the relaxation of the periodicity requirement allows L/SES to be used in sub-regions of the large-scale flows, thereby potentially allowing reduction in computational cost and access to large Reynolds numbers even in such idealized geometries.

We also described two alternative approaches for embedding a high-resolution calculation in a sub-region of interest of a larger turbulent flow.
%Overall, several different approaches towards obtaining high-fidelity information from a turbulent flow are contrasted. 
In particular, the L/SES-$\epsilon$ method extracts the rate of the inter-scale energy transfer in that sub-region and uses it as a time-dependent energy injection rate in the SES calculation. This approach is effectively a generalization of the traditional class of the volumetric spectral forcing techniques, albeit in which the energy injection rate is not specified as an ad hoc parameter, but instead directly determined from the LES. The L/SES-$\epsilon$ method, however, still reduces the entire large-scale flow structure to a single scalar parameter based only on the inter-scale kinetic energy transfer rate. Various obtained small-scale statistics suggest that an independent simulation, forced using such energy injection rate, results in a much lower Re$_\lambda$, thereby leading to much lower values of $U_{rms}$ as well as inaccurate statistics of the small-scale flow quantities, such as enstrophy or dissipation.

In this regard, a more specific observation can be made based on the {\it a priori} tests described above. In particular, one can consider an idealized case of the HIT turbulence and compare a DNS of a larger flow with an SES-$\epsilon$ corresponding to a smaller sub-region of such a DNS, with both calculations forced with the same volumetric energy injection rate, thereby leading to the same average dissipation rate. The flow solution in the SES-$\epsilon$ will not represent the corresponding sub-region of the larger flow even in the statistically averaged sense. In particular, while SES-$\epsilon$ can recover the same mean dissipation rate and kinetic energy spectrum representative of the large-scale flow in line with the Kolmogorov '41 theory, it fails to capture the correct total kinetic energy content in the region of interest or the small-scale statistics of the velocity gradients. This suggests that classical volumetrically forced DNS driven in a conventional way through energy injection at the largest scale with a constant rate cannot be considered to represent a small section of a larger external flow at the same Reynolds number.

In addition to the traditional volumetric forcing, we also contrasted the L/SES with a boundary forcing approach, in which a high-resolution region is directly embedded in a larger flow using static (or adaptive) mesh refinement. In this case, such refined region embedded in an LES interacts with the coarser regions directly through the coarse-fine interfaces. This direct embedding somewhat increases the resulting flow Reynolds number in the high-resolution region compared to the volumetric-only L/SES-$\epsilon$ approach, however agreement in Re$_\lambda$ and other small-scale flow statistics with the DNS is still worse than for the L/SES. %This shows that only boundary or volumetric forcing are insufficient, and rather their combination is important to achieve high solution accuracy in the embedded calculation.

%Therefore we consider statistics of SGS dissipation from a priori L/SES-M2 to see if an L/SES forced using a fully resolved DNS can reproduce statistics in other regions of the DNS itself. Instead of observing PDF, we study the variability in SGS dissipation by averaging the quantity over sub-regions of various different sizes within the DNS. This is done because an L/SES domain represents a finite-sized DNS sub-region.

%, especially because in order to act as a closure model itself, closure terms must be obtained exactly at each LES cell to ensure stable simulations \citep{liu1994properties}

Comparison of the accuracy of the L/SES approach with the L/SES-$\epsilon$ and L/SES-SMR raises a more fundamental question regarding the nature of the interaction between various scales of turbulence. The inability of L/SES-$\epsilon$ to match the Re$_\lambda$ of the large-scale flow shows that a single scalar, even representing accurately the inter-scale energy transfer rate, does not capture all the relevant information transmitted from the large to small scales. In particular, since this scalar is obtained at a specific filter scale $\Delta = L_{SES}$, which is equal to the SES-$\epsilon$ forcing scale, this indicates that using information only from scales close to the driving scale is insufficient to replicate the small-scale behavior of the underlying flow. To investigate this further, an L/SES simulation was performed, which used the band-pass-filtered forcing data extracted from scales in the range from $L_{SES}$ to $2L_{SES}$. This produced much worse results than the regular L/SES approach, which uses low-pass filtered forcing data, and thus incorporates all scales $> \Delta$. Results of this test are not shown here for brevity. This suggests that in a turbulent flow, the transfer of information does not occur only between the adjacent wavenumbers, but rather through a complex interaction of all scales \citep{domaradzki1990local,domaradzki2007analysis,yeung1991response}. This is further corroborated by a somewhat better accuracy of the L/SES-SMR approach, in which the refined region is allowed to interact with all larger scales of the ambient flow through the boundaries. At the same time, the fact that even L/SES-SMR was not able to achieve the same solution fidelity as the L/SES shows that capturing the small-scale complexity of a larger host flow cannot be achieved by merely driving a single wavenumber or forcing through just the domain boundaries. Forcing of the embedded calculation must be volumetric in nature affecting the entire domain, in both physical and spectral space.

%Large-scale flows can be studied using under-resolved numerical simulations, for example in an LES. Such simulations use closure models for the unresolved sub-grid scales (SGS) which are based on several assumptions such as those of homogeneity and equilibrium. Another objective for the present work is to obtain a method to derive exact SGS closures in specific LES sub-regions to probe and validate LES models in complex regimes. A discussion on current LES closure models is presented below.
%Inadequacies of traditional driving methods have shown that a single scalar such as the energy dissipation rate is often insufficient to capture all the complexity of an external flow. In the present work, we investigate how information is transferred from the large to the small scales, and what data from the large scales is required to appropriately drive a smaller, fully resolved simulation. We explore different methods to communicate to a small-scale simulation that is a part of a larger flow. This opens up the possibility of accessing progressively higher Reynolds number dynamics with only a limited number of scales in a small simulation.

%This further highlights shortcomings of not driving the intermediate scales (from $L_{SMR}$ to just less than $L_{LES}$) volumetrically, as the performance is bad at scales immediately smaller than this range, but begins to recover as consecutive small scales become available in the refined region. 

An additional comment should be made regarding the overall formulation of the L/SES approach. Governing equations for the small-scale flow field can be directly derived by subtracting the filtered LES equations from the Navier-Stokes equations. Such equations would directly describe the evolution of the fluctuating quantities on scales smaller than the desired filter scale. Depending on the definition of the fluctuating quantities, two forms of such equations can be obtained, which are given in Appendix~\ref{appB} (also see \citet{rah2018derivation}). Instead of the forcing described in \S~\ref{sec:lses}, either set of equations could be used to evolve the SES solution. Resulting formulation is directly dual to the LES equations as the small-scale quantities are evolved and the large-scale flow is represented with the filtered quantities appearing as closure terms on the right-hand side of the equations. Such closure terms can be obtained directly from the LES data, thus eliminating the need for the closed-form closure models for the large-scale flow.

At the same time, equations given in Appendix~\ref{appB} present several drawbacks compared to the proposed L/SES procedure. The first set of equations (\ref{eq:fluct1.1})-(\ref{eq:fluct1.3}) for the fluctuating primitive variables involves source terms in the mass conservation equation (\ref{eq:fluct1.1}), which is undesirable. This, however, can be remedied using the second set of equations (\ref{eq:fluct2.1})-(\ref{eq:fluct2.3}) at the expense of more complex definitions of the fluctuating quantities (eqs.~\ref{eq:Mfluct} and \ref{eq:Efluct}). Furthermore, both sets of equations include terms, which involve products of filtered and fluctuating quantities. For instance, the transport equation (\ref{eq:fluct2.2}) of the fluctuating momentum $M_i'$ depends not only on the advection of $M_i'$ by the fluctuating velocity $u_j''$, but also on the advection of the filtered momentum $\overline{M}_i$ by $u_j''$ and of $M_i'$ by the filtered velocity $\tilde{u}_j$. Furthermore, unclosed LES terms such as $R_{ij}$ are also present and must be obtained by first calculating the total velocity, taking the product of its components, and explicitly filtering the result. Therefore, the filtered and fluctuating fields cannot be completely isolated from each other and the above procedure does not have any advantage over the proposed L/SES method. Finally and most importantly, since LES filtered quantities appear in the source terms, they would affect all SES scales directly. This is in contrast with the proposed L/SES, in which the LES data is used only to drive the large SES scales, while the small scales are allowed to develop naturally. Due to these considerations, the transport equations for the fluctuating small-scale quantities were not considered further in this work.

Present work provides a demonstration of the L/SES method for a canonical flow, namely HIT. At the same time, several open issues merit further investigation. First, the key step of the approach, which separates information at various scales, is the filtering operation applied in the physical space. Since the method is intended for the arbitrary boundary conditions, perfect spectrally sharp filtering is not possible. Therefore, an important future step for the improvement of the overall method involves the development of the filters with more desirable properties, as discussed in \S~\ref{sssec:filter}. Such improved filters would not only increase the accuracy of the method, but they would also reduce the computational cost by reducing spectral leakage and thus maximizing the range of scales available for analysis.
%not be contaminated by spectral leakage
Furthermore, it is desirable to design filters, which can work with the smaller sponge and buffer regions. Reducing the extent of such regions is important when the chosen SES domain lies close to a physical boundary of the LES, e.g., near a wall, where additional sponge and buffer region data may not be available. %Future work should be directed towards a systematic exploration of the appropriate filter type and L/SES domain size.

We also note that since present work considered only the {\it a priori} tests, various aspects of the method related to the LES calculations were not explored. More specifically, this concerns the choice of the filter scale relative to the LES grid size, $\Delta/dx_{LES}$, as well as the overall accuracy of the LES solution and the potential effect of the SGS closure models. There would always exist a discrepancy between the flow field obtained in an LES and the true Navier-Stokes solution. Such discrepancy, however, would be the largest at the small LES scales, which are most affected by the lack of resolution. Therefore, for {\it a posteriori} SES, the filter scale $\Delta$ would need to be sufficiently larger than the smallest scales that the LES resolves, and of the order of the Taylor scale $\lambda$. Explicit filtering of the LES data at scale $\Delta$ would then ensure that such unphysical small scales are removed thereby minimizing their impact on the SES forcing. These aspects will be presented in detail in the follow-up paper focusing on the {\it a posteriori} tests for a range of Re$_{\lambda}$. Finally, it would be interesting in the future to analyze the sensitivity of the SES solution to the potential errors in the large-scale component of the LES flow field in the {\it a priori} sense by introducing noise or systematic bias in the large-scale DNS velocities. 
%Additionally, all present tests used LESes that were not too poorly resolved. However for coarser LES resolutions, if scales just larger than the filter scale contain some inaccuracies, they will be passed down to the L/SES and may affect its behaviour. Hence for implicitly filtered LESes, selection of an appropriate filter scale can be an important factor, which was not explored here. Also if the LES simulation uses a closure model, its effects would be observed in the forcing data, and the ability of L/SES to produce the correct small scales in such cases would need to be studied. Overall the effect of the accuracy and resolution of LES on the L/SES needs to be probed further. 

Another important extension of the L/SES approach concerns compressible turbulent regimes. In this work, the focus was on the weakly compressible HIT. As a result, forcing the velocity field (eq.~\ref{eq:deltau}) or the momentum (eq.~\ref{eq:deltamom}) yielded virtually identical results. This would not be the case for more complex flows. For instance, in compressible HIT, perturbing the momentum instead of velocity would be more appropriate. For other situations, such as boundary or shear layers, scalar mixing, or reacting flows, forcing of other quantities such as energy (or pressure) and scalars may be necessary to recover proper structure not only of the velocity field, but also of the fields of passive and active scalars. Such extension of the L/SES approach to compressible, as well as variable density flows is an important future step.

Notwithstanding these current limitations of the L/SES, proposed approach presents a number of important advantages. Since the large scales are directly imposed in the SES, all their complexity is injected into the small-scale calculation. This makes the L/SES a versatile forcing technique, which incorporates all the relevant large-scale features by design, eliminating the need for the flow-dependent modifications. It is also a fundamentally different approach from LES modelling, as it allows one to obtain closure terms in the region of interest directly from first principles, without any assumptions or approximations. %The ability to correctly obtain sub-filter quantities can be useful to validate new LES models or extend their applicability to unknown regimes. It can also serve as a useful discovery tool to probe specific LES regions with higher fidelity. 

Finally, one of the key advantages of the L/SES method is its ability to recover the detailed small-scale flow structure, including the highly intermittent statistics of the velocity-gradient-based quantities corresponding to the Reynolds number of the actual flow of interest. This opens the possibility for probing the high-Reynolds-number flow regimes, which are largely inaccessible in the DNS, with fully resolved simulations. Such calculations can be performed at a fraction of the computational cost of a full DNS due to the combination of a coarse LES and a well-resolved SES in a smaller sub-region. Detailed exploration of this aspect of the L/SES method, and the demonstration of the computational advantage of the L/SES in comparison with the DNS, will be presented in the follow-up paper, which will focus on the {\it a posteriori} tests for a wide range of Reynolds numbers.

\begin{acknowledgments}
%\section*{Acknowledgments}
%\label{Acknowledgments}

Authors acknowledge funding support from NASA under award 80NSSC22K0630 (technical officer: Dr. Sanaz Vahidinia), Air Force Office of Scientific Research (AFOSR) under award FA9550-21-1-0012 (program manager: Dr. Chiping Li), as well as the University of Connecticut Research Excellence Program. Computing resources were provided by the Department of Defense High Performance Computing Modernization Program (HPCMP). Authors are also grateful to Diego Donzis and George Matheou for valuable discussions.
\end{acknowledgments}

\section*{Data availability}
Raw data were generated at the Department of Defense HPCMP facilities. Derived data supporting the findings of this study are available from the corresponding author upon reasonable request.

\section*{Author declarations}
\subsection*{Author contributions}
\pmb{Arnab Moitro}: Conceptualization (equal); Data curation (lead); Formal analysis (lead); Investigation (lead); Methodology (equal); Software (lead); Validation (lead); Visualization (lead); Writing – original draft (lead); Writing – review \& editing (equal). \pmb{Sai Sandeep Dammati}: Data curation (supporting); Formal analysis (supporting); Software (supporting); Visualization (supporting); Writing – review \& editing (supporting). \pmb{Alexei Poludnenko}: Conceptualization (equal); Data curation (supporting); Funding acquisition; Supervision; Methodology (equal); Writing – original draft (supporting); Writing – review \& editing (equal).

\subsection*{Conflict of interest}
The authors have no conflicts to disclose.

%--------------------------------------------------------------------------------------------
\appendix
\section{LES subgrid-scale closures and Reynolds stress estimates}
\label{appA}

%We would like to see if the L/SES can be an effective tool for LES closure modelling. The actual desired unclosed terms are extracted from the DNS itself by filtering it at various scales ($4dx$ to $64dx$, representing different levels of LES resolution), and are compared with the respective L/SES results. The various unclosed terms in LES arise due to solving the filtered governing equations \citep{martin2000subgrid}
Here, for completeness, we include the LES equations along with the associated unclosed subgrid-scale (SGS) terms, in particular the Reynolds stress term, which is compared between the DNS and SES in \S~\ref{ssec:closures}. We also describe two reference closure models, namely Dynamic Smagorinsky \citep{moin1991dynamic} and Vreman \citep{vreman2004eddy} models, which are used in the comparisons in \S~\ref{ssec:closures}.

LES do not resolve the small scales of turbulent motion. Instead, they solve the low-pass filtered Navier-Stokes equations \citep{martin2000subgrid}
% \begin{eqnarray} \label{LES equations}
%    \label{LES equations1} \frac{\partial \bar{\rho}}{\partial t} + \frac{\partial \bar{\rho} \tilde{u}_j}{\partial x_j} & = & 0, \\ 
%     \label{LES equations2} \frac{\partial \bar{\rho} \tilde{u}_i}{\partial t} + \frac{\partial \bar{\rho} \tilde{u}_i  \tilde{u}_j}{\partial x_j} +\frac{\partial \bar{p}}{\partial x_i} - \frac{\partial \tilde{\sigma}_{ij}}{\partial x_j} & = &  - \frac{\partial \tau_{ij}}{\partial x_j}, \\ 
%     \label{LES equations3} \frac{\partial \bar{\rho}\tilde{E}}{\partial t} + \frac{\partial (\bar{\rho}\tilde{E} +\bar{p}) \tilde{u}_j}{\partial x_j} - \frac{\partial \tilde{\sigma}_{ij}\tilde{u}_i}{\partial x_j} + \frac{\partial \tilde{q}_j}{\partial x_j}& = & -\frac{\partial}{\partial x_j} (\gamma c_V Q_j+\frac{J_i}{2}-D_j).
% \end{eqnarray}
\begin{equation}
    \label{LES equations1} \frac{\partial \bar{\rho}}{\partial t} + \frac{\partial \bar{\rho} \tilde{u}_j}{\partial x_j} = 0,
\end{equation}
\begin{equation}
    \label{LES equations2} \frac{\partial \bar{\rho} \tilde{u}_i}{\partial t} + \frac{\partial \bar{\rho} \tilde{u}_i  \tilde{u}_j}{\partial x_j} +\frac{\partial \bar{p}}{\partial x_i} - \frac{\partial \tilde{\sigma}_{ij}}{\partial x_j} =  - \frac{\partial \tau_{ij}}{\partial x_j},
\end{equation}
\begin{equation}
    \label{LES equations3}
    \begin{split}
        \frac{\partial \bar{\rho}\tilde{E}}{\partial t} + \frac{\partial (\bar{\rho}\tilde{E} +\bar{p}) \tilde{u}_j}{\partial x_j} - \frac{\partial \tilde{\sigma}_{ij}\tilde{u}_i}{\partial x_j} + \frac{\partial \tilde{q}_j}{\partial x_j} = \\ -\frac{\partial}{\partial x_j} (\gamma c_V Q_j+\frac{J_i}{2}-D_j).
    \end{split}
\end{equation}
Here, calorically perfect gas is assumed, $\bar{\phi}$ and $\tilde{\phi}=\overline{\rho \phi}/\bar{\rho}$ represent Reynolds and Favre filtering of a quantity $\phi$, respectively. The $\tilde{q}_j=-\tilde{k}\partial \tilde{T}/\partial x_j$ and $E=e+u_ku_k/2$, where $e$ is the internal energy per unit mass and $c_V$ is the heat capacity at constant volume.

Filtering gives rise to the unclosed SGS terms on the right-hand sides of eqs.~(\ref{LES equations2})-(\ref{LES equations3}), which cannot be obtained directly from the filtered flow variables and instead need to be modelled. They have the following form
 \begin{eqnarray}
 \label{eq:tauij} \tau_{ij} & = & \bar{\rho} (\widetilde{u_iu_j}-\tilde{u}_i\tilde{u}_j), \\
 Q_j & = & \bar{\rho}(\widetilde{u_jT}-\tilde{u}_j\tilde{T}),                \\
 J_j & = & \bar{\rho} (\widetilde{u_ju_ku_k}-\tilde{u}_j\widetilde{u_ku_k}), \\
 D_j & = & \overline{\sigma_{ij}u_i}- \tilde{\sigma}_{ij} \tilde{u}_i.
 \end{eqnarray}
Here $\tau_{ij}$ is the Reynolds stress, $Q_j$ is the SGS heat flux, while $J_j$ and $D_j$ describe the SGS turbulent diffusion and viscous dissipation. The equation of state, $p=\rho R T$, has been used to represent the pressure gradient in terms of the SGS heat flux. Laminar diffusive fluxes, $\overline{k\frac{\partial T}{\partial x_j}}-\tilde{k}\frac{\partial \tilde{T}}{\partial x_j}$, are small compared to the turbulent transport and they typically can be ignored \citep{veynante2002turbulent}.
 
The Reynolds stress $\tau_{ij}$ is the most important closure term to consider in the simulations of low-Mach-number, weakly compressible turbulence, which is the case in this study. SGS closure terms in the energy conservation equation (\ref{LES equations3}) in such low-Mach regimes are orders of magnitude smaller than their filtered counterparts at all scales and thus can be ignored\citep{vreman1995subgrid,xie2018modified}. Assessment of their accuracy in the SES, along with the analysis of the L/SES method performance in compressible regimes, is the subject for future work.

For comparison with the direct calculation of $\tau_{ij}$ in the SES discussed in \S~\ref{ssec:closures}, we consider two widely used SGS models, namely Dynamic Smagorinsky (DS) \citep{moin1991dynamic} and Vreman models \citep{vreman2004eddy}. The `true' model coefficients are calculated using a pseudo-LES field formed by filtering the DNS, and then applying test filters directly on the resultant data.

The DS model closures are calculated as
%However, energy equation closures are also calculated and shown to be insignificant in comparison to the filtered terms.
% \begin{eqnarray} 
%     \label{eq: DS closure1} \tau_{kl} &-& \frac{1}{3}q^2\delta_{kl} = 2C\bar{\rho} \Delta^2 |\tilde{S}| (\tilde{S}_{kl}-\frac{1}{3} \tilde{S}_{mm} \delta_{kl}), \\
%      \tilde{S}_{kl} &=& \frac{1}{2} (\frac{\partial \tilde{u}_k}{\partial x_l} + \frac{\partial \tilde{u}_l}{\partial x_k}), \qquad  |\tilde{S}|=(2\tilde{S}_{kl}\tilde{S}_{kl})^{1/2}, \qquad q^2= 2C_I\bar{\rho}\Delta^2 |\tilde{S}|^2, \\
%      C_I &=& \frac{\langle L_{kk} \rangle}{\langle 2 \hat{\bar{\rho}} \hat{\Delta}^2 |\hat{\tilde{S}}|^2- 2\Delta^2 \widehat{\bar{\rho}|\tilde{S}|^2 }\rangle}, \qquad L_{kl}=\widehat{\bar{\rho} \tilde{u}_k \tilde{u}_l} - (1/\hat{\bar{\rho}}) ( \widehat{\bar{\rho} \tilde{u}_k} \widehat{\bar{\rho} \tilde{u}_l}), \\
%     \label{eq: DS closure_last} C &=& \frac{\langle L_{kl} \tilde{S}_{kl}-(1/3) \tilde{S}_{mm} L_{jj}  \rangle}{\langle -2 \hat{\bar{\rho}} \hat{\Delta}^2 |\hat{\tilde{S}}| (\hat{\tilde{S}}_{kl} \tilde{S}_{kl} -(1/3) \hat{\tilde{S}}_{mm} \tilde{S}_{jj} )  + 2\Delta^2( \widehat{\bar{\rho}|\tilde{S}|\tilde{S}_{kl}} \tilde{S}_{kl} -(1/3) \widehat{\bar{\rho}|\tilde{S}|\tilde{S}_{mm}} \tilde{S}_{jj})\rangle}.
% \end{eqnarray}
\begin{equation}
    \label{eq: DS closure1} \tau_{kl} - \frac{1}{3}q^2\delta_{kl} = 2C\bar{\rho} \Delta^2 |\tilde{S}| (\tilde{S}_{kl}-\frac{1}{3} \tilde{S}_{mm} \delta_{kl}),
\end{equation}
\begin{equation}
\begin{split}
    \tilde{S}_{kl} = \frac{1}{2} (\frac{\partial \tilde{u}_k}{\partial x_l} + \frac{\partial \tilde{u}_l}{\partial x_k}), \qquad  |\tilde{S}|=(2\tilde{S}_{kl}\tilde{S}_{kl})^{1/2}, \\ 
    \qquad q^2= 2C_I\bar{\rho}\Delta^2 |\tilde{S}|^2,
\end{split}
\end{equation}
\begin{equation}
\begin{split}
    C_I = \frac{\langle L_{kk} \rangle}{\langle 2 \hat{\bar{\rho}} \hat{\Delta}^2 |\hat{\tilde{S}}|^2- 2\Delta^2 \widehat{\bar{\rho}|\tilde{S}|^2 }\rangle}, \\
    \qquad L_{kl}=\widehat{\bar{\rho} \tilde{u}_k \tilde{u}_l} - (1/\hat{\bar{\rho}}) ( \widehat{\bar{\rho} \tilde{u}_k} \widehat{\bar{\rho} \tilde{u}_l}),
\end{split}
\end{equation}
\begin{equation}
\label{eq: DS closure_last} 
\begin{split}
    C = \frac{\langle L_{kl} \tilde{S}_{kl}-(1/3) \tilde{S}_{mm} L_{jj}  \rangle}{C_D}, \\
    C_D = \langle -2 \hat{\bar{\rho}} \hat{\Delta}^2 |\hat{\tilde{S}}| (\hat{\tilde{S}}_{kl} \tilde{S}_{kl} -(1/3) \hat{\tilde{S}}_{mm} \tilde{S}_{jj} )  \\
    + 2\Delta^2( \widehat{\bar{\rho}|\tilde{S}|\tilde{S}_{kl}} \tilde{S}_{kl} -(1/3) \widehat{\bar{\rho}|\tilde{S}|\tilde{S}_{mm}} \tilde{S}_{jj})\rangle
\end{split}
\end{equation}
Here, $\tilde{...}$ represents the Favre-filtered DNS variables mimicking the LES data, and $\hat{...}$ represents explicit test filtering of the filtered DNS variable. The test filter width $\hat{\Delta}$ is chosen to be twice the width of the scale, at which the closure is desired, i.e., $\hat{\Delta}=2\Delta$. Model coefficients obtained from the DNS were found to be $C=0.21$ and $C_I=0.014$, in agreement with the findings of \citet{fureby1997comparative}. 

Closures using the Vreman model are calculated as
%Revisit this description and provide more clarifications
% \begin{eqnarray}
%     \label{eq: Vre closure1} \langle \tau_{ij} \rangle &=& \langle -2\nu_e \tilde{S}_{ij}+\frac{2}{3}k_\tau \delta_{ij}\rangle, \qquad k_\tau=2\nu_e |\tilde{S}|, \\
%     \nu_e &=& c\sqrt{\frac{B_\beta}{\alpha_{ij}\alpha_{ij}}}, \qquad c=2.5C_S^2, \\
%     \alpha_{ij} &=& \frac{\partial \tilde{u}_j}{\partial x_i}, \qquad \beta_{ij}=\Delta^2\alpha_{mi}\alpha_{mj}, \\
%     \label{eq: Vre closure_last} B_\beta &=& \beta_{11}\beta_{22}-\beta_{12}^2+\beta_{11}\beta_{33}-\beta_{13}^2 +\beta_{22}\beta_{33}-\beta_{23}^2.
% \end{eqnarray}
\begin{equation}
        \label{eq: Vre closure1} \langle \tau_{ij} \rangle = \langle -2\nu_e \tilde{S}_{ij}+\frac{2}{3}k_\tau \delta_{ij}\rangle, \qquad k_\tau=2\nu_e |\tilde{S}|
\end{equation}
\begin{equation}
    \nu_e = c\sqrt{\frac{B_\beta}{\alpha_{ij}\alpha_{ij}}}, \qquad c=2.5C_S^2,
\end{equation}
\begin{equation}
    \alpha_{ij} = \frac{\partial \tilde{u}_j}{\partial x_i}, \qquad \beta_{ij}=\Delta^2\alpha_{mi}\alpha_{mj}, \\
\end{equation}
\begin{equation}
\begin{split}
    \label{eq: Vre closure_last} B_\beta = \beta_{11}\beta_{22} - \beta_{12}^2 + \beta_{11}\beta_{33} - \beta_{13}^2 \\ + \beta_{22}\beta_{33} - \beta_{23}^2.
\end{split}
\end{equation}
Here, the suggested value of $C_s=0.17$ is used\citep{vreman2004eddy}, and the rest of the symbols have the same meaning as above. 

%--------------------------------------------------------------------------------------------
%\subsection{Energy equation closure estimates}
%\label{app:EnergyClosures}

%\begin{figure}
%\includegraphics[width=\textwidth]{Figures/eclosure_unst.png}
%  \caption{Mean absolute value of summation of all energy equation sub-filter scale terms normalized by $\rho U_{rms,DNS}^3/L_{DNS}$ vs. time for NE-HIT at different filter scales. Corresponding sum of filtered scale terms also shown for reference.}
%\label{fig:eclos_unst}
%\end{figure}

%Closure terms for the energy equation were calculated for both a priori SESes and compared with the corresponding terms from the DNS (obtained directly by explicitly filtering the resolved field). For low Mach number flows, compressibility effects are not too significant. Therefore the closure terms in eq. \ref{LES equations3} are not treated individually, and only the mean of their summation is considered in fig. \ref{fig:eclos_unst} for NE-HIT. These closures from the SESes fare poorly against the DNS, differing by an order of magnitude in some cases. However these terms are orders of magnitude smaller than their filtered counterparts at all scales, and thus can be ignored, as suggested by \citet{vreman1995subgrid,xie2018modified}. A similar trend is also seen in the corresponding value from E-HIT and is not shown here. Due to the insignificance of this term, it is not compared against any existing LES model.

%--------------------------------------------------------------------------------------------
\section{Governing equations for fluctuating quantities}
\label{appB}

Here, we describe the evolution equations for the small-scale fluctuating component of the flow field, which provide an alternative formulation to the L/SES proposed above. These equations are obtained by subtracting the filtered LES equations (\ref{LES equations1})-(\ref{LES equations3}) from the Navier-Stokes equations. In particular, consider the fluctuating components of any quantity $\phi$, namely $\phi'=\phi-\overline{\phi}$ and $\phi''=\phi-\tilde{\phi}$, where $\overline{\phi}$ and $\tilde{\phi}$ represent Reynolds and Favre filtering, respectively.

For the fluctuating primitive variables, $\rho'$, $u''$, and $e''$, the governing equations take the form
% \begin{eqnarray}
%     \frac{\partial \rho'}{\partial t}+ \frac{\partial \rho'u_j''}{\partial x_j} = -\frac{\partial }{\partial x_j}(\overline{\rho} u_j'' + \rho' \tilde{u}_j),
%     \label{eq:fluct1.1} \\
%     \Big(\frac{\partial \rho'u_i''}{\partial t}+ \frac{\partial \rho'u_i''u_j''}{\partial x_j}\Big) + \Big(\frac{\partial \overline{\rho}u_i''}{\partial t}+ \frac{\partial} {\partial x_j}\big(\overline{\rho}u_i''(u_j''+\tilde{u}_j)\big)\Big) + \Big(\frac{\partial \rho' \tilde{u}_i}{\partial t}+ \frac{\partial}{\partial x_j} \big(\rho' \tilde{u}_i(u_j''+\tilde{u}_j)\big)\Big) = \nonumber \\
%     -\frac{\partial p'}{\partial x_i}- \frac{\partial R_{ij}}{\partial x_j}- \frac{\partial}{\partial x_j} (\overline{\rho} \tilde{u}_i u_j'' + \rho' u_i'' \tilde{u}_j),
%     \label{eq:fluct1.2} \\
%     \Big(\frac{\partial \rho'e''}{\partial t}+ \frac{\partial \rho'e''u_j''}{\partial x_j}\Big) + \Big(\frac{\partial \overline{\rho}e''}{\partial t}+ \frac{\partial} {\partial x_j}\big(\overline{\rho}e''(u_j''+\tilde{u}_j)\big)\Big) + \Big(\frac{\partial \rho' \tilde{e}}{\partial t}+ \frac{\partial}{\partial x_j} \big(\rho' \tilde{e}(u_j''+\tilde{u}_j)\big)\Big) = \nonumber \\
%     -p'\frac{\partial u_j''}{\partial x_j} +\frac{\partial q_j'}{\partial x_j}+ \frac{\partial Q_{j}}{\partial x_j} -\alpha_P^{SGS}- \Big(\overline{p}\frac{\partial u_j''}{\partial x_j} + p'\frac{\partial \tilde{u}_j}{\partial x_j}\Big) - \frac{\partial}{\partial x_j} \Big(\overline{\rho} \tilde{e} u_j'' + \rho' e'' \tilde{u}_j\Big).
%     \label{eq:fluct1.3}
% \end{eqnarray}
\begin{equation}
\label{eq:fluct1.1}
    \begin{split}
        \frac{\partial \rho'}{\partial t}+ \frac{\partial \rho'u_j''}{\partial x_j} = -\frac{\partial }{\partial x_j}(\overline{\rho} u_j'' + \rho' \tilde{u}_j),
    \end{split}
\end{equation}
\begin{equation}
    \begin{split}
        \Big(\frac{\partial \rho'u_i''}{\partial t}+ \frac{\partial \rho'u_i''u_j''}{\partial x_j}\Big) + \\
        \Big(\frac{\partial \overline{\rho}u_i''}{\partial t}+ \frac{\partial} {\partial x_j}\big(\overline{\rho}u_i''(u_j''+\tilde{u}_j)\big)\Big) + \\
        \Big(\frac{\partial \rho' \tilde{u}_i}{\partial t}+ \frac{\partial}{\partial x_j} \big(\rho' \tilde{u}_i(u_j''+\tilde{u}_j)\big)\Big) = \\
    -\frac{\partial p'}{\partial x_i}- \frac{\partial R_{ij}}{\partial x_j}- \frac{\partial}{\partial x_j} (\overline{\rho} \tilde{u}_i u_j'' + \rho' u_i'' \tilde{u}_j), 
    \end{split}
\label{eq:fluct1.2}
\end{equation}
\begin{equation}
\label{eq:fluct1.3}
    \begin{split}
        \Big(\frac{\partial \rho'e''}{\partial t}+ \frac{\partial \rho'e''u_j''}{\partial x_j}\Big) + \\
        \Big(\frac{\partial \overline{\rho}e''}{\partial t}+ \frac{\partial} {\partial x_j}\big(\overline{\rho}e''(u_j''+\tilde{u}_j)\big)\Big) + \\
        \Big(\frac{\partial \rho' \tilde{e}}{\partial t}+ \frac{\partial}{\partial x_j} \big(\rho' \tilde{e}(u_j''+\tilde{u}_j)\big)\Big) = \\
    -p'\frac{\partial u_j''}{\partial x_j} +\frac{\partial q_j'}{\partial x_j}+ \frac{\partial Q_{j}}{\partial x_j} -\alpha_P^{SGS}- \\
    \Big(\overline{p}\frac{\partial u_j''}{\partial x_j} + p'\frac{\partial \tilde{u}_j}{\partial x_j}\Big) - \frac{\partial}{\partial x_j} \Big(\overline{\rho} \tilde{e} u_j'' + \rho' e'' \tilde{u}_j\Big).
    \end{split}
\end{equation}
Here, the equation of state has the usual form $p'=(\gamma-1)\overline{\rho} e''$. The closure terms are
\begin{eqnarray}
    R_{ij} &=& - \overline{\rho}(\widetilde{u_iu_j}-\tilde{u}_i \tilde{u}_j), \\
    \alpha_P^{SGS} &=& -\Big(\overline{p\frac{\partial u_j}{\partial x_j} }- \overline{p}\frac{\partial \tilde{u}_j}{\partial x_j} \Big), \\
    Q_j & = & -\bar{\rho}(\widetilde{eu_j}-\tilde{e}\tilde{u}_j).
\end{eqnarray}
In this form, governing equations are quite complex, they do not have the usual conservative flux form, and furthermore even the mass continuity equation (\ref{eq:fluct1.1}) has a source term.

An alternative set of equations can be formulated if we introduce modified fluctuating momentum $M'_i$ and fluctuating internal energy per unit volume, which are defined as
\begin{eqnarray}
    M_i' & = & \overline{\rho} u_i''+ \rho' \tilde{u}_i + \rho' u_i'', 
    \label{eq:Mfluct} \\
    E'   & = & \overline{\rho} e''+ \rho' \tilde{e} + \rho' e''.
    \label{eq:Efluct}
\end{eqnarray}
In this case, governing equations take on the following form
% \begin{eqnarray}
%     \frac{\partial \rho'}{\partial t}+ \frac{\partial M_j'}{\partial x_j} &=& 0,
%     \label{eq:fluct2.1} \\
%     \frac{\partial M_i'}{\partial t}+ \frac{\partial M_i'u_j''}{\partial x_j} &=& -\frac{\partial p'}{\partial x_i}- \frac{\partial R_{ij}}{\partial x_j}- \frac{\partial}{\partial x_j} (\overline{M}_i u_j'' + M_i' \tilde{u}_j),
%     \label{eq:fluct2.2} \\
%     \frac{\partial E'}{\partial t}+ \frac{\partial (E'+p')u_j''}{\partial x_j} &=&  \frac{\partial q_j'}{\partial x_j}+ \frac{\partial Q_{j}}{\partial x_j}- \frac{\partial}{\partial x_j} (\overline{E} u_j'' + E' \tilde{u}_j) -\alpha_P^{SGS}- \Big(\overline{p}\frac{\partial u_j''}{\partial x_j} + p'\frac{\partial \tilde{u}_j}{\partial x_j}\Big).
%     \label{eq:fluct2.3}
% \end{eqnarray}
\begin{equation}
\label{eq:fluct2.1}
    \begin{split}
        \frac{\partial \rho'}{\partial t}+ \frac{\partial M_j'}{\partial x_j} = 0,
    \end{split}
\end{equation}
\begin{equation}
\label{eq:fluct2.2}
    \begin{split}
        \frac{\partial M_i'}{\partial t}+ \frac{\partial M_i'u_j''}{\partial x_j} = -\frac{\partial p'}{\partial x_i}- \frac{\partial R_{ij}}{\partial x_j} \\
        - \frac{\partial}{\partial x_j} (\overline{M}_i u_j'' + M_i' \tilde{u}_j),
    \end{split}
\end{equation}
\begin{equation}
\label{eq:fluct2.3}
    \begin{split}
        \frac{\partial E'}{\partial t}+ \frac{\partial (E'+p')u_j''}{\partial x_j} =  \frac{\partial q_j'}{\partial x_j}+ \frac{\partial Q_{j}}{\partial x_j} \\
        - \frac{\partial}{\partial x_j} (\overline{E} u_j'' + E' \tilde{u}_j) -
        \alpha_P^{SGS}- \Big(\overline{p}\frac{\partial u_j''}{\partial x_j} + p'\frac{\partial \tilde{u}_j}{\partial x_j}\Big).
    \end{split}
\end{equation}
Here, filtered conserved variables are defined as $\overline{M}_i = \overline{\rho} \tilde{u}_i$ and $\overline{E} = \overline{\rho} \tilde{e}$. The equation of state becomes $p'=(\gamma-1)E'=(\gamma-1)(\overline{\rho} e''+ \rho' \tilde{e} + \rho' e'')$. In this formulation, governing equations take on a simpler and more familiar conservative flux form. They also avoid the source terms in the mass continuity equation at the expense of more complex definitions of the fluctuating momentum and energy, which involve both fluctuating and filtered primitive variables. Finally, note that viscous terms are not included in both sets of equations because we only intend to demonstrate the complexity of such an approach rather than formulate the full set of governing equations.

% Give eq. of state

%Differences between SMR1 and the other SMR simulations are also quantitatively non-negligible, indicating the relevance of volumetric forcing even when there is a large separation between the largest and smallest scales.

%\nocite{*}

\bibliography{ms}% Produces the bibliography via BibTeX.

\end{document}